\documentclass[12pt]{article}
\usepackage[T1]{fontenc}
\usepackage{amsmath}
\usepackage{amsfonts}
\usepackage{graphicx}
\usepackage[mathscr]{euscript}
\usepackage{rotating}
\usepackage{pdflscape}
\usepackage[hmargin=2.54cm,vmargin=2.54cm]{geometry}
\usepackage{booktabs,caption}
\usepackage{natbib}
\usepackage{setspace}
\usepackage{bbm}
\usepackage{amsthm}
\usepackage{enumerate}
\usepackage{threeparttable}
\usepackage{verbatim}
\usepackage{caption}
\usepackage{url}
\usepackage{amssymb}
\usepackage{subcaption,array}
\usepackage{adjustbox}
\usepackage[titletoc,title]{appendix}

\usepackage[hidelinks,breaklinks]{hyperref}
\hypersetup{colorlinks=true,allcolors=blue}

\usepackage{framed}
\usepackage[dvipsnames]{xcolor}
\usepackage{bbm}

\newtheorem{prop}{Proposition}

\newtheorem{lemma}{Lemma}

\theoremstyle{definition}

\usepackage{xcolor}

\begin{document}

\title{Identification of Average Marginal Effects \\in Fixed Effects Dynamic Discrete Choice Models\thanks{We would like to thank Manuel Arellano, Dmitry Arkhangelsky, Stephane Bonhomme, Irene Botosaru, Juanjo Dolado, Juan Carlos Escanciano, Christophe Gaillac, Jiaying Gu, Francis Guiton, Bo Honore, Chris Muris, and Enrique Sentana for helpful comments and discussions. We have also benefited from the comments of seminar and conference participants at CEMFI, Toronto, Toulouse, 2020 Econometric Society World Congress, 2021 IAAE Congress, the Panel Data Workshop at Nuffield College, and the Conference in Honor of Manuel Arellano at the Bank of Spain. We thank Susumu Imai for generously sharing the dataset that we use in our empirical application. This research is supported by funding from the Social Sciences and Humanities Research Council. The second author gratefully acknowledges support from the \textit{Agencia Estatal de Investigaci\'{o}n del Gobierno de Espa\~{n}a}, grant RTI2018-095231-B-I00, and Comunidad de Madrid (Spain), grant EPUC3M11 (VPRICIT).}}

\author{Victor Aguirregabiria\thanks{Department of Economics, University of Toronto. 150 St. George Street, Toronto, ON, M5S 3G7, Canada, \href{mailto: victor.aguirregabiria@utoronto.ca}{victor.aguirregabiria@utoronto.ca}.}\\ \emph{University of Toronto, CEPR}\and Jes\'{u}s M. Carro\thanks{Department of Economics, Universidad Carlos III de Madrid. C./ Madrid, 126, 28903 Getafe (Madrid), Spain,
\href{mailto: jcarro@eco.uc3m.es}{jcarro@eco.uc3m.es}.}\\ \emph{Universidad Carlos III de Madrid}}

\date{July 1, 2024}

\maketitle

\begin{abstract}
    In nonlinear panel data models, fixed effects methods are often criticized because they cannot identify average marginal effects (AMEs) in short panels. The common argument is that identifying AMEs requires knowledge of the distribution of unobserved heterogeneity, but this distribution is not identified in a fixed effects model with a short panel. In this paper, we derive identification results that contradict this argument. In a panel data dynamic logit model, and for $T$ as small as three, we prove the point identification of different AMEs, including causal effects of changes in the lagged dependent variable or the last choice's duration. Our proofs are constructive and provide simple closed-form expressions for the AMEs in terms of probabilities of choice histories. We illustrate our results using Monte Carlo experiments and with an empirical application of a dynamic structural model of consumer brand choice with state dependence.

    \vspace{0.4cm} 
    
    
    \vspace{0.4cm} 
    
\end{abstract}

\thispagestyle{empty}

\newpage 

\setcounter{page}{1}

\begin{doublespacing}

\section{Introduction\label{sec:intro}}

Ignoring the correlation between unobserved heterogeneity and pre-determined explanatory variables in dynamic panel data models can generate significant biases in estimating dynamic causal effects. The literature distinguishes two approaches to deal with this issue. The \textit{random effects (RE)} approach integrates over the unobserved heterogeneity using a parametric assumption on the distribution of this heterogeneity conditional on
the initial values of the predetermined explanatory variables. Typically, this distribution cannot be identified nonparametrically in short panels, and random
effects approaches are not robust to the misspecification of parametric
restrictions. This issue is the so-called \textit{initial conditions problem}
(\citeauthor{heckman_1981}, \citeyear{heckman_1981}). In contrast,
\textit{fixed effects (FE)} approaches do not restrict this
distribution such that identifying parameters of interest is robust
to the misspecification of this primitive.

A limitation of FE methods in dynamic discrete choice models with short panels is that they cannot identify the distribution of the time-invariant unobserved heterogeneity. This limitation arises because the data comprises a finite number of probabilities—equivalent to the number of possible choice histories—while the distribution of unobserved heterogeneity has an infinite dimension. This identification problem has generated a more substantial criticism of FE approaches. The applied researcher is often interested in estimating the average marginal effects (AME) of changes in explanatory variables or structural parameters. Since these AMEs are expectations over the distribution of the unobserved heterogeneity, and this distribution is not identified, the common wisdom is that FE approaches cannot (point) identify AMEs.\footnote{Examples of recent papers describing this common wisdom are
\citeauthor{abrevaya_hsu_2021} (\citeyear{abrevaya_hsu_2021}) (on page 5:
\textit{"For `pure'\ fixed effects models, where the conditional distribution is left unspecified, identification of the partial effects described above would generally require $T\rightarrow \infty$."}) and \citeauthor{honore_depaula_2021} (\citeyear{honore_depaula_2021}) (on page 2:
\textit{"It is important to recognize that knowing $\beta$ [slope parameters] is typically not sufficient for calculating counterfactual distributions or marginal effects. Those will depend on the distribution of $\alpha_{i}$
[incidental parameters] as well as on $\beta$ and they are typically not point-identified even if $\beta$ is."})}

This paper presents new results on the point identification of AMEs in FE dynamic logit models. We prove the identification of the AME for a change in the lagged dependent variable, a crucial parameter in dynamic models that measures the causal effect of an agent's past decision on their current decision. Our constructive proofs provide simple closed-form expressions for AMEs based on the probabilities of choice histories in panels where the time dimension can be as small as $T = 3$.

In Section \ref{sec:binary}, we focus exclusively on binary choice models. Proposition \ref{prop_1_ident_AME} presents a straightforward identification result for dynamic binary logit models. However, it is limited to AMEs where exogenous covariates remain constant over time or are absent from the model. The main contributions of this paper are presented in Propositions \ref{prop_2_nec_suf_cond} and \ref{prop_3_iden_ame}. Proposition \ref{prop_2_nec_suf_cond} provides necessary and sufficient conditions for identifying a general type of AME in a broad class of discrete choice models. Proposition \ref{prop_3_iden_ame} applies these conditions to the dynamic binary logit model, showing the identification of AMEs without requiring covariates to remain constant over time, unlike the more straightforward result in Proposition \ref{prop_1_ident_AME}. In Section \ref{subsec:binary_exten}, we expand our analysis to include identification results for average transition probabilities (Proposition \ref{new_prop_4_iden_tranprob}), n-periods forward AMEs (Proposition \ref{new_prop_5}), and AMEs in models with duration dependence (Proposition \ref{new_prop_6_duration}). Section \ref{sec:mnl} presents our identification and non-identification results for dynamic multinomial logit models (Propositions \ref{new_prop_7_mnl} and \ref{new_prop_8_mnl_noiden}) and dynamic ordered logit models (Proposition \ref{iden_ordered_logit}).

This paper is related to a large literature on FE estimation of panel data discrete choice models pioneered by \citeauthor{rasch_1961}
(\citeyear{rasch_1961}), \citeauthor{andersen_1970}
(\citeyear{andersen_1970}), and \citeauthor{chamberlain_1980}
(\citeyear{chamberlain_1980}) for static models, and by \citeauthor{chamberlain_1985} (\citeyear{chamberlain_1985}) and
\citeauthor{honore_kyriazidou_2000} (\citeyear{honore_kyriazidou_2000}) for dynamic models. Most papers in this literature focus on identifying and estimating slope parameters and do not present identification results on AMEs. Some important exceptions are \citeauthor{bonhomme_2011} (\citeyear{bonhomme_2011}), 
\citeauthor{hoderlein_white_2012} (\citeyear{hoderlein_white_2012}), and \citeauthor{chernozhukov_fernandez_2013}
(\citeyear{chernozhukov_fernandez_2013}), and, more recently, \citeauthor{dobronyi_gu_2021} (\citeyear{dobronyi_gu_2021}), \citeauthor{davezies_2022} (\citeyear{davezies_2022}), \citeauthor{pakel_weidner_2023} (\citeyear{pakel_weidner_2023}), and \citeauthor{botosaru_muris_2024} (\citeyear{botosaru_muris_2024}).\footnote{\citeauthor{chamberlain_1984} (\citeyear{chamberlain_1984}),
\citeauthor{hahn_2001} (\citeyear{hahn_2001}), and more recently
\citeauthor{arellano_bonhomme_2017} (\citeyear{arellano_bonhomme_2017}), show
the identification of a few AMEs in FE nonlinear panel data models. However, these are AMEs for a particular subpopulation of individuals defined by the data. In contrast, we focus on identifying marginal effects defined as 
averages over the whole population of individuals.}

In an unpublished manuscript, \citeauthor{bonhomme_2011} (\citeyear{bonhomme_2011})  examines the identification of AMEs within a broad class of non-linear static panel data models, encompassing those with discrete or continuous dependent variables. His work establishes a sufficient condition for identifying AMEs in such models: an injective operator linking the distribution of unobserved heterogeneity to the distribution of the observed dependent variable. However, this condition does not hold in discrete choice models with short panels and continuous distribution of the unobserved heterogeneity. On the other hand, our Proposition \ref{prop_2_nec_suf_cond} establishes both a necessary and sufficient condition for identifying AMEs in a general class of dynamic panel data discrete choice models. Notably, our condition does not necessitate an injective relationship between the distributions of unobserved heterogeneity and the dependent variable. Moreover, we show that our condition holds in dynamic models where the injectivity condition is not satisfied. This includes models such as the autoregressive binary logit (Proposition \ref{prop_3_iden_ame}), binary logit with duration dependence (Proposition \ref{new_prop_6_duration}), multinomial logit (Proposition \ref{new_prop_7_mnl}), and ordered logit (Proposition \ref{iden_ordered_logit}).

\citeauthor{chernozhukov_fernandez_2013} (\citeyear{chernozhukov_fernandez_2013}) -- henceforth, CFHN -- study the identification of AMEs within nonparametric and semiparametric binary choice models. Their semiparametric model assumes that the transitory shock has a known distribution -- e.g., logit model -- and corresponds to the model we consider in this paper. They propose a computational method to estimate the bounds in the identified set of the AME. Through numerical examples, they reveal that while the bounds for AME can be notably wide within fully nonparametric models, they rapidly narrow as $T$ increases within the semiparametric model. In contrast to CFHN, our paper adopts a sequential identification strategy, the first step of which is the identification of slope parameters.\footnote{The sequential approach that we consider in this paper has been used, among others, by  \citeauthor{honore_depaula_2021}  (\citeyear{honore_depaula_2021}) [on page 2 of their paper]: \textit{"It seems that point- or set-identifying and estimating $\beta$ is a natural first step if one is interested in bounding, say, average marginal effects"}. Previous results on the identification of slope parameters in dynamic logit models include \citeauthor{chamberlain_1985} (\citeyear{chamberlain_1985}), \citeauthor{honore_kyriazidou_2000} (\citeyear{honore_kyriazidou_2000}), \citeauthor{magnac_2000} (\citeyear{magnac_2000}, \citeyear{magnac_2004}), \citeauthor{aguirregabiria_gu_2021} (\citeyear{aguirregabiria_gu_2021}), \citeauthor{honore_weidner_2020} (\citeyear{honore_weidner_2020}), \citeauthor{dobronyi_gu_2021} (\citeyear{dobronyi_gu_2021}), and
\citeauthor{honore_muris_2021} (\citeyear{honore_muris_2021}).} In the second stage, we take the slope parameters as known to the researcher and consider the identification of AMEs. Our Proposition \ref{prop_2_nec_suf_cond} establishes necessary and sufficient conditions for AME point identification. These conditions consist of infinite equations with a finite number of unknowns. Nonetheless, under the logistic structure, we demonstrate that the infinite system of equations can be effectively encapsulated within a finite framework. Specifically, the unobserved heterogeneity enters through a finite-order polynomial, and the infinite system of equations can be represented as a finite system in terms of the coefficients of each monomial term. We illustrate that this system of equations admits a solution and, through straightforward manipulations, yields closed-form expressions for the targeted AMEs. While CFHN's approach is computationally intensive, owing to the vast dimensionality of unobserved heterogeneity distribution, our method stands out for its computational simplicity as it provides closed-form expressions for AMEs. 

Some identification results in our paper are linked to those outlined in \citeauthor{dobronyi_gu_2021} (\citeyear{dobronyi_gu_2021}) (referred to as DGK hereafter). Within the framework of an autoregressive fixed effects binary logit model, DGK elucidate the sharp identified set of slope parameters, the probability distribution of unobserved heterogeneity, and various functionals derived from these parameters, such as average treatment effects. Their findings stem from a comprehensive examination of the full likelihood function of the model, revealing a polynomial structure in terms of the fixed effects. This polynomial structure also serves as the foundation for our identification results for average treatment effects.\footnote{As DGK acknowledge in section 4.1.1 of their paper, our finding of this polynomial structure predates their work.} However, it is essential to note that our derivations diverge from those presented in DGK. In particular, our Proposition \ref{prop_2_nec_suf_cond} establishes a necessary and sufficient condition for the point identification of average treatment effects in a general class of panel data discrete choice models beyond logit models. This condition consists of the existence of a solution to a system of equations with analytical expressions. Unlike DGK, we focus on point identification results and a sequential identification approach, where slope parameters are initially identified. In contrast, DGK explore both point and set identification and delineate the identified set of average marginal effects, even in cases where slope parameters lack point identification. Moreover, while DGK only investigate autoregressive binary logit models, our scope extends to identifying average marginal effects in multinomial and ordered models, where the unobserved heterogeneity is non-scalar. Additionally, we examine models with duration dependence, broadening the applicability of our identification framework across diverse model specifications.

\citeauthor{davezies_2022} (\citeyear{davezies_2022}), \citeauthor{pakel_weidner_2023}(\citeyear{pakel_weidner_2023}), and \citeauthor{botosaru_muris_2024} (\citeyear{botosaru_muris_2024}) investigate the estimation of AMEs within binary choice logit models, mainly when these parameters are only partially identified. Their research introduces computationally efficient inference methods regarding the identified set. Recognizing the complexities inherent in estimating the sharp identified set, these authors propose diverse approaches for inference on outer bounds for this set.

The remainder of the paper is organized as follows. First, in Section \ref{sec:binary},  we present our identification results within binary choice logit models. Next, in Section \ref{sec:mnl}, we investigate the identification of AMEs in multinomial and ordered discrete choice models. To illustrate our results, we conduct Monte Carlo experiments, detailed in Section \ref{sec:montecarlos}, and apply our methodology to a dynamic demand model using consumer scanner data, discussed in Section \ref{sec:application}. We summarize and conclude in section \ref{sec:conclusion}.

\section{Binary choice models \label{sec:binary}} 

\subsection{Model \label{subsec:binary_model}} 

Consider a panel dataset $\{y_{it},\mathbf{x}_{it}: i=1, 2, ..., N; t=1, 2, ..., T\}$ where $t$ represents time, $i$ denotes individuals, and $y_{it}$ can assume two values, $y_{it}\in \mathcal{Y} =\{0,1\}$. Our focus lies in the investigation of dynamic logit models using panel data. The following equation describes the binary choice model:
\begin{equation}
    y_{it} = 
    \mathbbm{1} \left\{  
        \text{ } \alpha_{i} + 
        \beta \text{ } y_{i,t-1} +
        \mathbf{x}_{it}^{\prime} 
        \text{ } \boldsymbol{\gamma} +
        \varepsilon_{it}
        \geq 0
        \text{ }
    \right \}  \text{.} 
    \label{eq_binary_model}
\end{equation}
where $\mathbbm{1}\{ . \}$ is the indicator function, $\beta$ and $\boldsymbol{\gamma}$ are parameters of interest, and $\boldsymbol{\alpha} \equiv (\alpha_{i}: i=1, 2, ...N)$ are incidental parameters. The unobservable $\varepsilon_{it}$ is i.i.d. over time and across individuals with a Logistic distribution. The explanatory variables in the $K \times 1$
vector $\mathbf{x}_{it}$ are strictly exogenous with respect to the transitory shock $\varepsilon_{it}$. In other words, for any pair of periods $(t,s)$, the
variables $\mathbf{x}_{it}$ and $\varepsilon_{is}$ are independently distributed.

The parameter $\alpha_{i}$ represents permanent unobserved heterogeneity in primitives affecting individuals' decisions, such as preferences or productivity. Its marginal distribution is denoted as $f_{\alpha}(\alpha_{i})$, and $f_{\alpha| \mathbf{x}^{\{1,T\}}}(\alpha
_{i}|\mathbf{x}_{i}^{\{1,T\}})$ represents the distribution of $\alpha_{i}$ conditioned on the history of $\mathbf{x}$ variables denoted as $\mathbf{x}_{i}^{\{1,T\}}=$ $(\mathbf{x}_{i1},\mathbf{x}_{i2},...,\mathbf{x}_{iT})$. These distributions are not subject to any constraints. Likewise, the probability of the initial choice $y_{i1}$ conditioned on $\alpha_{i}$ and $\mathbf{x}_{i}^{\{1,T\}}$ -— represented as $p^{\ast}(y_{i1}|\alpha_{i},\mathbf{x}_{i}^{\{1,T\}})$ -— is unrestricted. Following the standard
setting in fixed effect (FE) approaches, our identification results do not rely on any restrictions on the initial conditions. Assumption 1-BC summarizes the conditions in this binary choice model.

\bigskip

\noindent \textbf{ASSUMPTION 1-BC.} 
\textit{
    (A) (Logit) $\varepsilon_{it}$ is $i.i.d.$ over $(i,t)$ with Logistic distribution; (B) (Strict exogeneity of $\mathbf{x}_{it}$) variable $\varepsilon_{it}$ is independent of $\left(  \alpha_{i},\mathbf{x}_{i}^{\{1,T\}}\right)$; and (C) (Fixed effects) the density functions $f_{\alpha}(\alpha_{i})$, $f_{\alpha| \mathbf{x}^{\{1,T\}}}(\alpha_{i}|\mathbf{x}_{i}^{\{1,T\}})$, and $p^{\ast}(y_{i1}|\alpha_{i},\mathbf{x}_{i}^{\{1,T\}})$ are unrestricted.
} $\qquad \blacksquare$

\subsection{Average marginal effects \label{subsec:binary_ame}} 

For the definition of average marginal effects and other parameters of interest, it is convenient to define transition probabilities and their average versions. For $j,k\in \mathcal{Y} = \{0,1\}$, define the individual-specific transition probabilities:
\begin{equation}
    \pi_{kj}(\alpha_{i},\mathbf{x}) \equiv 
    \mathbb{P}\left(  
        y_{it}=j \text{ } | \text{ } \alpha_{i}, \text{ } y_{i,t-1}=k, \text{ } \mathbf{x}_{it}=\mathbf{x} 
    \right),
\label{indiv_trans_prob}
\end{equation}
where $\mathbf{x}$ is an arbitrary value chosen by the econometrician. In the binary choice model defined by equation \eqref{eq_binary_model} and Assumption 1-BC, we have that:
\begin{equation}
    \begin{array}[c]{rcl}  
    \pi_{11}(\alpha_{i},\mathbf{x}) =
    \Lambda 
    \left( 
        \alpha_{i} + \beta + 
        \mathbf{x}^{\prime} \boldsymbol{\gamma}
    \right) 
    & \text{ and } &   
    \pi_{01}(\alpha_{i},\mathbf{x}) =
    \Lambda 
    \left( 
        \alpha_{i} + 
        \mathbf{x}^{\prime} \boldsymbol{\gamma}
    \right)
    \end{array}
\end{equation}
where $\Lambda(u)$ is the Logistic function $e^{u}/[1+e^{u}]$. Let $\Delta(\alpha_{i}, \mathbf{x})$ be the \textit{individual-specific} causal effect on $y_{it}$ of a change in $y_{i,t-1}$ from $0$ to $1$.
\begin{equation}
    \begin{array}[c]{rcl}
        \Delta(\alpha_{i}, \mathbf{x})
        & \equiv & 
        \mathbb{E}\left(  
            y_{it} \text{ } | \text{ } \alpha_{i}, 
            \text{ } y_{i,t-1}=1, \text{ }
            \mathbf{x}_{it} = \mathbf{x}
        \right) -
        \mathbb{E}\left(  
            y_{it} \text{ } | \text{ } \alpha_{i}, \text{ } y_{i,t-1}=0, \text{ }
            \mathbf{x}_{it} = \mathbf{x}
        \right)  \\
        &  & \\
        & = & 
        \pi_{11}(\alpha_{i}, \mathbf{x}) - 
        \pi_{01}(\alpha_{i}, \mathbf{x}) = 
        \Lambda(\alpha_{i} + \mathbf{x}^{\prime} \boldsymbol{\gamma} + \beta) - 
        \Lambda(\alpha_{i} + \mathbf{x}^{\prime} \boldsymbol{\gamma}) \text{.}
    \end{array}
\label{def delta_i}
\end{equation}
This individual-specific causal effect represents the impact of $y_{i,t-1}$ on $y_{it}$. It measures the persistence of individual $i$ in state $1$ generated by \textit{true dynamics} or state dependence.

It is well-established in the literature that parameters $\beta$ and $\boldsymbol{\gamma}$ are identified in short panels (see section \ref{subsec:binary_iden_slopes} below). However, the individual effects $\alpha_{i}$ are not identified because of the incidental parameters problem (\citeauthor{neyman_scott_1948},
\citeyear{neyman_scott_1948}; \citeauthor{heckman_1981},
\citeyear{heckman_1981}; \citeauthor{lancaster_2000},
\citeyear{lancaster_2000}). Consequently, the individual-specific treatment effects are not identified. Instead, we study the identification of \textit{Average Marginal Effects} (AMEs) defined by integrating the individual effects $\Delta(\alpha_{i}, \mathbf{x})$ over the distribution of $\alpha_{i}$. 

In the model without $\mathbf{x}$ covariates (i.e., $\boldsymbol{\gamma}= \boldsymbol{0}$), the individual-specific effect is $\Delta(\alpha_{i}) = \Lambda(\alpha_{i} + \beta) - \Lambda(\alpha_{i})$, and the Average Marginal Effect is:
\begin{equation}
    AME = 
    {\displaystyle \int} 
    \left[ \Lambda(\alpha_{i} + \beta) - \Lambda(\alpha_{i}) \right] \text{ }
    f_{\alpha}( \alpha_{i} ) \text{ } d\alpha_{i}
\label{eq_def_AME_without_x}
\end{equation}
The sign of the parameter $\beta$ informs us about the sign of $AME$. However, the absolute magnitude of $\beta$ provides very limited insight into the magnitude of $AME$. Specifically, for any positive value $\beta$, we have that $AME$ can take virtually any value within the interval $(0,1)$ depending on the location of the distribution of $\alpha_{i}$. This limitation underscores the crucial importance of identifying Average Marginal Effects.

\bigskip

\noindent \textbf{EXAMPLE.} Suppose that $y_{it}$ is the indicator for firm $i$ being active in the market during period $t$. Consider the following thought experiment. At period $t-1$, firms are randomly split into two groups, say groups 0 and 1. Firms in group 0 are designated to be inactive, while those in group 1 are designated to be active. After one period, we examine the proportion of active firms in each group. The Average Marginal Effect defined in equation \eqref{eq_def_AME_without_x} is the proportion of active firms in group 1 minus the proportion of active firms in group 0. $\qquad \blacksquare$

\bigskip

We can extend this type of AME to the model with covariates (when  $\boldsymbol{\gamma} \neq \boldsymbol{0}$). In this scenario, the distribution of $\alpha_{i}$ depends on the individual's covariate history. Consequently, the AME is
not solely determined by the value of $\mathbf{x}$ at which we assess the individual effect $\Delta(\alpha_{i}, \mathbf{x})$, but also on the values of the covariates that we consider when integrating over the distribution of the individual effect. We provide identification results for two versions of these AMEs. The first AME is defined as an average conditional on a value of the whole history of the covariates between periods $1$ and $T$. For any chosen values of $\mathbf{x}$ and $\mathbf{x}^{\{1,T\}}$, determined by the researcher, we define the following AME:
\begin{equation}
    AME(\mathbf{x}, \mathbf{x}^{\{1,T\}}) \equiv 
    {\displaystyle \int} 
    \Delta(\alpha_{i}, \mathbf{x}) \text{ }
    f_{\alpha| \mathbf{x}^{\{1,T\}}}( \alpha_{i} \vert 
    \mathbf{x}_{i}^{\{1,T\}} = \mathbf{x}^{\{1,T\}}) \text{ } d\alpha_{i} 
\label{eq_def_AME}
\end{equation}
This parameter denotes the average value of the marginal effect $\Delta(\alpha_{i}, \mathbf{x})$ across the population of individuals with a covariate history $\mathbf{x}^{\{1,T\}}$. 

In Proposition \ref{prop_1_ident_AME}, we provide a straightforward proof of the identification of this AME under two specific conditions: either the model excludes covariates ($\boldsymbol{\gamma}=0$), or the covariate history $\mathbf{x}^{\{1,T\}}$ is such that the values remain constant $\mathbf{x}_{2} = \ldots = \mathbf{x}_{T} = \mathbf{x}$. Proposition \ref{prop_2_nec_suf_cond} establishes necessary and sufficient conditions for identifying parameter $AME(\mathbf{x}, \mathbf{x}^{\{1,T\}})$ within a broad class of models. By applying these conditions to the binary choice logit model, in Proposition \ref{prop_3_iden_ame}, we substantially extend the result of Proposition \ref{prop_1_ident_AME} to a broader context, demonstrating the identification of $AME(\mathbf{x}, \mathbf{x}^{\{1,T\}})$ for any arbitrary selection of $\mathbf{x}$ and $\mathbf{x}^{\{1,T\}}$ as determined by the researcher, provided that $\mathbf{x}_{T} = \mathbf{x}$.

Often, we are not particularly interested in the causal effect of $y_{t-1}$ on $y_{t}$ for a subpopulation of individuals with a specific history of covariates, but rather in the average of these effects across the entire distribution of covariate histories. Specifically, the following parameter represents the average value of the marginal effect $\Delta(\alpha_{i}, \mathbf{x})$ across the population of individuals with a value of the covariates equal to $\mathbf{x}$ at period $t$:\footnote{Since the joint distribution of $\alpha_{i}$ and $\mathbf{x}_{it}$ can change over time, this AME can vary with $t$.  We can use the identified time-specific AMEs to test the null hypothesis of stationarity of the distribution of $(\alpha_{i},\mathbf{x}_{it})$.} 
\begin{equation}
    AME_{t}(\mathbf{x}_{it} = \mathbf{x}) \equiv 
    {\displaystyle \int}
    \Delta(\alpha_{i}, \mathbf{x_{it}}) 
    \text{ }
    f_{\alpha \vert \mathbf{x}_{t}}(\alpha_{i} 
    \text{ } \vert \text{ } 
    \mathbf{x}_{it} = \mathbf{x}) 
    \text{ } d\alpha_{i}
\label{eq_def_AME_prop_2_3}
\end{equation}
Similarly, the following causal parameter is the average value of $\Delta(\alpha_{i}, \mathbf{x}_{it})$ across the joint distribution of $\alpha_{i}$ and $\mathbf{x}_{it}$ at period $t$:
\begin{equation}
    \widetilde{AME}_{t} \text{ } \equiv \text{ }
    {\displaystyle \int}
    \Delta(\alpha_{i}, \mathbf{x_{it}}) 
    \text{ }
    f_{\alpha,\mathbf{x}_{t}}(\alpha_{i}, \mathbf{x}_{it}) 
    \text{ } d\alpha_{i} \text{ } d \mathbf{x}_{it}
\label{eq_integrated_ame}
\end{equation}
Importantly, the AMEs defined in equations \eqref{eq_def_AME_prop_2_3} and \eqref{eq_integrated_ame} can be derived by integrating the AME presented in equation \eqref{eq_def_AME} over the empirical distribution of the covariate history. Specifically:
\begin{equation}
    \begin{array}[c]{rcl}
        AME_{t}(\mathbf{x}_{it}=\mathbf{x}) 
        & = &
        \mathbb{E}_{ \mathbf{x}_{i}^{\{1,T\}} \vert \mathbf{x}_{it}=\mathbf{x}}
        \left[ 
            AME(\mathbf{x}_{it}, \mathbf{x}_{i}^{\{1,T\}})
        \right]
        \\ \\
        \widetilde{AME}_{t} & = &
        \mathbb{E}_{ \mathbf{x}_{i}^{\{1,T\}} }
        \left[ 
            AME(\mathbf{x}_{it}, \mathbf{x}_{i}^{\{1,T\}})
        \right]
    \end{array}
\label{eq_integrated_AMEs}
\end{equation}
where $\mathbb{E}_{ \mathbf{x}_{i}^{\{1,T\}} } \left[ . \right]$ represents the expectation operator over the distribution of $\mathbf{x}_{i}^{\{1,T\}}$. Similarly, $\mathbb{E}_{ \mathbf{x}_{i}^{\{1,T\}} \vert \mathbf{x}_{it}=\mathbf{x}} \left[ . \right]$ is the expectation over the distribution of $\mathbf{x}_{i}^{\{1,T\}}$ conditional on $\mathbf{x}_{it}=\mathbf{x}$.

\subsection{Identification: Slope parameters \label{subsec:binary_iden_slopes}} 

We consider a sequential approach to the identification of AMEs. Slope parameters $\boldsymbol{\theta} \equiv (\beta, \boldsymbol{\gamma})$ are identified in a first step. In the second step, we treat $\boldsymbol{\theta}$ as known and focus on establishing the identification of AMEs. The identification of slope parameters in fixed effects dynamic logit models has been well-established in prior literature.\footnote{Specifically, these identification results on slope parameters include: \citeauthor{chamberlain_1985} (\citeyear{chamberlain_1985}) for the binary choice AR(1) model without exogenous regressors;
\citeauthor{magnac_2000} (\citeyear{magnac_2000}) for multinomial AR(1) models; \citeauthor{honore_kyriazidou_2000}
(\citeyear{honore_kyriazidou_2000}) for binary and multinomial models with exogenous regressors; \citeauthor{aguirregabiria_gu_2021}
(\citeyear{aguirregabiria_gu_2021}) for models with duration dependence;
\citeauthor{honore_weidner_2020} (\citeyear{honore_weidner_2020}) and
\citeauthor{dobronyi_gu_2021} (\citeyear{dobronyi_gu_2021}) for binary $AR(p)$
models with $p\geq2$; and \citeauthor{honore_muris_2021}
(\citeyear{honore_muris_2021}) for the dynamic ordered logit.} A crucial prerequisite for identification is that the panel must include at least three periods in addition to the initial condition. In models with covariates $\mathbf{x}_{it}$, the identification of $\beta$ necessitates the constancy of these covariates for a minimum of two consecutive periods, that is, $Pr(\mathbf{x}_{i2} - \mathbf{x}_{i3} = 0) > 0$. For the identification of $\boldsymbol{\gamma}$, an additional condition is necessary: the variance-covariance matrix $Var(\mathbf{x}_{i1} - \mathbf{x}_{i2} | \mathbf{x}_{i2} - \mathbf{x}_{i3}=0)$ must be full-column rank.

When the support of $\mathbf{x}_{it}$ is discrete, the conditions for consistent estimation of the slope parameters through a Conditional Maximum Likelihood (CML) method remain the same as those for identification. When the support of $\mathbf{x}_{it}$ is continuous, the conditions become somewhat more stringent, as elucidated in Theorem 1 by \citeauthor{honore_kyriazidou_2000} (\citeyear{honore_kyriazidou_2000}). For consistent estimation of $\beta$, we need the density function of $\mathbf{x}_{i2} - \mathbf{x}_{i3}$ to be strictly positive and continuous in a neighbourhood of zero (condition C3 in Theorem 1, \citeauthor{honore_kyriazidou_2000}, \citeyear{honore_kyriazidou_2000}). Achieving consistent estimation of $\boldsymbol{\gamma}$ in continuous scenarios necessitates the additional condition that the variance-covariance matrix $Var(\mathbf{x}_{i1} - \mathbf{x}_{i2} | \mathbf{x}_{i2} - \mathbf{x}_{i3} = u)$ is full-column rank for every value $u$ in a neighbourhood of zero (condition C6 in Theorem 1, \citeauthor{honore_kyriazidou_2000}, \citeyear{honore_kyriazidou_2000}).

\subsection{Identification of AME: A simple result \label{subsec:binary_iden_simple}} 

The distribution of the fixed effects is not non-parametrically identified in discrete choice models with short panels. Consequently, conventional wisdom dictates that functions dependent on this distribution, such as AMEs, remain non-identifiable. Contrary to this prevailing belief, our research challenges this notion, showing specific instances where AMEs are point-identified without necessitating full knowledge or imposing constraints on the fixed effects distribution. We begin by presenting a straightforward yet significant and novel result. Section \ref{subsec:binary_iden_ns} introduces a method grounded in novel necessary and sufficient conditions, yielding more general identification results.

Consider the binary choice model in equation \eqref{eq_binary_model}. Our proof of identification exploits a relationship between the individual effect $\Delta(\alpha_{i},\mathbf{x})$, the transition probabilities $\pi_{01}(\alpha_{i}, \mathbf{x})$ and $\pi_{11}(\alpha_{i}, \mathbf{x})$, and the parameter $\beta$ in the Logit model. The following Lemma 1 establishes this relationship.

\begin{lemma}
\label{new_lemma_1} 
\textit{
    In the binary choice model defined by equation \eqref{eq_binary_model} and Assumption 1-BC, the following conditions hold:
    \begin{equation}
        \Delta(\alpha_{i}, \mathbf{x}) =
        \left[  
            e^{\beta}  -1 
        \right]  \text{ } 
        \pi_{01}(\alpha_{i}, \mathbf{x}) 
        \text{ } 
        \pi_{10}(\alpha_{i}, \mathbf{x})
        \text{,}
    \label{eq_lemma_1_condition_1}
    \end{equation}
    and
    \begin{equation}
        e^{\beta}  =
        \dfrac{\pi_{11}(\alpha_{i}, \mathbf{x})
        \text{ } \pi_{00}(\alpha_{i}, \mathbf{x})}{\pi_{10}(\alpha_{i}, \mathbf{x})
        \text{ } \pi_{01}(\alpha_{i}, \mathbf{x})}
        \text{.}
        \qquad \blacksquare 
        \label{eq_lemma_1_condition_2}
    \end{equation}
}
\end{lemma}

\bigskip

\noindent \textit{Proof of Lemma \ref{new_lemma_1}.} See Appendix \ref{appendix_proof_lemma_1}.

\bigskip

Proposition \ref{prop_1_ident_AME} establishes the identification of $AME$ in the dynamic binary choice model without covariates, and of $AME(\mathbf{x}, [\mathbf{x}_{1}, \mathbf{x}, \mathbf{x}])$ in the model with covariates.

\bigskip

\begin{prop} 
\label{prop_1_ident_AME}
    Consider the binary choice model defined by equation \eqref{eq_binary_model}, under Assumption 1-BC, with a fixed $T \geq 3$ and known $\beta$. In the model without covariates (i.e., $\boldsymbol{\gamma}=0$), $AME$ is identified by the following expression:
    \begin{equation}
        AME = 
        \left[  
            e^{\beta}  -1
        \right]  
        \left[ 
            \mathbb{P}_{0,1,0} +
            \mathbb{P}_{1,0,1}
        \right]
    \label{eq_prop_1_iden_AME_no_x}
    \end{equation}
    where $\mathbb{P}_{y_{1},y_{2},y_{3}}$ is the probability of the choice history $(y_{1},y_{2},y_{3})$. In the model with covariates (i.e., $\boldsymbol{\gamma} \neq 0$), for any arbitrary values $\mathbf{x}$ and $\mathbf{x}_{1}$ chosen by the econometrician, the causal effect parameter $AME(\mathbf{x}, \mathbf{x}^{\{1,3\}} = [\mathbf{x}_{1}, \mathbf{x}, \mathbf{x}])$ is point identified by the following expression
    \begin{equation}
        AME(\mathbf{x}, \mathbf{x}^{\{1,3\}} = [\mathbf{x}_{1}, \mathbf{x}, \mathbf{x}]) = 
        \left[  
            e^{\beta}  -1
        \right]  
        \left[ 
            \mathbb{P}_{0,1,0 \vert \mathbf{x}_{1}, \mathbf{x}, \mathbf{x}} +
            \mathbb{P}_{1,0,1 \vert \mathbf{x}_{1}, \mathbf{x}, \mathbf{x}}
        \right]
    \label{eq_prop_1_iden_AME}
    \end{equation}
    where $\mathbb{P}_{y_{1},y_{2},y_{3} \vert \mathbf{x}^{\{1,3\}}}$ is the probability of the choice history $(y_{1},y_{2},y_{3})$ conditional on the covariate history $\mathbf{x}^{\{1,3\}}$. In equation \eqref{eq_prop_1_iden_AME}, $\mathbf{x}_{1}$ is free to take any value, while $\mathbf{x}_{2} = \mathbf{x}_{3} = \mathbf{x}$.
    $\qquad \blacksquare$
\end{prop}

\noindent \textit{Proof of Proposition
\ref{prop_1_ident_AME}.} See Appendix \ref{appendix_proof_prop_1}.

\bigskip

\noindent \textbf{Remark 1.1.} The proof of identification of $AME(\mathbf{x}, [\mathbf{x}_{1}, \mathbf{x}, \mathbf{x}])$ in Proposition \ref{prop_1_ident_AME} relies on the restriction $\mathbf{x}_{i2} = \mathbf{x}_{i3} = \mathbf{x}$. This constraint arises from the application of Lemma \ref{new_lemma_1}, which establishes $\Delta(\alpha_{i}, \mathbf{x}) = \left[ e^{\beta} -1 \right] \text{ } \pi_{01}(\alpha_{i}, \mathbf{x}) \text{ } \pi_{10}(\alpha_{i}, \mathbf{x})$. To relate this Lemma to the probability of a choice history $(y_{i1}, y_{i2}, y_{i3})$, it is necessary that $\mathbf{x}_{i2} = \mathbf{x}_{i3} = \mathbf{x}$. 

\bigskip

\noindent \textbf{Remark 1.2.} Since Proposition \ref{prop_1_ident_AME} applies to arbitrary values of $\mathbf{x}_{1}$ and $\mathbf{x}$, it enables the construction of the following integrated AME: 
\begin{equation}
    \widetilde{AME}_{\mathbf{x}_{2} =\mathbf{x}_{3}} 
    \text{ } = \text{ } 
    \displaystyle \int
    AME(\mathbf{x}, \mathbf{x}_{i}^{\{1,3\}}) \text{ }
    p\left( \mathbf{x}_{i}^{\{1,3\}} \vert \mathbf{x}_{i2} =\mathbf{x}_{i3} \right) \text{ }
    d \mathbf{x}_{i}^{\{1,3\}}
\label{eq_integrated_AME}
\end{equation}
where $p\left( \mathbf{x}_{i}^{\{1,3\}} \vert \mathbf{x}_{i2} =\mathbf{x}_{i3} \right)$ represents the empirical distribution of the covariate history conditional on $\mathbf{x}_{i2} =\mathbf{x}_{i3}$. While this is a relevant average causal parameter, its interest is limited by the constraint of covariate constancy between two consecutive periods, making it applicable only to that specific sub-population. In cases where $\mathbf{x}$ includes continuous covariates, this integrated AME applies to a subpopulation of mass zero. To address this limitation and obtain an AME without the restrictive condition $\mathbf{x}_{i2} = \mathbf{x}_{i3}$, we introduced a causal parameter denoted as $AME_{t}(\mathbf{x}_{it} = \mathbf{x})$, as specified in equation \eqref{eq_def_AME_prop_2_3}, and its integrated counterpart, $\widetilde{AME}_{t}$, as defined in equation \eqref{eq_integrated_ame}. Proposition \ref{prop_3_iden_ame} below establishes the identification of $AME_{t}(\mathbf{x}_{it} = \mathbf{x})$ and of $\widetilde{AME}_{t}$, providing an average causal effect applicable to the entire population.

\bigskip

\noindent \textbf{Remark 1.3} \textit{(Over-identification).} For panels where $T \geq 4$, there are over-identifying restrictions on $AME(\mathbf{x}, [\mathbf{x}_{1}, \mathbf{x}, \mathbf{x}])$ because we can construct the empirical distribution of 3-period histories using various groups of periods. As an illustration, when $T=4$, we can consider two groups: $(y_{1}, \mathbf{x}_{1}, y_{2}, \mathbf{x}_{2}, y_{3}, \mathbf{x}_{3})$ and $(y_{2}, \mathbf{x}_{2}, y_{3}, \mathbf{x}_{3}, y_{4}, \mathbf{x}_{4})$. For each of these groups, a distinct estimator of $AME(\mathbf{x}, [\mathbf{x}_{1}, \mathbf{x}, \mathbf{x}])$ can be derived, such that this AME is over-identified.

\bigskip

\noindent \textbf{Remark 1.4} \textit{(Estimation).} Equation \eqref{eq_prop_1_iden_AME} suggests a simple analog or plug-in estimator for the AMEs. In a first step, we estimate $\beta$ using Conditional Maximum Likelihood (CML) and the probabilities $\mathbb{P}_{y_{1},y_{2},y_{3} \vert \mathbf{x}^{\{1,3\}}}$ using a frequency estimator. Then, we plug these estimates in equation \eqref{eq_prop_1_iden_AME} to obtain estimates of
$AME(\mathbf{x}, [\mathbf{x}_{1}, \mathbf{x}, \mathbf{x}])$. When the support of $\mathbf{x}$ is discrete, this estimator is root$-N$ consistent and asymptotically normal. This estimator integrates slope parameter estimates with frequency estimates of choice histories, and when $T$ is greater than three, the set of three-period histories used to derive this estimator may contain overlapping observations. For these reasons, the bootstrap method emerges as a straightforward and convenient approach for inference on $AME(\mathbf{x}, [\mathbf{x}_{1}, \mathbf{x}, \mathbf{x}])$. Indeed, this is the methodology employed in our empirical application in Section \ref{sec:application}. For the estimation of the integrated $\widetilde{AME}_{\mathbf{x}_{2} =\mathbf{x}_{3}}$ defined in equation \eqref{eq_integrated_AME}, we can use a kernel-weighted estimator in line with \citeauthor{honore_kyriazidou_2000} (\citeyear{honore_kyriazidou_2000}). 

\subsection{Identification of AME: Necessary and sufficient conditions \label{subsec:binary_iden_ns}} 

The identification findings in the preceding section indicate that, despite the inherent challenges arising from the non-identification of fixed effects distribution, there exist instances where average marginal effects are identified with fixed $T$. However, these results prompt several questions. The proof establishes a specific linear combination (weighted sum) of probability choices that equates to the AME. The question then arises: How can one determine these weights? Is there a systematic procedure to assess the existence of a linear combination that identifies the AME in other models?

In this section, we derive necessary and sufficient conditions, developing a constructive approach to obtain AMEs within a broad class of dynamic logit models. Employing this method, we establish the identification of various AMEs or causal effects and unveil instances of non-identification for specific AMEs.

Let $\mathbf{y}^{\{1,T\}}_{i} \in \mathcal{Y}^{T}$ be the vector with individual $i$'s observed choice history, and let $\mathbf{x}^{\{1,T\}}_{i}\in \mathcal{X}^{T}$ represent the history of the exogenous variables. We use $\mathbf{y}_{i}^{\{2,T\}} \in \mathcal{Y}^{T-1}$ to denote the sub-history from period $2$ to $T$. For arbitrary histories $\mathbf{y}^{\{1,T\}}$ and $\mathbf{x}^{\{1,T\}}$, let $\mathbb{P}_{\mathbf{y}^{\{1,T\}}|\mathbf{x}^{\{1,T\}}}$ represent the probability of a choice history conditional on a covariate history. This probability is identified from the data. Let
$\boldsymbol{P}_{\mathcal{Y}^{T}|\mathcal{X}^{T}}$ be the vector with the probabilities $\mathbb{P}_{\mathbf{y}^{\{1,T\}}|\mathbf{x}^{\{1,T\}}}$ for every possible value of $\mathbf{y}^{\{1,T\}}$ and $\mathbf{x}^{\{1,T\}}$ in  $\mathcal{Y}^{T} \times \mathcal{X}^{T}$. Vector
$\boldsymbol{P}_{\mathcal{Y}^{T}|\mathcal{X}^{T}}$ contains all the information in the data that is relevant to identify slope parameters, the distribution of $\alpha$, and any AME of interest.

According to the model, probability $\mathbb{P}_{\mathbf{y}^{\{1,T\}}|\mathbf{x}^{\{1,T\}}}$ has the following structure:
\begin{equation}
    \mathbb{P}_{\mathbf{y}^{\{1,T\}} | \mathbf{x}^{\{1,T\}}} \text{ } = \text{ } 
    \int G\left(  
            \mathbf{y}^{\{2,T\}} 
            \text{ } | \text{ } 
            y_{1},  \mathbf{x}^{\{1,T\}}, 
            \alpha_{i}, \boldsymbol{\theta}
        \right)  
        \text{ } p^{\ast}(y_{1} | \alpha_{i}, \mathbf{x}^{\{1,T\}})
        \text{ } f_{\alpha| \mathbf{x}^{\{1,T\}}}(\alpha_{i} | \mathbf{x}^{\{1,T\}}) \text{ } d \alpha_{i},
\label{Observed Probability}
\end{equation}
where $G(\mathbf{y}^{\{2,T\}}| y_{1}, \mathbf{x}^{\{1,T\}}, \alpha_{i}, \boldsymbol{\theta
})$ is the probability of sub-history $\mathbf{y}^{\{2,T\}}$ predicted by the model. For instance, in the binary logit model:
\begin{equation}
    G\left(  \mathbf{y}^{\{2,T\}}| y_{1},  \mathbf{x}^{\{1,T\}}, \alpha_{i},
    \boldsymbol{\theta} \right)  
    \text{ } \equiv \text{ } 
    \prod_{t=2}^{T}
    \Lambda \left(  
        [2 y_{t} - 1] \text{ }
        [ \alpha_{i} + \beta y_{t-1} + \mathbf{x}_{t}^{\prime} \boldsymbol{\gamma}]
    \right)
    \label{eq_g_function_binary}
\end{equation}

In general, we can say that this $AME(\mathbf{x}, \mathbf{x}^{\{1,T\}})$ is point identified if there is a function $h(\boldsymbol{P}_{\mathcal{Y}^{T} | \mathcal{X}^{T}},\boldsymbol{\theta})$ such that $AME(\mathbf{x}, \mathbf{x}^{\{1,T\}}) = h(\boldsymbol{P}_{\mathcal{Y}^{T} | \mathcal{X}^{T}},\boldsymbol{\theta})$. Proposition \ref{prop_2_nec_suf_cond} states a necessary and sufficient condition for the point identification of this AME.

\bigskip

\begin{prop}  \label{prop_2_nec_suf_cond}
    Consider a fixed effects model represented by the probability function $G$. Let $AME(\mathbf{x}, \mathbf{x}^{\{1,T\}})$ denote the causal parameter as defined in equation \eqref{eq_def_AME} for arbitrary values of $\mathbf{x}$ and $\mathbf{x}^{\{1,T\}}$ chosen by the researcher. This $AME$ is point-identified if and only if there exists a weighting function $w_{\mathbf{y}^{\{1,T\}}, \mathbf{x}^{\{1,T\}}, \boldsymbol{\theta}}$, mapping from $\mathcal{Y}^{T} \times \mathcal{X}^{T} \times \Theta$ to $\mathbb{R}$, which satisfies the following equation:
    \begin{equation}
        {\displaystyle \sum 
        \limits_{\mathbf{y}^{\{2,T\}} \in \mathcal{Y}^{T-1}}}
        w_{y_{1},\mathbf{y}^{\{2,T\}}, \mathbf{x}^{\{1,T\}}, \boldsymbol{\theta}}
        \text{ }
        G\left(  \mathbf{y}^{\{2,T\}}| y_{1}, \mathbf{x}^{\{1,T\}}, \alpha_{i}, \boldsymbol{\theta} \right)  
        \text{ } = \text{ } 
        \Delta(\alpha_{i}, \mathbf{x}),
    \label{eq_prop_2_system_restrictions}
    \end{equation}
    for every $y_{1} \in \mathcal{Y}$ and $\alpha_{i} \in \mathbb{R}$. Furthermore, this condition implies the following form for the function that identifies $AME(\mathbf{x}, \mathbf{x}^{\{1,T\}})$:
    \begin{equation}
        AME(\mathbf{x}, \mathbf{x}^{\{1,T\}})
        \text{ } = \text{ } 
        {\displaystyle \sum
        \limits_{\mathbf{y}^{\{1,T\}} \in \mathcal{Y}^{T}}} 
        w_{\mathbf{y}^{\{1,T\}}, \mathbf{x}^{\{1,T\}}, \boldsymbol{\theta}}
        \text{ } \mathbb{P}_{\mathbf{y}^{\{1,T\}} |\mathbf{x}^{\{1,T\}}}
        \qquad \blacksquare 
        \label{eq_prop_2_weighted_sum}
    \end{equation}
\end{prop}

\noindent \textit{Proof of Proposition
\ref{prop_2_nec_suf_cond}.} See Appendix \ref{appendix_proof_prop_2}.

\bigskip

\noindent \noindent \textbf{Remark 2.1.} Proposition \ref{prop_2_nec_suf_cond} places no constraints on the form of function $G(.)$. Specifically, it does not require the structure outlined in equation 
\eqref{eq_g_function_binary}. Consequently, Proposition \ref{prop_2_nec_suf_cond} applies to a broad class of fixed-effects dynamic discrete choice models, extending beyond the logit class. 

\bigskip

\noindent \noindent \textbf{Remark 2.2.}
\citeauthor{bonhomme_2011} (\citeyear{bonhomme_2011}) delves into the identification of AMEs within a broad range of non-linear static panel data models, including those with discrete or continuous dependent variables. His work establishes a sufficient condition, albeit not necessary, for identifying AMEs in such models: the existence of an injective operator connecting the distribution of unobserved heterogeneity to the distribution of the observed dependent variable. However, this condition fails in discrete choice models with short panels and unobserved heterogeneity with continuous support. In contrast, our Proposition \ref{prop_2_nec_suf_cond} establishes both a necessary and sufficient condition for identifying AMEs in a general class of panel data discrete choice models. Notably, our condition does not require an injective relationship between the distributions of unobserved heterogeneity and the dependent variable.

\bigskip

\noindent \noindent \textbf{Remark 2.3.} Looking at the structure of the system of equations in  \eqref{eq_prop_2_system_restrictions}, one may be inclined to perceive that meeting the necessary and sufficient conditions for identification established in this Proposition seems quite unlikely. Specifically, note that \eqref{eq_prop_2_system_restrictions} establishes an infinite system of equations -- as many equations as values of $\alpha_{i}$. The researcher knows the closed-form expressions for functions $G(.)$ and
$\Delta(.)$. The unknown variables in
this system of equations consist of the weights $w_{\mathbf{y}^{\{1,T\}}, \mathbf{x}^{\{1,T\}}, \boldsymbol{\theta}}$ for every $\mathbf{y}^{\{1,T\}} \in \mathcal{Y}^{T}$. Given that the set $\mathcal{Y}^{T}$ is finite, we are in a scenario where the system presents an infinite number of restrictions with only a finite set of unknowns. In the absence of a specific structure, this system lacks a solution. 

\bigskip

Proposition \ref{prop_3_iden_ame} below establishes that, in the binary choice logit model defined by equation \eqref{eq_binary_model} and Assumption 1-BC, the structure of functions $G(.)$ and $\Delta(.)$ is such that equation \eqref{eq_prop_2_system_restrictions} can be represented as a finite order polynomial in the variable $e^{\alpha_{i}}$. This result
implies that there is a solution to the system if and only if the coefficients multiplying every monomial term in this polynomial are all equal to zero. This property transforms the infinite system of equations into a finite linear system with finite unknowns. Furthermore, if a solution exists, this solution
implies a closed-form expression for the weights $w_{\mathbf{y}^{\{1,T\}}, \mathbf{x}^{\{1,T\}}, \boldsymbol{\theta}}$, and therefore, for the expression that identifies the $AME$.

\bigskip

\begin{prop}
\label{prop_3_iden_ame} 
    In the binary choice model defined by equation \eqref{eq_binary_model} and Assumption 1-BC:
    
    \medskip
    (A) In Proposition \ref{prop_2_nec_suf_cond}, we can express equation \eqref{eq_prop_2_system_restrictions} as a finite-order polynomial in the variable $e^{\alpha_{i}}$. This formulation leads to a system characterized by a finite number of linear equations, where the unknowns are the weights $w_{\mathbf{y}^{\{1,T\}}, \mathbf{x}^{\{1,T\}}, \boldsymbol{\theta}}$ for every $\mathbf{y}^{\{1,T\}} \in \mathcal{Y}^{T}$. 

    \medskip
    (B) For $T \geq 3$, the causal parameter $AME(\mathbf{x}, \mathbf{x}^{\{1,T\}})$, defined in equation \eqref{eq_def_AME}, is identified for any value $\mathbf{x}$ and $\mathbf{x}^{\{1,T\}}$ chosen by the researcher with $\mathbf{x}_{T}=\mathbf{x}$. For instance, for $T=3$:
    \begin{equation}
        \begin{array}[c]{rcl}
            AME(\mathbf{x}, \mathbf{x}^{\{1,3\}})
            & = & 
            w_{0,0,1,\mathbf{x}^{\{1,3\}}}
            \text{ } 
            \mathbb{P}_{ 0,0,1 | \mathbf{x}^{ \{1,3\} } } +
            w_{0,1,0,\mathbf{x}^{\{1,3\}}}
            \text{ } 
            \mathbb{P}_{ 0,1,0 | \mathbf{x}^{ \{1,3\} } } 
            \\
            & + & 
            w_{1,0,1,\mathbf{x}^{\{1,3\}}}
            \text{ } 
            \mathbb{P}_{ 1,0,1 | \mathbf{x}^{ \{1,3\} } } +
            w_{1,1,0,\mathbf{x}^{\{1,3\}}}
            \text{ } 
            \mathbb{P}_{ 1,1,0 | \mathbf{x}^{ \{1,3\} } }
            \end{array}
    \label{eq_prop_3_ident_no_integrated_AME}
    \end{equation}
    where $\mathbb{P}_{ y_{1},y_{2},y_{3} | \mathbf{x}^{ \{1,3\} }}$
    is the probability of  $(y_{1},y_{2},y_{3})$ conditional on $\mathbf{x}^{\{1,3\}}$. The weights are:
    \begin{equation}
    \begin{array}[c]{ll}
        w_{0,0,1,\mathbf{x}^{\{1,3\}}} = 
        -1 +
        e^{ 
            [\mathbf{x}_{2}-\mathbf{x}_{3}]^{\prime} 
            \boldsymbol{\gamma} 
            } 
        \text{ } ; \text{ } 
        w_{0,1,0,\mathbf{x}^{\{1,3\}}} = 
        -1 +
        e^{ 
            \beta + 
            [\mathbf{x}_{3}-\mathbf{x}_{2}]^{\prime} 
            \boldsymbol{\gamma} 
            } 
        \text{ } ; \text{ } 
        & \\ & \\
        w_{1,0,1,\mathbf{x}^{\{1,3\}}} = 
        -1 +
        e^{ 
            \beta + 
            [\mathbf{x}_{2}-\mathbf{x}_{3}]^{\prime} 
            \boldsymbol{\gamma} 
            } 
        \text{ } ; \text{ }
        w_{1,1,0,\mathbf{x}^{\{1,3\}}} = 
        -1 +
        e^{ 
            [\mathbf{x}_{3}-\mathbf{x}_{2}]^{\prime} 
            \boldsymbol{\gamma} 
            } 
        \text{.} & \qquad \blacksquare
    \end{array}
    \label{eq_prop_3_weights}
    \end{equation}    
\end{prop}

\noindent \textit{Proof of Proposition
\ref{prop_3_iden_ame}.} See Appendix \ref{appendix_proof_prop_3}.

\bigskip

\noindent \textbf{Remark 3.1.} Two significant corollaries of Proposition \ref{prop_3_iden_ame} are the identification of the AMEs $AME_{t}(\mathbf{x}_{it}=\mathbf{x})$ and $\widetilde{AME}_{t}$, defined in equations \eqref{eq_def_AME_prop_2_3} and \eqref{eq_integrated_ame}, respectively. As established in equation \eqref{eq_integrated_AMEs}, both AMEs can be derived by integrating the AME identified in Proposition \ref{prop_3_iden_ame} over a distribution of observed covariate histories. Specifically, for $T=3$ and $t=3$:
\begin{equation}
    \begin{array}[c]{rcl}
        AME_{3}(\mathbf{x}_{i3}=\mathbf{x}) & = &
        \mathbb{E}_{ \mathbf{x}_{i}^{\{1,3\}} \vert \mathbf{x}_{i3}=\mathbf{x}}
        \left[ 
            AME(\mathbf{x}_{i3}, \mathbf{x}_{i}^{\{1,3\}})
        \right]
        \\ \\
        \widetilde{AME}_{3} & = &
        \mathbb{E}_{ \mathbf{x}_{i}^{\{1,3\}} }
        \left[ 
            AME(\mathbf{x}_{i3}, \mathbf{x}_{i}^{\{1,3\}})
        \right]
    \end{array}
\label{eq_prop_3_ident_integrated_AME}
\end{equation}

\bigskip

\noindent \textbf{Remark 3.2.} It is noteworthy to highlight the broader scope of the identification result presented in Proposition \ref{prop_3_iden_ame} compared to that in Proposition \ref{prop_1_ident_AME}. While Proposition \ref{prop_1_ident_AME} enables the identification of $AME(\mathbf{x}, \mathbf{x}^{\{1,3\}} = [\mathbf{x}_{1}, \mathbf{x}, \mathbf{x}])$ for any given value of $\mathbf{x}_{1}$ and $\mathbf{x}_{2} = \mathbf{x}_{3} = \mathbf{x}$ and allows us to integrate these AMEs across all values of $\mathbf{x}_{1}$ and $\mathbf{x}$, this "integrated AME" still carries the restriction of covariates at two consecutive periods, rendering this causal effect applicable only to a specific sub-population. The identified AME in Proposition \ref{prop_3_iden_ame}, and the implied identification of $AME_{t}(\mathbf{x}_{it}=\mathbf{x})$ and $\widetilde{AME}_{t}$, are not subject to that constraint.

\bigskip

\noindent \textbf{Remark 3.3.} When $T=3$ and the history of the exogenous covariates is such that 
$\mathbf{x}_{2} = \mathbf{x}_{3} = \mathbf{x}$, the identification result in Proposition \ref{prop_3_iden_ame}(B) is the one in Proposition \ref{prop_1_ident_AME}. That is, the weights that solve the system of equations \eqref{eq_prop_2_system_restrictions}  lead to the expression $AME(\mathbf{x})= \left[  e^{\beta}  -1\right]
$ $\left[  \mathbb{P}_{0,1,0|[\mathbf{x}_{1}, \mathbf{x}, \mathbf{x}]} + \mathbb{P}_{1,0,1|[\mathbf{x}_{1}, \mathbf{x}, \mathbf{x}]} \right]$. 

\bigskip

\noindent \textbf{Remark 3.4.} \textit{(Over-identification).} For panels with $T \geq 4$, there are over-identifying restrictions on the parameter $AME(\mathbf{x}, \mathbf{x}^{\{1,3\}})$. In Appendix \ref{appendix_proof_prop_3_Tgt3}, we employ the same procedure to derive the closed-form expression for the weights when $T$ exceeds three. Notably, for panels with $T \geq 4$, there emerge over-identifying restrictions on $AME(\mathbf{x}, \mathbf{x}^{\{1,T\}})$.

\bigskip

\noindent \textbf{Remark 3.5.} Proposition \ref{prop_3_iden_ame} places no constraints on the stochastic process of $\mathbf{x}_{it}$, except that it is strictly exogenous concerning the transitory shock $\varepsilon_{it}$. Furthermore, this identification result applies to scenarios where $\mathbf{x}_{it}$ includes continuous variables.

\bigskip

\noindent \textbf{Remark 3.6.} In the dynamic binary logit model, Proposition \ref{prop_3_iden_ame}(A) can be extended to AMEs where $\Delta(\mathbf{x}, \alpha_{i})$ represents the product, ratio, or a polynomial function of logit transition probabilities, provided these probabilities are evaluated at the same covariate values $\mathbf{x}$. Proposition \ref{prop_3_iden_ame}(B) addresses the existence of a solution to the finite system of equations, which needs to be determined individually for each case.

\subsection{Extensions \label{subsec:binary_exten}} 

\subsubsection{Other average treatment effects} 

In a binary choice model without covariates, for $k, j \in \{0,1\}$, we define $\Pi_{kj}$ as the average transition probability from $k$ to $j$ that results from integrating the individual-specific transition probability over the distribution of $\alpha_{i}$.
\begin{equation}
    \Pi_{kj} \equiv 
    \int \pi_{kj}(\alpha_{i}) 
    \text{ }
    f_{\alpha}(\alpha_{i}) 
    \text{ } d\alpha_{i} 
\label{average_trans_prob_no_x}
\end{equation}
For instance, $\Pi_{01} = \int \Lambda(\alpha_{i}) \text{ } f_{\alpha}(\alpha_{i}) \text{ } d\alpha_{i}$, and $\Pi_{11} = \int \Lambda(\alpha_{i} + \beta) \text{ } f_{\alpha}(\alpha_{i}) \text{ } d\alpha_{i}$. In the model with covariates, we define $\Pi_{kj}(\mathbf{x}, \mathbf{x}^{\{1,T\}})$ as the average transition probability from $k$ to $j$ with the current value of the covariates is $\mathbf{x}$ and integrated over the distribution of $\alpha_{i}$ conditional on the covariate history $\mathbf{x}^{\{1,T\}}$:
\begin{equation}
    \Pi_{kj}(\mathbf{x}, \mathbf{x}^{\{1,T\}}) 
    \equiv 
    \int \pi_{kj}(\alpha_{i}, \mathbf{x}) 
    \text{ }
    f_{\alpha \vert \mathbf{x}^{\{1,T\}}}
    (\alpha_{i} \text{ } \vert \text{ } \mathbf{x}^{\{1,T\}}_{i} = \mathbf{x}^{\{1,T\}}) 
    \text{ } d\alpha_{i} 
\label{average_trans_prob_with_x}
\end{equation}
Proposition \ref{new_prop_4_iden_tranprob} establishes the identification of these average transition probabilities.

\bigskip

\begin{prop}
\label{new_prop_4_iden_tranprob} 
    Consider the binary choice model defined by equation \eqref{eq_binary_model}, under Assumption 1-BC, with a fixed $T \geq 3$ and known $\beta$. In the model without covariates, all the average transition probabilities $\Pi_{00}$, $\Pi_{01}$, $\Pi_{10}$, and $\Pi_{11}$ by the following expressions:
    \begin{equation}
    \left \{
        \begin{array}[c]{rcl}
            \Pi_{11}
            & = & 
            \mathbb{P}_{1,1} + 
            \mathbb{P}_{0,1,1} + 
            e^{\beta} \text{ } \mathbb{P}_{0,1,0}
            \\ & & \\
            \Pi_{00} & = &  
            \mathbb{P}_{0,0} + \mathbb{P}_{1,0,0} + 
            e^{\beta} \text{ } \mathbb{P}_{1,0,1}
        \end{array}
    \right.
    \label{eq_new_prop_4_tranprob_no_x}
    \end{equation}
    and by definition, $\Pi_{10} = 1 -\Pi_{11}$ and $\Pi_{01} = 1 -\Pi_{00}$. In the model with covariates, for any arbitrary values $\mathbf{x}$ and $\mathbf{x}_{1}$ chosen by the econometrician, the average transition probabilities $\Pi_{00}(\mathbf{x}, [\mathbf{x}_{1}, \mathbf{x}, \mathbf{x}])$, $\Pi_{01}(\mathbf{x}, [\mathbf{x}_{1}, \mathbf{x}, \mathbf{x}])$, $\Pi_{10}(\mathbf{x}, [\mathbf{x}_{1}, \mathbf{x}, \mathbf{x}])$, and $\Pi_{11}(\mathbf{x}, [\mathbf{x}_{1}, \mathbf{x}, \mathbf{x}])$ are point identified by the following expressions:    
    \begin{equation}
    \left \{
        \begin{array}[c]{rcl}
            \Pi_{11}(\mathbf{x}, [\mathbf{x}_{1}, \mathbf{x}, \mathbf{x}]) 
            & = & 
            \mathbb{P}_{1,1 \vert \mathbf{x}_{1}, \mathbf{x}} + 
            \mathbb{P}_{0,1,1 \vert \mathbf{x}_{1}, \mathbf{x}, \mathbf{x}} + 
            e^{\beta} \text{ } \mathbb{P}_{0,1,0 \vert 
            \mathbf{x}_{1}, \mathbf{x}, \mathbf{x}}
            \\ & & \\
            \Pi_{00}(\mathbf{x}, [\mathbf{x}_{1}, \mathbf{x}, \mathbf{x}]) 
            & = &  
            \mathbb{P}_{0,0 \vert \mathbf{x}_{1}, \mathbf{x}} + \mathbb{P}_{1,0,0 \vert \mathbf{x}_{1}, \mathbf{x}, \mathbf{x}} + 
            e^{\beta} \text{ } \mathbb{P}_{1,0,1 \vert \mathbf{x}_{1}, \mathbf{x}, \mathbf{x}}
        \end{array}
    \right.
    \label{eq_new_prop_4_tranprob}
    \end{equation}
    where $\mathbf{x}_{1}$ is free to take any value, while $\mathbf{x}_{2} = \mathbf{x}_{3} = \mathbf{x}$. 
    $\qquad \blacksquare$
\end{prop}

\noindent \textit{Proof of Proposition
\ref{new_prop_4_iden_tranprob}.} See Appendix \ref{appendix_proof_prop_4}.

\bigskip

\noindent \textbf{Remark 4.1.} Using the definition of $AME$, it is straightforward to show that $AME = \Pi_{11} -\Pi_{01}$. Identifying the average transition probabilities is an alternative way of establishing the identification of the AME. This identification not only elucidates the AME but also extends to other relevant causal effects, encompassing the ratio $\Pi_{11}/\Pi_{01}$, the percentage change $[(\Pi_{11}-\Pi_{01})/\Pi_{01}]$, the additive effect $\Pi_{01}+\Pi_{11}$, a weighted sum of $\Pi_{01}$ and $\Pi_{11}$, or, more broadly, any well-defined function involving these parameters.

\bigskip

\noindent \textbf{Remark 4.2.} Proposition \ref{new_prop_4_iden_tranprob} implies also the identification of two interesting average treatment effects ($ATEs$). For concreteness, we describe these $ATEs$ using the application in the \textit{Example} above. Consider a policy experiment where firms in the experimental group are assigned to the active status at period $t-1$. For instance, they receive a large temporary subsidy to operate in the market. Firms in the control group are left in their observed status at period $t-1$. Then, at period $t$, the researcher observes the proportion of firms that remain active in the experimental and control groups. The difference between these two proportions is the average effect of this policy treatment, which we can denote as $ATE_{11,t}$. According to the model, this
average treatment effect is $ATE_{11,t} \equiv \Pi_{11} - \mathbb{E}(y_{it}|t)$, where $\mathbb{E}(y_{it}|t)$ is the mean value of $y$ in the actual distribution of this variable at period $t$. Since this distribution may change over time, this ATE may also vary with $t$. We can consider a similar experiment where firms in the experimental group are assigned to be inactive at period $t-1$ -- they receive a large temporary subsidy for being inactive. We use $ATE_{01,t}$ to denote the average effect of this other policy treatment. By definition, $ATE_{01,t} \equiv \Pi_{01} - \mathbb{E}(y_{it}|t)$.

\subsubsection{n-periods forward AME} 

Researchers can be interested in the response to a treatment over multiple periods. Let $\Delta^{(n)}(\alpha_{i})$ denote the individual-specific causal effect on $y_{i,t+n}$ resulting from a change in $y_{it}$ from 0 to 1.
\begin{equation}
    \Delta^{(n)}(\alpha_{i}) \equiv 
    \mathbb{E}\left(  y_{i,t+n} 
    \text{ } | \text{ }
    \alpha_{i}, y_{it}=1 \right)  - 
    \mathbb{E}\left(  y_{i,t+n} 
    \text{ } | \text{ } 
    \alpha_{i}, y_{it}=0 \right) = 
    \pi^{(n)}_{11}(\alpha_{i}) - 
    \pi^{(n)}_{01}(\alpha_{i}) \text{.} 
\label{def delta(n)_i}
\end{equation}
As previously mentioned in the context of $\Delta(\alpha_{i})$, the individual effect projected forward over $n$ periods remains unidentifiable in a short panel. Our focus, however, lies in the average the average of this effect:
\begin{equation}
    AME^{(n)} \equiv 
    {\displaystyle \int} 
    \Delta^{(n)}(\alpha_{i}) \text{ } 
    f_{\alpha}(\alpha_{i}) \text{ } d\alpha_{i} = {\displaystyle \int} 
    \left[
        \pi_{11}^{(n)}(\alpha_{i}) -
        \pi_{01}^{(n)}(\alpha_{i}) 
    \right]  
    f_{\alpha}(\alpha_{i}) \text{ } d\alpha_{i}
    \text{.}
\label{eq_def_AME_n}
\end{equation}

This $n$-periods forward average causal effect differs from raising the 1-period AME to the power of $n$: specifically, $AME^{(n)} \neq \lbrack AME \rbrack ^{n}$. Consequently, identifying $AME^{(n)}$ is not a straightforward implication from identifying the singular $AME$.

In the binary choice model with $\mathbf{x}$ covariates, we can define $n$-periods forward
AMEs conditional on a constant value of $\mathbf{x}$:
\begin{equation}
    \begin{array}[c]{ccl}
        AME^{(n)}(\mathbf{x}^{\{1,T\}}=[ \mathbf{x}, ..., \mathbf{x}]) & = & 
        {\displaystyle \int} 
        \left[
            \pi_{11}^{(n)}(\alpha_{i}, \mathbf{x} ) -
            \pi_{01}^{(n)}(\alpha_{i}, \mathbf{x}) 
        \right]  
        \text{ } 
        f_{\alpha|\mathbf{x}}
        (\alpha_{i}|\mathbf{x}_{i}^{\{1,T\}} =[\mathbf{x},...,\mathbf{x}])
        \text{ } d\alpha_{i}
    \end{array}
\end{equation}
Our proof of the identification of $AME^{(n)}$ builds on Lemma \ref{new_lemma_1} and the following Lemma \ref{new_lemma_2}.

\bigskip

\begin{lemma}
\label{new_lemma_2} 
    In the binary choice model defined by equation \eqref{eq_binary_model} and Assumption 1-BC, the $n$-periods forward individual causal effect with constant-over-time $\mathbf{x}$ satisfies the following equation:
    \begin{equation}
        \Delta^{(n)}(\alpha_{i}, \mathbf{x}) =
        \left[  e^{\beta} - 1 \right]  ^{n} \text{ }
        \left[
            \pi_{10}(\alpha_{i}, \mathbf{x})
        \right]  ^{n} \text{ } 
        \left[  
            \pi_{01}(\alpha_{i}, \mathbf{x})
        \right]^{n}
        \text{.}
        \qquad \blacksquare 
    \label{eq_lemma_indiv_nper}
    \end{equation}
\end{lemma}

\bigskip

\noindent \textit{Proof of Lemma \ref{new_lemma_2}}. See Appendix \ref{appendix_proof_lemma_2_nperiod}.

\bigskip

\begin{prop}
\label{new_prop_5} 
    Consider the binary choice model defined by equation \eqref{eq_binary_model} and Assumption 1-BC. Let $n$ be any positive integer, and let $\widetilde{\mathbf{10}}^{n}$ be the choice history that consists of the n-times repetition of the sequence (1,0), e.g., for $n=2$, we have that $\widetilde{\mathbf{10}}^{2}=(1,0,1,0)$. If $\mathit{T\geq2n+1}$, then parameter $AME^{(n)}(\mathbf{x}^{\{1,T\}}=[ \mathbf{x}, ..., \mathbf{x}])$ is identified as:
    \begin{equation}
        AME^{(n)}(\mathbf{x}^{\{1,T\}}=[ \mathbf{x}, ..., \mathbf{x}]) = 
        \left[ e^{\beta}-1 \right]^{n}
        \text{ } 
        \left[
            \mathbb{P}_{0, \widetilde{\mathbf{10}}^{n} \vert \mathbf{x}, ..., \mathbf{x}} + \mathbb{P}_{\widetilde{\mathbf{10}}^{n},1 \vert \mathbf{x}, ..., \mathbf{x}}
        \right]  
    \label{Proposition 2: Identification AME(n)}
    \end{equation}
    where $\mathbb{P}_{0,\widetilde{\mathbf{10}}^{n} \vert \mathbf{x}, ..., \mathbf{x}}$ and $\mathbb{P}_{\widetilde{\mathbf{10}}^{n},1 \vert \mathbf{x}, ..., \mathbf{x}}$ are the probabilities of choice histories $(0,\widetilde{\mathbf{10}}^{n})$ and $(\widetilde{\mathbf{10}}^{n},1)$, respetively, conditional on $\mathbf{x}_{1} = ... = \mathbf{x}_{T} = \mathbf{x}$. 
    \qquad$\blacksquare$
\end{prop}

\bigskip

\noindent \textit{Proof of Proposition \ref{new_prop_5}.} See Appendix \ref{appendix_proof_prop_5_AMEn}.

\subsubsection{Duration dependence} 

There are many applications of dynamic models where the dependent variable has duration dependence. For instance, in a model of individual employment, where $y=1$ represents employment and $y=0$ unemployment, a worker's productivity often rises with job experience, suggesting that the likelihood of employment increases over the duration in that state. Similarly, in a market entry and exit model, where $y=1$ denotes a firm's activity in the market, and $y=0$ denotes inactivity, a firm's profit may rise with its market experience, increasing the probability that the firm stays active.

Let $d_{it}\in \{0, 1, 2, ...\}$ represent the
duration in the choice of alternative $y=1$. This duration variable can be defined recursively as $d_{it} = y_{i,t-1} \text{ } (d_{i,t-1}+1)$. We consider the following dynamic binary choice logit model with duration dependence:
\begin{equation}
    \begin{array}[c]{ccl}
        y_{it} & = & 
        1 \left \{  
            \text{ } \alpha_{i} +
            \beta \text{ } y_{i,t-1} +
            \delta \text{ } d_{it} +
            \varepsilon_{it} \geq 0 
        \right \},
    \end{array}
\label{eq_binary_duration_model}
\end{equation}
where $\beta$ and $\delta$ are parameters of interest.\footnote{We can extend our identification results of AMEs for models with duration dependence to allow for a more flexible specification with a different parameter for each possible value of duration greater than $1$. That is, we can replace the term $\delta \text{ } d_{it}$ with $\delta_{2} \text{ } \mathbbm{1}\{d_{it} = 2\} + \delta_{3} \text{ } \mathbbm{1}\{d_{it} = 3\} + ... $ where  $\mathbbm{1}\{ . \}$ is the indicator function, and $\delta_{2}$, $\delta_{3}$, ... are parameters.} 

We are interested in the causal effect on $y_{it}$ of a change in the duration variable $d_{it}$. For instance, in a market entry/exit model, we can be interested in the causal effect of one more year of experience on the probability of being active in the market. Let $\Delta_{d}(\alpha_{i},d)$ be the \textit{individual-specific} causal effect on $y_{it}$ of a unit change in duration from $d$ to $d+1$.
\begin{equation}
    \begin{array}[c]{ccl}
        \Delta_{d}(\alpha_{i},d)
        & \equiv & 
        \mathbb{E}\left(
            y_{it} \text{ } | \text{ } \alpha_{i}, d_{it} = d+1 
        \right)  -
        \mathbb{E}\left(  
            y_{it} \text{ } | \text{ } \alpha_{i}, d_{it} = d 
        \right) 
        \\ &  & \\
        & = & 
        \pi_{11}(\alpha_{i}, d+1) -
        \pi_{11}(\alpha_{i}, d) = 
        \Lambda \left(
            \alpha_{i} + \beta + \delta \text{ } (d+1)
        \right) -
        \Lambda \left(
            \alpha_{i} + \beta + \delta \text{ } d
        \right)
        \text{.}
    \end{array}
\label{eq_def_delta_dura_i}
\end{equation}
where $\pi_{11}(\alpha_{i}, d) \equiv \mathbb{E}(y_{it}|\alpha_{i}, y_{i,t-1}=1, d_{it}=d)$. We are interested in the identification of the following \textit{AME}:
\begin{equation}
    \begin{array}[c]{ccl}
        AME_{d}(d) & \equiv & 
        {\displaystyle \int} 
        \Delta_{d}(\alpha_{i},d) \text{ } 
        f_{\alpha}(\alpha_{i}) \text{ }  d\alpha_{i} = {\displaystyle \int} 
        \left[  
            \pi_{11}(\alpha_{i}, d+1) -
            \pi_{11}(\alpha_{i}, d)
        \right] \text{ }
        f_{\alpha}(\alpha_{i}) \text{ } d\alpha_{i}
    \end{array}
\label{eq_def_AME_with_duration}
\end{equation}

In this binary choice model with duration dependence, we need to slightly modify Assumption 1-BC to consider that the initial condition of a choice history includes not only $y_{i1}$ but also the initial duration $d_{i1}$. Therefore, the density function of the initial condition is $p^{\ast}(y_{i1}, d_{i1} | \alpha_{i})$. 

Proposition \ref{new_prop_6_duration} establishes the identification of $AME_{d}(d)$.

\bigskip 

\begin{prop}
\label{new_prop_6_duration} Consider the binary choice model with duration dependence defined by equation \eqref{eq_binary_duration_model} and Assumption 1-BC. For $T \geq 3 + d$, the causal effect $AME_{d}(d)$ is identified for any $d \geq 1$. For instance, with $T\geq 4$, the causal effect $AME_{d}(1)$ is identified as:
\begin{equation}
    \begin{array}[c]{lll}
        AME_{d}(1) & = & 
        \displaystyle \frac{e^{\beta + 2 \delta} -e^{\beta + \delta}}{2} \text{ }
        \left[      
            \mathbb{P}_{0,0,1,0} +
            \mathbb{P}_{0,1,0,0}
        \right] +
        \displaystyle \frac{e^{\beta + 2 \delta} -e^{\beta + \delta}}{e^{\beta + \delta}} 
        \text{ } 
        \mathbb{P}_{0,0,1,1} 
        \\ & & \\ 
        & + & 
        \left(  
            \displaystyle 
            \frac{
                e^{\beta + 2 \delta}
                \left(1-e^{\beta + 2 \delta}\right)
            }
            {e^{\beta + \delta}} + 
            e^{\beta + 2 \delta}-1 
        \right)  
        \mathbb{P}_{0,1,1,0}
        \\ & & \\ 
        & + & 
        \left(  
            1 - \displaystyle 
            \frac{e^{\beta + \delta}}
            {e^{\beta + 2 \delta}}
        \right)
        \left[
            \mathbb{P}_{1,0,1,0} + 
            \mathbb{P}_{1,0,1,1}
        \right]  +
        \left(
            \displaystyle 
            \frac{e^{\beta + 2 \delta}-1}
            {e^{\beta + \delta}} - 1 +
            \displaystyle 
            \frac{1}{e^{\beta + 2 \delta}}
        \right)  
        \mathbb{P}_{1,1,0,0}. 
        \qquad \blacksquare
    \end{array}
\label{AME_d=1_to_d=2}
\end{equation}
 
\end{prop}

\bigskip

\noindent \textit{Proof of Proposition \ref{new_prop_6_duration}.} See Appendix \ref{sec:proof_prop_5}.

\section{Multinomial and ordered choice models \label{sec:mnl}} 

\subsection{Multinomial choice model \label{subsec:mnl}} 

Consider a panel dataset $\{ y_{it}, \mathbf{x}_{it}: i=1, 2, ..., N; t=1, 2,
..., T \}$ where $y_{it}$ can take $J+1$ values: $y_{it}\in \mathcal{Y} = \{0,1,...,J\}$. We can interpret the dependent variable as the choice alternative that maximizes a utility or payoff function. More specifically:
\begin{equation}
    y_{it} = \arg \max_{j\in \mathcal{Y}} \text{ } 
    \left \{  
        \alpha_{i}(j) +
        \beta_{j} \text{ } 
        \mathbbm{1}\{ y_{i,t-1} = j \} +
        \mathbf{x}_{it}^{\prime} \text{ } \boldsymbol{\gamma}_{j} +
        \varepsilon_{it}(j)\right 
    \}  \text{,} 
\label{eq_MNL_model}
\end{equation}
where $\{ \beta_{j}: j \in \mathcal{Y} \}$ and $\{ \boldsymbol{\gamma}_{j}: j \in \mathcal{Y}\}$ are parameters of interest, and $\boldsymbol{\alpha}_{i} \equiv \{ \alpha_{i}(j): j\in \mathcal{Y}\}$ are incidental parameters. The unobservables $\{ \varepsilon_{it}(j):j\in \mathcal{Y}\}$ are i.i.d. type 1 extreme value.  The explanatory variables in the $K \times 1$ vector $\mathbf{x}_{it}$ are strictly exogenous concerning the transitory shocks $\varepsilon_{it}(j)$. 

Parameter $\beta_{j}$ represents the impact on the utility of staying in the same choice as in the previous period. It can be interpreted as the effect of habits or as a negative switching cost. The $\beta_{j}$ and $\gamma_{j}$ parameters can be identified only relative to the value of these parameters for a baseline choice alternative. We adopt the conventional normalization of $\beta_{0} = 0$ and $\gamma_{0}=0$, such that the identified parameters are actually $\beta_{j} - \beta_{0}$ and $\gamma_{j} - \gamma_{0}$. The vector $\boldsymbol{\alpha}_{i} \equiv [\alpha_{i}(0), \alpha_{i}(1), ..., \alpha_{i}(J)]$ represents unobserved individual heterogeneity in payoffs. Assumption 1-MNL summarizes
the conditions in this model.

\bigskip

\noindent \textbf{Assumption 1-MNL.} 
\textit{
    (A) (Logit) $\varepsilon_{it}(j)$ is $i.i.d.$ over $(i,t,j)$ with type 1 extreme value distribution; (B) (Strict exogeneity of $\mathbf{x}_{it}$) the variables $\varepsilon_{it}(j)$ are independent of $\left(  \boldsymbol{\alpha}_{i}, \mathbf{x}_{i}^{\{1,T\}} \right)$; and (C) (Fixed effects) the density functions $f_{\alpha}(\boldsymbol{\alpha}_{i})$, $f_{\alpha| \mathbf{x}^{\{1,T\}}}(\boldsymbol{\alpha}_{i} | \mathbf{x}_{i}^{\{1,T\}})$, and $p^{\ast}(y_{i1} | \boldsymbol{\alpha}_{i}, \mathbf{x}_{i}^{\{1,T\}})$ are unrestricted. $\qquad \blacksquare$
}

\bigskip

In this multinomial model, we can define individual-specific transition probabilities in the same way as we did above in equation \eqref{indiv_trans_prob} for the binary choice model. For any $j,k \in \mathcal{Y}$, and any value $\mathbf{x} \in \mathcal{X}$, $\pi_{kj}(\boldsymbol{\alpha}_{i}, \mathbf{x}) \equiv 
\mathbb{P}\left(  y_{i,t+1}=j \text{ } | \text{ } \boldsymbol{\alpha}_{i}, \text{ } y_{it}=k, \text{ } \mathbf{x}_{i,t+1}=\mathbf{x} \right)$. For this multinomial logit, the transition probability has the following form:
\begin{equation}
    \pi_{kj}(\boldsymbol{\alpha}_{i}, \mathbf{x})
    =
    \frac{
    \exp\{ 
        \alpha_{i}(j) + 
        \beta_{j} \text{ } \mathbbm{1}\{j=k\} + 
        \mathbf{x}^{\prime} \boldsymbol{\gamma}_{j}
    \}
    }
    { \sum_{\ell=0}^{J} 
    \exp\{ 
        \alpha_{i}(\ell) + 
        \beta_{\ell} \text{ } \mathbbm{1}\{\ell=k\} + 
        \mathbf{x}^{\prime} \boldsymbol{\gamma}_{\ell}
    \}
    }
\label{eq_trans_prob_mnl}
\end{equation}
We also define $\Pi_{kj}(\mathbf{x})$ as the \textit{Average Transition Probability} (ATP) from $k$ to $j$ in the same way as we did for the binary choice model in equation \eqref{average_trans_prob_with_x}. For any $j,k \in \mathcal{Y}$, any value $\mathbf{x} \in \mathcal{X}$, and any history of covariates $\mathbf{x}^{\{1,T\}}$, we have that $\Pi_{kj}(\mathbf{x}, \mathbf{x}^{\{1,T\}}) \equiv \int \pi_{kj}(\boldsymbol{\alpha}_{i}, \mathbf{x}) \text{ } f_{\alpha| \mathbf{x}^{\{1,T\}}}( \boldsymbol{\alpha}_{i} \text{ } | \text{ } \mathbf{x}_{i}^{\{1,T\}}=\mathbf{x}^{\{1,T\}}) 
\text{ } d \boldsymbol{\alpha}_{i}$.

For any values $j, k, \ell \in \mathcal{Y}$ with $k \neq \ell$, let $\Delta_{j, k \rightarrow \ell}(\boldsymbol{\alpha}_{i}, \mathbf{x})$ be the
\textit{individual-specific} causal effect on the probability of $y_{it}=j$ of a change in $y_{i,t-1}$ from $k$ to $\ell$ given that $\mathbf{x}_{it} = \mathbf{x}$. 
\begin{equation}
    \begin{array}[c]{rcl}
    \Delta_{j, k \rightarrow \ell}(\boldsymbol{\alpha}_{i}, \mathbf{x}) 
    & \equiv & 
    \mathbb{E}\left( 
        \mathbbm{1}\{y_{it}=j\} \text{ } | \text{ }
        \boldsymbol{\alpha}_{i}, \mathbf{x}, 
        y_{i,t-1}=\ell
    \right)  -
    \mathbb{E}\left(  
        \mathbbm{1}\{y_{it}=j \} \text{ } | \text{ }
        \boldsymbol{\alpha}_{i}, \mathbf{x}, 
        y_{i,t-1}=k
    \right)  
    \\ & & \\
    & = &
    \pi_{\ell j}(\boldsymbol{\alpha}_{i}, \mathbf{x}) - \pi_{kj}(\boldsymbol{\alpha}_{i}, \mathbf{x})
    \text{.} 
    \end{array}
\label{eq_def_delta_i_mnl}
\end{equation}
We are interested in the identification of the following AMEs:
\begin{equation}
    \begin{array}[c]{rcl}
        AME_{j, k\rightarrow \ell}
        ( \mathbf{x}, \mathbf{x}^{\{1,T\}}) 
        & \equiv &
        {\displaystyle \int} 
        \Delta_{j, k \rightarrow \ell}
        (\boldsymbol{\alpha}_{i}, \mathbf{x}) \text{ } 
        f_{\alpha| \mathbf{x}^{\{1,T\}}}(\boldsymbol{\alpha}_{i} | 
        \mathbf{x}_{i}^{\{ 1,T\}} = \mathbf{x}^{\{ 1,T\}}) 
        \text{  } d \boldsymbol{\alpha}_{i} 
        \\
        & = & 
        \Pi_{\ell j}(\mathbf{x}, \mathbf{x}^{\{ 1,T\}})  -
        \Pi_{k j}(\mathbf{x}, \mathbf{x}^{\{ 1,T\}})
    \end{array}
\label{eq_def_AME_mnl}
\end{equation}

\subsection{Identification in multinomial model
\label{sec: ident PI_jj}}

To obtain our identification result for the multinomial model in Proposition \ref{new_prop_7_mnl}, we apply the following Lemma \ref{new_lemma_3}, which is an extension of Lemma \ref{new_lemma_1} to the multinomial case.

\bigskip

\begin{lemma}
\label{new_lemma_3} 
    In the model defined by equation  \eqref{eq_MNL_model} and Assumption 1-MNL, for any triple of choice alternatives $j$, $k$, $\ell$, with $k \neq \ell$,  the following condition holds:
    \begin{equation}
        e^{ 
            \beta_{j} \text{ }
            \left[
                1 - \mathbbm{1}\{k=j\} - 
                \mathbbm{1}\{\ell=j\}
            \right]
        }  = 
        \dfrac{\pi_{k \ell}(\boldsymbol{\alpha}_{i}, \mathbf{x}) \text{ } 
        \pi_{j j}(\boldsymbol{\alpha}_{i}, \mathbf{x})} {\pi_{k j}(\boldsymbol{\alpha}_{i}, \mathbf{x}) \text{ }
        \pi_{j \ell}(\boldsymbol{\alpha}_{i}, \mathbf{x})} \text{.} \qquad \blacksquare
    \label{eq_lemma_3_mnl}
    \end{equation}
\end{lemma}

\bigskip

\noindent \textit{Proof of Lemma \ref{new_lemma_3}.} See Appendix \ref{appendix_proof_lemma_3_mnl}.

\bigskip

Proposition \ref{new_prop_7_mnl} establishes the identification of the average transition probabilities $\Pi_{jj}(\mathbf{x}, \mathbf{x}^{\{1,T\}})$ for covariate histories with $\mathbf{x}_{2} = ... = \mathbf{x}_{T}$.

\bigskip

\begin{prop}
\label{new_prop_7_mnl} 
    In the model defined by equation  \eqref{eq_MNL_model} and Assumption 1-MNL, for any choice alternative $j$, any $\mathbf{x} \in \mathcal{X}$, and covariate histories with $\mathbf{x}_{2} = ... = \mathbf{x}_{T} = \mathbf{x}$, if $T \geq3$, the average transition probability $\Pi_{jj}(\mathbf{x}, \mathbf{x}^{\{1,T\}})$ is identified. For instance, for $T=3$:
    \begin{equation}
        \begin{array}[c]{rcl}
            \Pi_{jj}(\mathbf{x}, \mathbf{x}^{\{1,3\}}) 
            & = & 
            \mathbb{P}_{jj | \mathbf{x}^{\{1,2\}}} +
            {\displaystyle \sum
            \limits_{k\neq j}} 
            \left[   
                \mathbb{P}_{kjj | \mathbf{x}^{\{1,3\}} } +
                {\displaystyle 
                \sum \limits_{\ell \neq j}} 
                e^{   
                    \beta_{j} \text{ }
                    \left[
                        1 - \mathbbm{1}\{k=j\} - 
                        \mathbbm{1}\{\ell=j\}
                    \right]
                }
                 \text{ } 
                 \mathbb{P}_{kj\ell | \mathbf{x}^{\{1,3\}} } 
            \right]  \mathit{,}
        \end{array}
    \label{eq:iden_PI_jj}
    \end{equation}
    where $\mathbf{x}^{\{1,3\}}$ is such that $\mathbf{x}_{2} = \mathbf{x}_{3} = \mathbf{x}$. $\qquad \blacksquare$
\end{prop}

\bigskip

\noindent \textit{Proof of Proposition \ref{new_prop_7_mnl}.}  See Appendix \ref{appendix_prop_prop_7_mnl_tranprob}.

\bigskip

Unfortunately, the procedure described in the proof of Proposition \ref{new_prop_7_mnl} does not provide an identification result for the parameters $\Pi_{jk}$ with $j \neq k$ when the number of choice alternatives is greater than two. Based on the necessary and sufficient conditions in Proposition \ref{prop_2_nec_suf_cond}, the following Proposition \ref{new_prop_8_mnl_noiden} establishes that this identification is not possible when $J+1 \geq 3$.

\bigskip 

\begin{prop}
\label{new_prop_8_mnl_noiden} 
    In the multinomial model defined by equation  \eqref{eq_MNL_model} and Assumption 1-MNL, with $J+1 \geq 3$ and $\boldsymbol{\theta}$ known to the researcher, there is no function $h\left(  \mathbf{P}_{\mathcal{Y}|\mathcal{X}}, \boldsymbol{\theta} \right)$ that equal the average transition probability $\Pi_{kj}$ for $k \neq j$. Consequently, no average causal effect $AME_{j, k \rightarrow \ell}$ is identified. $\qquad \blacksquare$
\end{prop}

\bigskip

\noindent \textit{Proof of Proposition \ref{new_prop_8_mnl_noiden}.}  See Appendix \ref{appendix_proof_prop_8_mnl_noiden}.

\subsection{Ordered logit model \label{sec: iden ordered}}

Consider the following ordered logit model with $J+1$ choice alternatives.\footnote{For simplicity, we do not include covariates $\mathbf{x}$. It is straightforward to extend the identification result in Proposition \ref{iden_ordered_logit} below to an AME conditional on $\mathbf{x}_{2} = \mathbf{x}_{2}$, similarly as our results in Propositions and \ref{prop_1_ident_AME} and \ref{new_prop_7_mnl}.}
\begin{equation}
    \begin{array}[c]{lllll}
        y_{it} & = & j & if & y_{it}^{\ast} \in 
        \left(  \lambda_{j-1},\lambda_{j} \right], 
        \text{ with } j=0,1,...,J
        \\ 
        y_{it}^{\ast} & = & 
        \multicolumn{3}{l}{
            \alpha_{i}+{\displaystyle 
            \sum \limits_{k=0}^{J-1}}\beta_{k} 
            \text{ } 
            \mathbbm{1}\{ y_{i,t-1}=k \} \text{ } 
            - \varepsilon_{it} \text{.}
            }
        \end{array}
\label{Ordered_model}
\end{equation}
where $\lambda$'s are parameters with $\lambda_{-1}=-\infty$ and $\lambda_{J+1}=+\infty$, and $\varepsilon
_{it}$ follows a logistic distribution. This model implies the following transition probability function:
\begin{equation}
    \pi_{kj}(\alpha_{i}) =
    \Lambda \left(  
        \alpha_{i} + \beta_{k} -\lambda_{j-1}
    \right)  -
    \Lambda \left(  
        \alpha_{i} + \beta_{k} -\lambda_{j}
    \right)  
\label{Ordered_probability}
\end{equation}
\citeauthor{honore_muris_2021} (\citeyear{honore_muris_2021}) establish the
identification of the $\mathbf{\beta}$ and $\mathbf{\lambda}$ parameters. The following proposition establishes the identification of all the Average Transition Probabilities $\Pi_{kj}$ and the AME $AME_{j, k \rightarrow \ell} = \Pi_{\ell j} - \Pi_{k j}$ for any values of $k$, $\ell$, and $j$. 

\begin{prop}
\label{iden_ordered_logit} Consider the ordered logit model as defined in equation \eqref{Ordered_model} with $J+1=3$ and given the
values of $\boldsymbol{\beta}$ and $\boldsymbol{\lambda}$ parameters. If $T \geq 3$, the
average transition probabilities $\{ \Pi_{kj}: k, j \in \left \{  0,1,2\right \}  \}$
are identified. For instance, for $j=0$:
\begin{equation}
    \begin{array}[c]{rcl}
    \displaystyle
    \Pi_{00}  &  = &
    \displaystyle \sum \limits_{k=0}^{J}
    \mathbb{P}_{k,0,0} +
    \sum \limits_{k=0}^{J}
    \sum \limits_{\ell=1}^{J}
    e^{\beta_{k}-\beta_{0}} \text{ }
    \mathbb{P}_{k,0,\ell}
    \\ \\
    \Pi_{10}  &  = &
    \displaystyle
    \sum \limits_{k=0}^{J}
    \sum \limits_{\ell=0}^{1}
    \mathbb{P}_{k,0,\ell} +
    \sum \limits_{k=0}^{1}
    e^{\beta_{k}-\beta_{1}} \text{ }
    \mathbb{P}_{k,0,2} + \mathbb{P}_{2,0,2} + 
    \left( e^{\beta_{2}-\beta_{1}} -1 \right)
    \mathbb{P}_{2,2,0}
    \\
    &  + &
    \displaystyle
    \sum \limits_{k=0}^{J}
    \frac{
    \left(  e^{\beta_{1}-\lambda_{0}}
    - e^{\beta_{k}-\lambda_{1}} \right)
    \left(  e^{\beta_{k}} - e^{\beta_{1}} \right)
    }
    {
    \left(  e^{\beta_{k}-\lambda_{0}}
    - e^{\beta_{k}-\lambda_{1}} \right)
    e^{\beta_{1}}
    }
    \mathbb{P}_{k,1,0}
    \\ \\
    \Pi_{20}  & = & 
    \displaystyle
    \mathbb{P}_{0,0,0}+\mathbb{P}_{0,0,1} +
    e^{\lambda_{1} - \lambda_{0}} \text{ }
    \mathbb{P}_{0,0,2} + 
        \left(  1 - \frac{e^{\beta_{2}-\lambda_{0}} }
        {e^{\beta_{1}-\lambda_{1}} }
    \right)  \mathbb{P}_{0,2,0}
    \\
    &  + &
    \displaystyle
    \sum \limits_{k=0}^{l}
    {\displaystyle \sum \limits_{\ell=0}^{J}}
    \mathbb{P}_{1,k,\ell} + 
    \left(  1-
        \frac{e^{\beta_{2}-\lambda_{0}} }
        {e^{\beta_{1}-\lambda_{1}}  }
    \right)  \mathbb{P}_{1,2,0} +
    \sum \limits_{k=0}^{J-1}
    \mathbb{P}_{2,0,k}
    \end{array}
\label{eq_proposition_ordered_logit}
\end{equation}
This implies the identification of $AME_{j, k \rightarrow \ell} = \Pi_{\ell j} - \Pi_{k j}$.
$\qquad \blacksquare$
\end{prop}

\bigskip

\noindent \textit{Proof of Proposition \ref{iden_ordered_logit}.}  See Appendix \ref{proof_iden_ordered}.

\section{Monte Carlo experiments}

\label{sec:montecarlos}

These Monte Carlo experiments serve two main purposes. First, we compare the bias and variance of the FE estimator of $AME$ to those from RE estimators imposing restrictions that we typically find in applications of RE models. Second, we evaluate the power of two testing procedures designed to reject a misspecified RE model: the standard Hausman test, which relies on RE and FE estimators of slope parameters, and an alternative Hausman test based on the RE and FE estimators of AMEs.

The DGP is the binary choice AR1 model described in equation \eqref{eq_binary_model} without exogenous covariates. The model for the initial condition is $y_{i1} = \mathbbm{1}\{ \alpha_{i}+u_{i}\geq0\}$ where $u_{i}$ is i.i.d. Logistic and independent of $\alpha_{i}$ and $\varepsilon_{it}$. The number of periods is $T=4$. We implement experiments for two sample sizes $N$, $1,000$ and $2,000$. We consider six DGPs based on two values of parameter $\beta$ (i.e., $\beta= -1$ and $\beta= 1$) and three distributions of the unobserved heterogeneity $\alpha_{i}$: no heterogeneity, such that $\alpha_{i}=0$ for every $i$; finite mixture with two points of support, $\alpha_{i}=-1$ with probability $0.3$, and $\alpha_{i}=0.5$ with probability $0.7$; and a mixture of two normal random variables: $\alpha_{i} \sim N\left(  -1,3\right)  $ with probability $0.3$, and $\alpha_{i} \sim N\left(  0.5,3\right)  $ with probability $0.7$.

Table \ref{table_1} summarizes the six DGPs, the labels we use to represent them, and the corresponding value of $AME$ in the population. When we change the distribution of the unobserved heterogeneity, keeping the value of parameter $\beta$ constant, the value of $AME$ can vary substantially. For instance, when $\beta=1$, $AME$ equals $0.23$ in the DGP without unobserved heterogeneity, $0.20$ for the finite mixture, and $0.11$ for a mixture of normal distributions.

\bigskip

\begin{table}
\caption{\textbf{DGPs and true value of AME} \label{table_1}}
\centering
\begin{tabular}[c]{cc|ccc}
\hline \hline
&  & \multicolumn{3}{c}{\textit{Distribution of }$\alpha_{i}$}
\\ 
\multicolumn{2}{c|}{\textit{Value of }$\beta$} 
& \textit{No }$\alpha_{i}$ &
\multicolumn{1}{|c}{\textit{Finite mixture}} &
\multicolumn{1}{|c}{\textit{Mixture of normals}}
\\ \hline
\multicolumn{1}{l}{} & \multicolumn{1}{r|}{$\beta=-1$} &
\multicolumn{1}{|c|}{$
    \begin{array}[c]{c}
        \text{DGP \textit{NoUH(-1)}}\\
        AME=-0.2311
        \end{array}
    $} & \multicolumn{1}{|c}{$
    \begin{array}[c]{c}
        \text{DGP \textit{FinMix(-1)}}\\
        AME=-0.2164
    \end{array}
    $} & \multicolumn{1}{|c}{$
    \begin{array}[c]{c}
        \text{DGP \textit{MixNor(-1)}}\\
        AME=-0.113
    \end{array}
    $} \\
    \hline
    \multicolumn{1}{l}{} & \multicolumn{1}{r|}{$\beta=1$} & \multicolumn{1}{|c|}{$
    \begin{array}[c]{c}
        \text{DGP \textit{NoUH(+1)}}\\
        AME=0.2311
    \end{array}
    $} & \multicolumn{1}{|c}{$
    \begin{array}[c]{c}
        \text{DGP \textit{FinMix(+1)}}\\
        AME=0.2059
    \end{array}
    $} & \multicolumn{1}{|c}{$
    \begin{array}[c]{c}
        \text{DGP \textit{MixNor(+1)}}\\
        AME=0.1108
    \end{array}
    $} 
    \\
    \hline \hline
\end{tabular}
\end{table}

For each DGP, we simulate $1,000$ random samples with $N$ individuals (with $N=1,000$ or $N=2,000$) and $T=4$. For each sample, we calculate three estimators of $\beta$ and $AME$: (1) a FE estimator, that we denote \textit{FE-CMLE};\footnote{The FE estimator of $\beta$ is the CMLE proposed by Chamberlain (1985). For the estimation of parameter $AME$, we use a plug-in estimator based on the formula for the identified $AME$ when $T=4$, which we present in Table \ref{table_7} in the Appendix. In this formula, we replace parameter $\beta$ with its CML estimate and the probabilities of choice histories with frequency estimates.} (2) a maximum likelihood estimator that assumes that the distribution of $\alpha_{i}$ is discrete with two mass points, that we denote \textit{RE-MLE}; and (3) a maximum likelihood estimator that assumes there is no unobserved heterogeneity, that we denote \textit{NoUH-MLE}.\footnote{For the DGPs without unobserved heterogeneity (i.e., \textit{NoUH(-1)} and\textit{\ NoUH(+1)}), we do not report results for the RE MLE. The reason is that, for these DGPs, the finite mixture (two types) RE model is not identified, and the estimates of $\beta$ are extremely poor. As expected, the estimate of the mixing probability in the mixture is close to zero, but the points in the support set of $\alpha_{i}$ are poorly identified, and they take extreme values. This problem also affects the estimation of $\beta$, which presents a huge bias and variance. For this reason, we have preferred not to present results for this combination of estimator and DGP. However, it is important to note that avoiding these numerical/identification problems in estimating the distribution of $\alpha$ is a key advantage of FE estimation.} Table \ref{table_2} presents
results from experiments with $N=1000$.\footnote{The results for sample size $N=2000$ are qualitatively very similar except that, as one would expect, all the estimators have lower variance when the sample size increases. Therefore, we present only results from experiments with $N=1000$ here.}

\bigskip

\noindent \textit{(i) Bias of FE estimators relative to MLE.} The mean bias of the FE estimator is very small: between $0.1\%$ and $0.7\%$ of the true value for $\beta$, and between $0.2\%$ and $1.4\%$ for $AME$. In this FE
approach, the estimation of $AME$ does not involve a substantially larger bias than that of $\beta$. This bias is of a similar magnitude as the ones of \textit{NoUH-MLE} and \textit{RE-MLE} estimators when these estimators are consistent (i.e., when the DGPs are \textit{NoUH} and
\textit{FinMix}, respectively).

\bigskip

\noindent \textit{(ii) Variance of FE estimators relative to RE-MLE.} As a percentage of the true value, the standard deviation of the FE estimator is between $10\%$ and $20\%$ for the estimator of $\beta$, and between $7\%$ and $30\%$ for the estimator of $AME$. These ratios are substantially
smaller for the RE-MLE estimator: between $9\%$ and $13\%$ for the estimator of $\beta$, and between $8\%$ and $23\%$ for the estimator of $AME$. As expected, the FE estimators have a larger variance than the RE-MLE estimators. The loss of precision associated with FE estimation is of a similar magnitude when estimating $AME$ than when estimating $\beta$.

The variance of the FE estimator is substantially larger when $\beta$ is positive than when it is negative, but this is not the case for the RE-MLE estimators. This result has a clear explanation. The histories that contribute to the identification of the parameters $\beta$ and $AME$ involve some alternation of the two choices over time, e.g., $\{0,1,0,1\}$ or $\{0,0,1,1\}$. These histories occur more frequently when $\beta$ is negative than when it is positive. Identifying negative state dependence is simpler than identifying positive state dependence. This is because positive state dependence has implications similar to unobserved heterogeneity, whereas negative state dependence does not.

\begin{table}
\caption{\textbf{Monte Carlo Experiments with sample size N=1,000} \label{table_2}}
\centering
\begin{tabular}[c]{cc|c|c|c|c|c|c|c} 
\hline \hline
&  & \multicolumn{7}{c}{\textit{Statistics}}\\
&  & \textit{True} & \textit{Mean} & \textit{Std} & \textit{True} &
\textit{Mean} & \textit{Std} & \textit{RMSE}\\
&  & $\beta$ & $\widehat{\beta}$ & $\widehat{\beta}$ & $AME$ & $\widehat{AME}
$ & $\widehat{AME}$ & $\widehat{AME}$\\ \hline
\textbf{DGP} & \multicolumn{1}{r|}{\textit{FE-CMLE}} & {\footnotesize -1.0} &
{\footnotesize -1.0074} & {\footnotesize 0.1310} & {\footnotesize -0.2311} &
{\footnotesize -0.2314} & {\footnotesize 0.0235} & {\footnotesize 0.0235}\\
\textbf{NoUH(-1)} & \multicolumn{1}{r|}{\textit{RE-MLE}} &
{\footnotesize -1.0} & {\footnotesize NA} & {\footnotesize NA} &
{\footnotesize -0.2311} & {\footnotesize NA} & {\footnotesize NA} &
{\footnotesize NA}\\
& \multicolumn{1}{r|}{\textit{NoUH-MLE}} & {\footnotesize -1.0} &
{\footnotesize -0.9998} & {\footnotesize 0.0798} & {\footnotesize -0.2311} &
{\footnotesize -0.2309} & {\footnotesize 0.0175} & {\footnotesize 0.0175}%
\\ \hline
\textbf{DGP} & \multicolumn{1}{r|}{\textit{FE-CMLE}} & {\footnotesize -1.0} &
{\footnotesize -1.0012} & {\footnotesize 0.1338} & {\footnotesize -0.2164} &
{\footnotesize -0.2160} & {\footnotesize 0.0222} & {\footnotesize 0.0221}\\
\textbf{FinMix(-1)} & \multicolumn{1}{r|}{\textit{RE-MLE}} &
{\footnotesize -1.0} & {\footnotesize -1.0036} & {\footnotesize 0.1214} &
{\footnotesize -0.2164} & {\footnotesize -0.2165} & {\footnotesize 0.0215} &
{\footnotesize 0.0214}\\
& \multicolumn{1}{r|}{\textit{NoUH-MLE}} & {\footnotesize -1.0} &
{\footnotesize -0.5979} & {\footnotesize 0.0781} & {\footnotesize -0.2164} &
{\footnotesize -0.1430} & {\footnotesize 0.0183} & {\footnotesize 0.0757}%
\\ \hline
\textbf{DGP} & \multicolumn{1}{r|}{\textit{FE-CMLE}} & {\footnotesize -1.0} &
{\footnotesize -1.0136} & {\footnotesize 0.2160} & {\footnotesize -0.1113} &
{\footnotesize -0.1110} & {\footnotesize 0.0178} & {\footnotesize 0.0176}\\
\textbf{MixNor(-1)} & \multicolumn{1}{r|}{\textit{RE-MLE}} &
{\footnotesize -1.0} & {\footnotesize -0.3604} & {\footnotesize 0.1825} &
{\footnotesize -0.1113} & {\footnotesize -0.0470} & {\footnotesize 0.0218} &
{\footnotesize 0.0679}\\
& \multicolumn{1}{r|}{\textit{NoUH-MLE}} & {\footnotesize -1.0} &
{\footnotesize 1.7190} & {\footnotesize 0.1028} & {\footnotesize -0.1113} &
{\footnotesize 0.4022} & {\footnotesize 0.0214} & {\footnotesize 0.5139}%
\\ \hline
\textbf{DGP} & \multicolumn{1}{r|}{\textit{FE-CMLE}} & {\footnotesize 1.0} &
{\footnotesize 1.0013} & {\footnotesize 0.1654} & {\footnotesize 0.2311} &
{\footnotesize 0.2344} & {\footnotesize 0.0526} & {\footnotesize 0.0527}\\
\textbf{NoUH(+1)} & \multicolumn{1}{r|}{\textit{RE-MLE}} & {\footnotesize 1.0}
& {\footnotesize NA} & {\footnotesize NA} & {\footnotesize 0.2311} &
{\footnotesize NA} & {\footnotesize NA} & {\footnotesize NA}\\
& \multicolumn{1}{r|}{\textit{NoUH-MLE}} & {\footnotesize 1.0} &
{\footnotesize 0.9980} & {\footnotesize 0.0778} & {\footnotesize 0.2311} &
{\footnotesize 0.2305} & {\footnotesize 0.0176} & {\footnotesize 0.0176}%
\\ \hline
\textbf{DGP} & \multicolumn{1}{r|}{\textit{FE-CMLE}} & {\footnotesize 1.0} &
{\footnotesize 0.9982} & {\footnotesize 0.1841} & {\footnotesize 0.2059} &
{\footnotesize 0.2089} & {\footnotesize 0.0539} & {\footnotesize 0.0539}\\
\textbf{FinMix(+1)} & \multicolumn{1}{r|}{\textit{RE-MLE}} &
{\footnotesize 1.0} & {\footnotesize 0.9864} & {\footnotesize 0.1296} &
{\footnotesize 0.2059} & {\footnotesize 0.2034} & {\footnotesize 0.0315} &
{\footnotesize 0.0316}\\
& \multicolumn{1}{r|}{\textit{NoUH-MLE}} & {\footnotesize 1.0} &
{\footnotesize 1.4100} & {\footnotesize 0.0843} & {\footnotesize 0.2059} &
{\footnotesize 0.3212} & {\footnotesize 0.0183} & {\footnotesize 0.1168}%
\\ \hline
\textbf{DGP} & \multicolumn{1}{r|}{\textit{FE-CMLE}} & {\footnotesize 1.0} &
{\footnotesize 1.0055} & {\footnotesize 0.2873} & {\footnotesize 0.1108} &
{\footnotesize 0.1169} & {\footnotesize 0.0511} & {\footnotesize 0.0515}\\
\textbf{MixNor(+1)} & \multicolumn{1}{r|}{\textit{RE-MLE}} &
{\footnotesize 1.0} & {\footnotesize 1.4863} & {\footnotesize 0.1828} &
{\footnotesize 0.1108} & {\footnotesize 0.2120} & {\footnotesize 0.0367} &
{\footnotesize 0.1078}\\
& \multicolumn{1}{r|}{\textit{NoUH-MLE}} & {\footnotesize 1.0} &
{\footnotesize 3.2453} & {\footnotesize 0.1194} & {\footnotesize 0.1108} &
{\footnotesize 0.6645} & {\footnotesize 0.0166} & {\footnotesize 0.5541}%
\\ \hline \hline
\end{tabular}

\end{table}

\noindent \textit{(iii) Bias of RE-MLE estimators due to misspecification.} The biases due to the misspecification of the RE model are substantial. The bias in the estimation of $\beta$ from ignoring unobserved heterogeneity, when present, is between $41\%$ of the true value (with the finite mixture DGP) and $270\%$ (with the mixture of normals DGP). The bias is even more significant in estimating $AME$: $60\%$ of the true value in the finite mixture DGP and more than $500\%$ in the mixture of normals DGP. The bias is also substantial for the RE-MLE that accounts for heterogeneity but misspecifies
its distribution: between $50\%$ and $65\%$ in the estimation of $\beta$; and between $58\%$ and $93\%$ for $AME$. As a result, the FE estimator dominates the RE-MLE in terms of Root Mean Square Error (RMSE) in the cases where the RE model is misspecified.

\bigskip

\noindent \textit{(iv) Testing for misspecification of RE models.} A common approach to test the validity of a RE model consists of using a Hausman test that compares the FE estimator of $\beta$ (consistent under the null and the alternative) and the RE-MLE of $\beta$ (efficient under the null but inconsistent under the alternative). See \citeauthor{hausman_1978} (\citeyear{hausman_1978}) and \citeauthor{hausman_taylor_1981} (\citeyear{hausman_taylor_1981}). Given our identification results, we can define a similar Hausman test using the FE and RE estimators of $AME$ instead of $\beta$. Therefore, we have two different Hausman statistics to test for the validity of a RE model. The statistic based on the estimators of $\beta$ is:
\begin{equation}
\begin{array}[c]{rcc}
    HS_{\beta} = \dfrac{ \left(  \widehat{\beta}_{FE} - \widehat{\beta}_{RE} \right)  ^{2}} {\widehat{Var} \left(  \widehat{\beta}_{FE} \right)  -
    \widehat{Var} \left(  \widehat{\beta}_{RE}\right)  } & \text{under }H_{0} &
    \sim \chi_{1}^{2}%
\end{array}
\label{hausman_test_beta}
\end{equation}
And the statistic based on the estimators of $AME$ is:
\begin{equation}
\begin{array}[c]{rcc}
    HS_{AME} = \dfrac{ \left(  \widehat{AME}_{FE} - \widehat{AME}_{RE} \right)^{2}} {\widehat{Var} \left(  \widehat{AME}_{FE} \right)  - \widehat{Var}
    \left(  \widehat{AME}_{RE}\right)  } & \text{under }H_{0} & \sim \chi_{1}^{2}%
\end{array}
\label{hausman_test_ame}
\end{equation}

The Hausman test based on $AME$ has several advantages over the test based on $\beta$. First, the researcher can be particularly interested in the causal effect implied by the model and not in the slope parameter itself. Second, and more substantially, the parameter $\beta$ test may suffer a scaling problem that does not affect the test on the $AME$. That is, the parameter $\beta$ depends on the variance of the transitory shock $\varepsilon_{it}$, which depends on the RE model's specification. For instance, a reason why the estimators $\widehat{\beta}_{FE}$ and $\widehat{\beta}_{NoUH-MLE}$ are different is that, in the model ignoring unobserved heterogeneity, the actual error term is $\alpha_{i}+\varepsilon_{it}$, and the variance of this variable is larger than the variance of $\varepsilon_{it}$. The estimation of $AME$ -- using either FE or RE approaches -- is not affected by this scaling problem.

\clearpage

\begin{center}
\textbf{Figures 1 to 6: Empirical distribution of p-values of Hausman tests}

\bigskip

\textbf{Figure 1. \qquad \qquad \qquad \qquad \qquad \qquad \qquad Figure 2}

\scalebox{0.6}{
\includegraphics{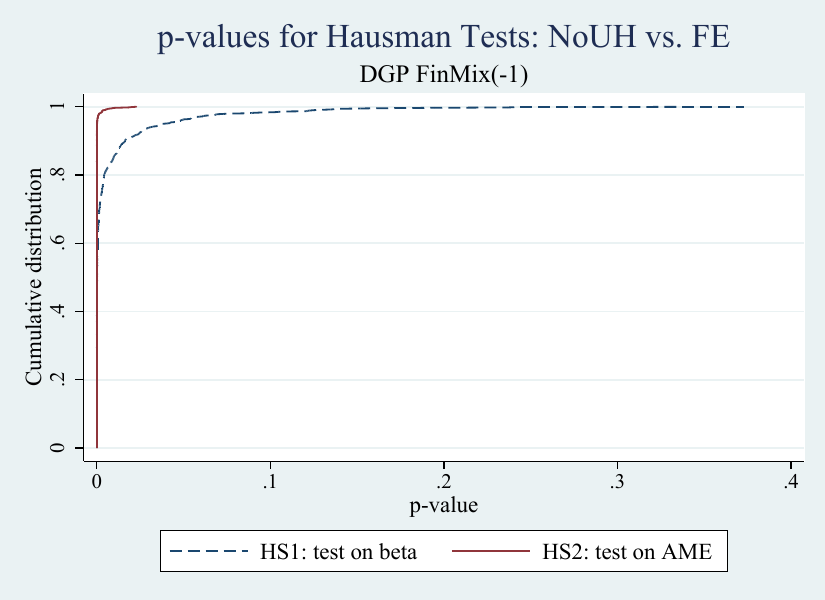}}\scalebox{0.6}{
\includegraphics{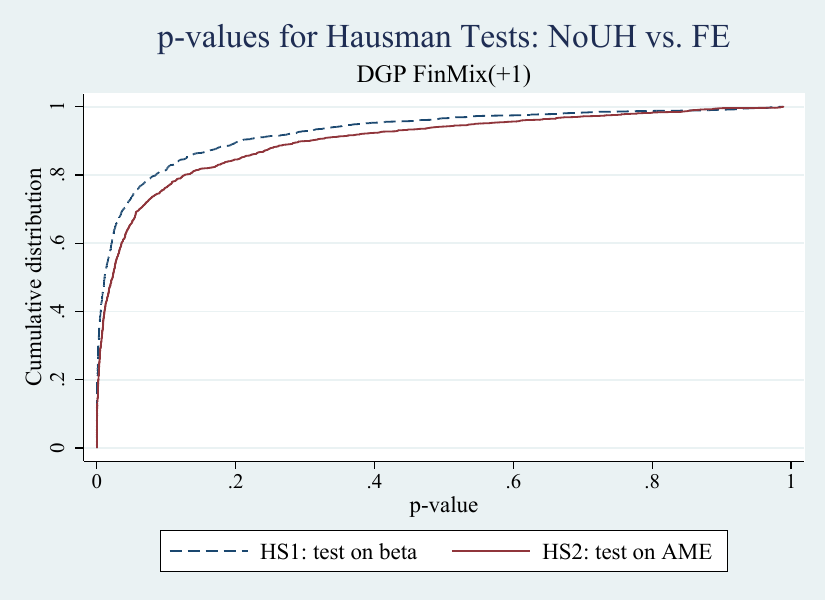}}

\textbf{Figure 3\qquad \qquad \qquad \qquad \qquad \qquad \qquad Figure 4}

\scalebox{0.6}{
\includegraphics{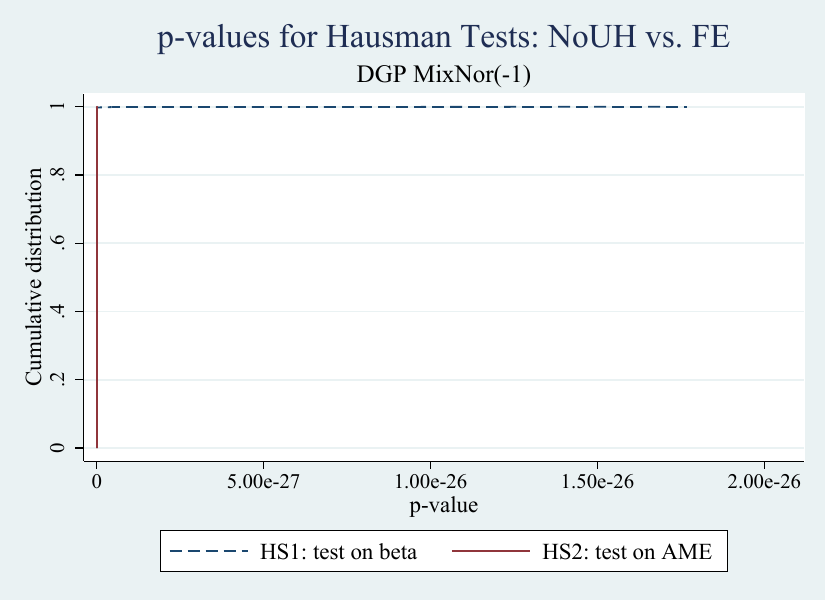}}\scalebox{0.6}{
\includegraphics{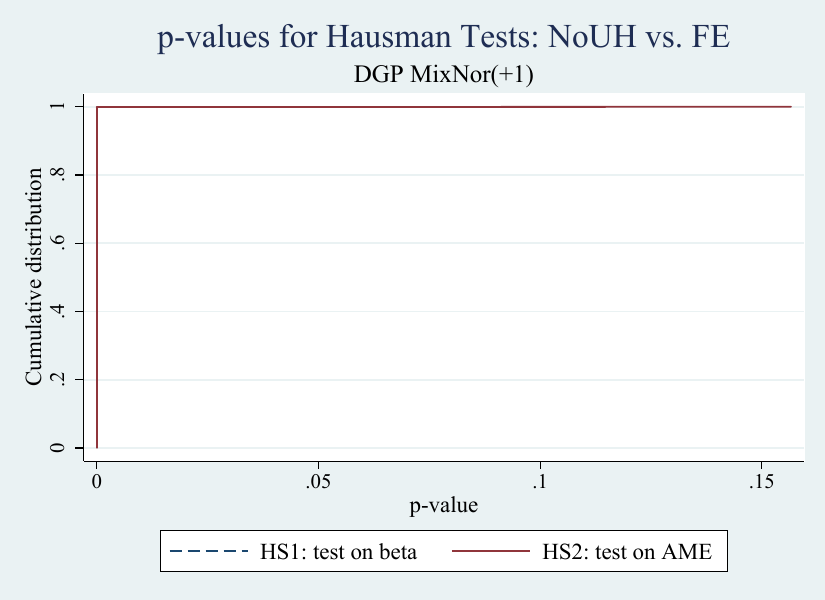}}

\bigskip

\textbf{Figure 5\qquad \qquad \qquad \qquad \qquad \qquad \qquad Figure 6}

\scalebox{0.6}{
\includegraphics{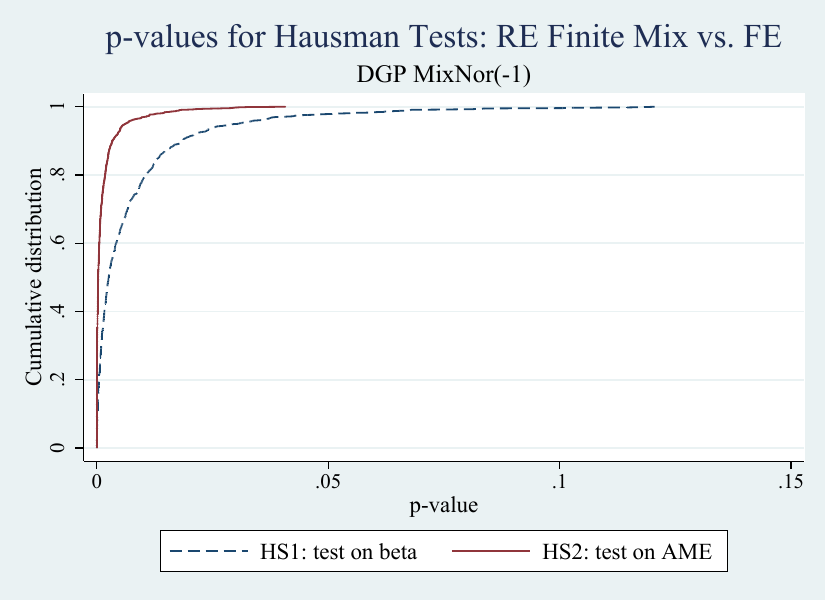}}\scalebox{0.6}{
\includegraphics{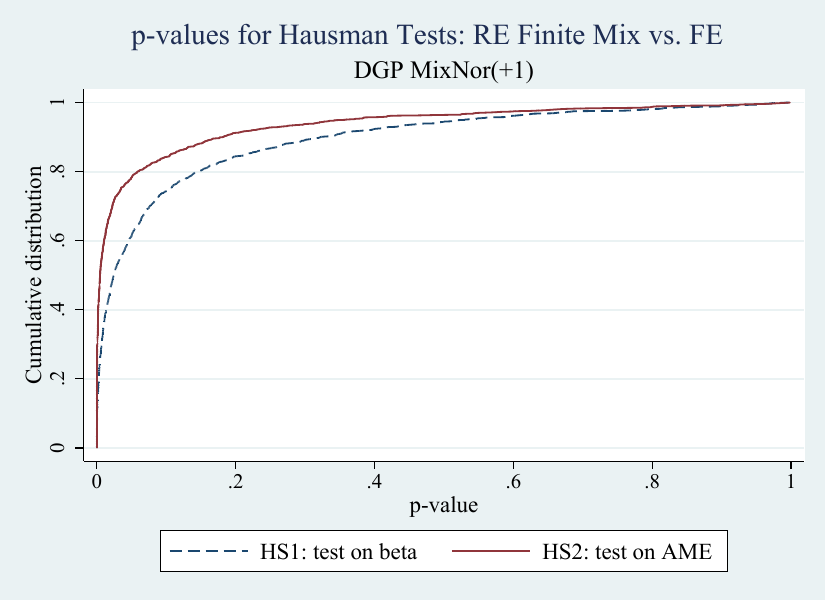}}
\end{center}

\clearpage

We compare the power of these two tests using our Monte Carlo experiments. Figures 1 to 6 summarize our results. Each figure corresponds to one DGP and presents the cumulative distribution function of the p-value for the null hypothesis that the RE model is valid. More specifically: \textit{Figure 1:} DGP is \textit{FinMix(-1)} and null hypothesis is no unobserved heterogeneity; \textit{Figure 2:} DGP is \textit{FinMix(+1)} and null hypothesis is no unobserved heterogeneity; \textit{Figure 3:} DGP is \textit{MixNor(-1)} and null hypothesis is no unobserved heterogeneity; \textit{Figure 4:} DGP is \textit{MixNor(+1)} and null hypothesis is no unobserved heterogeneity; \textit{Figure 5:} DGP is \textit{MixNor(-1)} and null hypothesis is the \textit{finite mixture} model; \textit{Figure 6:} DGP is \textit{MixNor(+1)} and null hypothesis is the \textit{finite mixture} model.

Figures 3 and 4 show that both tests have strong power to reject the null of no unobserved heterogeneity when the DGP is a mixture of normals. In Figures 1 and 5, the two tests also have strong power when the true value of $\beta$ is negative. The relevant comparison appears in Figures 2 and 6. In the DGP with a mixture of normals (Figure 6), the $HS_{AME}$ test has substantially larger power than the test $HS_{\beta}$. In particular, $HS_{\beta}$ has a serious low-power problem. For this test, with a 5\% significance level, we do not reject the null for more than half of the samples. In contrast, the $HS_{AME}$ test has reasonable power. With a 5\% significance level for this test, we can reject the null for $80\%$ of the samples. In Figure 2, the $HS_{\beta}$ test has more power than the $HS_{AME}$ test. However, the differences in power are much smaller than in Figure 6, and neither of the two tests has a low power problem. Overall, the $HS_{AME}$ test has more power than the $HS_{\beta}$. This test seems a useful byproduct of identifying AMEs in FE models.

\section{State Dependence in Consumer Brand Choice \label{sec:application}}

We apply our identification results to measure state dependence on consumer brand choices. There is substantial literature on testing and measuring state dependence on consumer brand choices, with seminal papers by \citeauthor{erdem_1996} (\citeyear{erdem_1996}), \citeauthor{keane_1997} (\citeyear{keane_1997}), and \citeauthor{roy_chintagunta_1996} (\citeyear{roy_chintagunta_1996}).\footnote{Other contributions to this literature are \citeauthor{seetharaman_ainslie_1999} (\citeyear{seetharaman_ainslie_1999}), \citeauthor{erdem_imai_2003} (\citeyear{erdem_imai_2003}), \citeauthor{seetharaman_2004} (\citeyear{seetharaman_2004}), \citeauthor{dube_hitsch_2010} (\citeyear{dube_hitsch_2010}), and \citeauthor{osborne_2011} (\citeyear{osborne_2011}), among others. There is also a related literature on the implications of brand-choice state dependence on market competition (see \citeauthor{viard_2007}, \citeyear{viard_2007}, and \citeauthor{pakes_porter_2021} \citeyear{pakes_porter_2021}.} These applications use consumer scanner panel data and estimate dynamic discrete choice models with persistent unobserved heterogeneity in consumer brand preferences and state dependence generated by purchasing/consumption habits or/and brand switching costs. The main goal is to determine the relative contribution of unobserved heterogeneity and state dependence to explain the
observed time persistence of consumer brand choices. Disentangling the contribution of these two factors has important implications for demand elasticities, competition, consumer welfare, and the evaluation of mergers.\footnote{See \citeauthor{erdem_imai_2003} (\citeyear{erdem_imai_2003}) for a detailed discussion of the important economic implications of distinguishing between unobserved heterogeneity and state dependence on consumer demand.}

All these previous studies estimate Random Effects (RE) models. In this application, we consider a FE model, estimate average transition probabilities $\Pi_{jj}$ and average treatment effects $ATE_{jj} \equiv \Pi_{jj} - \mathbb{E}(\mathbbm{1}\{y_{it}=j\})$, and use them to measure the contribution of state dependence to brand-choice persistence.

\subsection{Data}

The dataset comes from A.C. Nielsen scanner panel data for the ketchup product category in the geographic market of Sioux Falls, South Dakota.\footnote{Our sample comes from \citeauthor{erdem_imai_2003} (\citeyear{erdem_imai_2003}). We thank the authors for sharing the data with us.} It contains $996$ households and covers a 123-week period from mid-1986 to mid-1988.\footnote{The raw data contains $2797$ households. Here we use the same working sample of $996$ households as in \citeauthor{erdem_imai_2003} (\citeyear{erdem_imai_2003}). This sample focuses on households who are regular ketchup users. See page 30 of that paper for a description of the selection of this working sample.} For our analysis, a period is a \textit{household purchase occasion}. Periods $t=1, 2, ...$ represent a household's first, second, and so on ketchup purchases during the sample period. This timing is common in this literature (e.g., \citeauthor{erdem_1996}, \citeyear{erdem_1996}; \citeauthor{keane_1997}, \citeyear{keane_1997}). $T_{i}$ is the number of purchase occasions for household $i$. The total number of observations or purchase occasions in this sample is $\sum_{i=1}^{N}T_{i} = 9,562$. Table \ref{table_3} presents the distribution of $T_{i}$.

\begin{table}
\caption{\textbf{Distribution of the number of purchase occasions
($T_{i}$)} \label{table_3}}
\centering
\begin{tabular}[c]{c|c|c|c|c|c|c}
\hline \hline
\textit{Minimum} & \textit{5\%} & \textit{25\%} & \textit{Median} &
\textit{75\%} & \textit{95\%} & \textit{Maximum}\\ \hline
\text{3} & \text{4} & \text{5} & \text{8} & \text{12} & \text{21} &
\text{52}\\ \hline \hline
\end{tabular}

\end{table}

There are four brands in this market: three national brands, Heinz, Hunt's, and Del Monte; and a store brand. We ignore the quantity purchased and focus on brand choice. Table \ref{table_4} presents brands' market shares, constructed using the number of purchasing events, and the matrix of transition probabilities between the four brands. Heinz is the leading brand, with 66\% share of purchases, followed by Hunt's at 16\%, Del Monte at 12\%, and Store brands at 5\%. 

A measure of \textit{choice persistence for brand $j$} is the difference between the transition probability $Pr(y_{i,t+1}=j | y_{it}=j)$ and the unconditional probability or market share $Pr(y_{it}=j)$. This measure shows choice persistence for all the brands, with the largest for \textit{Del Monte} and \textit{Store brands} with $21.88\%$ and $21.66\%$, respectively, followed by \textit{Hunt's} with $16.67\%$, and \textit{Heinz} with $12.30\%$. The persistence observed may stem from consumer taste heterogeneity and from state dependence. Our main objective in this application is to distinguish the impact of these two factors and examine how they differ across brands.

\begin{table}
\caption{\textbf{Matrix of Transition Probabilities of Brand Choices (\%)} \label{table_4}}
\centering
\begin{tabular}[c]{r|c|c|c|c|c}
\hline \hline
& \multicolumn{4}{c}{\textit{Brand choice at $t+1$}} &
\multicolumn{1}{c}{\textit{Total}}\\
\textit{Brand choice at $t$} & \textit{Heinz} & \textit{Hunt's} & \textit{Del
Monte} & \textit{Store} & \\
& \textit{($j=0$)} & \textit{($j=1$)} & \textit{($j=2$)} & \textit{($j=3$)} &
\\ \hline
\textit{Heinz ($j=0$)} & \textbf{78.95} & \text{10.67} & \text{6.98} &
\text{3.40} & \text{100.00}\\
\textit{Hunt's ($j=1$)} & \text{45.16} & \textbf{32.30} & \text{15.76} &
\text{6.78} & \text{100.00}\\
\textit{Del Monte ($j=2$)} & \text{41.11} & \text{18.98} & \textbf{34.07} &
\text{5.83} & \text{100.00}\\
\textit{Store ($j=3$)} & \text{42.32} & \text{17.11} & \text{13.38} &
\textbf{27.19} & \text{100.00}\\ \hline
\textit{Market share ($\mathbb{P}_{j}$}) & \text{66.65} & \text{15.63} &
\text{12.19} & \text{5.53} & \text{100.00}\\
\textit{Choice persistence ($\mathbb{P}_{j|j}-\mathbb{P}_{j}$}) & \text{12.30}
& \text{16.67} & \text{21.88} & \text{21.66} & \text{}\\ \hline \hline
\end{tabular}

\end{table}

\subsection{Model}

Let $y_{it} \in \{0,1,2,3\}$ be the brand choice of household $i$ at purchase occasion $t$. We consider the following brand choice model with \textit{habit formation}:
\begin{equation}
    y_{it} = \arg \max_{j \in \{0,1,2,3\}} 
    \text{ } 
    \left \{  \text{ } \alpha_{i}(j)
    + \beta_{jj} \text{ } \mathbbm{1}\{y_{i,t-1}=j\} +\varepsilon_{it}(j) \text{ } \right \}
    \text{.} 
\label{eq:model_application}
\end{equation}
Parameter $\beta_{jj}$ represents habits in purchasing brand
$j$: the additional utility from buying the same brand as in the previous purchase. Parameter $\beta_{00}$ (for Heinz) is normalized to zero. Variable $\alpha_{i}(j)$ represents the household's time-invariant taste for brand $j$. For simplicity, we ignore duration dependence. We also omit prices.\footnote{In this dataset, supermarkets follow High-Low pricing, and prices can stay at the high (regular) level for relatively long periods. Omitting prices in our model can be interpreted using choice histories where prices remain constant.}

Following \citeauthor{aguirregabiria_gu_2021} (\citeyear{aguirregabiria_gu_2021}), we can interpret equation (\ref{eq:model_application}) as a model where households are forward-looking. The fixed effects $\alpha_{i}(j)$ can be interpreted as the sum of two components: a fixed effect in the current utility of choosing brand $j$ and the continuation value (expected and discounted future utility) of choosing brand $j$ today. In this model, these continuation values depend on the current choice $j$ but not on the state variable $y_{i,t-1}$ or the current $\varepsilon_{it}$.

\subsection{Estimation}

To illustrate our method using a short panel, we split the purchasing histories in the original sample into subs histories of length $T$, for $T=6$ and $T=8$. Table \ref{table_5} presents our Fixed Effect estimates of the \textit{brand habit} parameters $\beta_{jj}$. We use the Conditional Maximum Likelihood estimator. We obtain standard errors using a bootstrap method that resamples the $996$ purchasing histories in the original dataset.\footnote{Using the original sample of 996 purchasing histories, we resample independently and with replacement 996 histories. Then, we generate all the possible sub-histories of length $T$ from these histories. We also obtained asymptotic standard errors. Bootstrap standard errors are a bit larger (at the second or third significant digit) than the asymptotic ones.} Parameter estimates with $T=6$ and $T=8$ are very similar. They are significantly greater than zero at a 5\% significance level, showing evidence of state dependence on brand choice. The magnitude of the parameter estimate is not monotonically related to the brand's market share or the degree of brand choice persistence shown in Table \ref{table_4}. However, we need to take into account that a larger value of $\beta_{jj}$ does not imply a larger degree of state dependence as measured by the Average Transition Probabilities or by $ATE_{jj}$.

\begin{table}
\caption{\textbf{Conditional Maximum Likelihood Estimates} \label{table_5}}
\centering
\begin{tabular}[c]{r|cc|cc}
\hline \hline
\multicolumn{5}{c}{\textbf{of Brand Habit ($\beta_{jj}$) Parameters}}\\ \hline
\textit{Parameter} & \multicolumn{2}{c}{\textbf{$T=6$ sub-histories}} &
\multicolumn{2}{|c}{\textbf{$T=8$ sub-histories}}\\
\textit{$\beta_{jj}$} & \textit{Estimate} & \textit{(s.e.)} &
\textit{Estimate} & \textit{(s.e.)}\\ \hline
\textit{Heinz} & $0.00$ & $(.)$ & $0.00$ & $(.)$\\
\textit{Hunt's} & $0.2312$ & $(0.0590)$ & $0.2566$ & $(0.0570)$\\
\textit{Del Monte} & $0.1155$ & $(0.0718)$ & $0.1191$ & $(0.0722)$\\
\textit{Store} & $0.3245$ & $(0.1166)$ & $0.4675$ & $(0.1106)$\\ \hline
\textit{\# histories of length $T$} & \multicolumn{2}{c}{$4,764$} &
\multicolumn{2}{|c}{$3,396$}\\ \hline \hline
\multicolumn{5}{l}{{\footnotesize {$(1)$ Standard errors $(s.e)$ are obtained
using a bootstrap method. We generate}}}\\
\multicolumn{5}{l}{{\footnotesize {1,000 resamples (independent, with
replacement, and with $N=996$) from the}}}\\
\multicolumn{5}{l}{{\footnotesize {996 purchasing histories in the original
dataset. Then, we split each history}}}\\
\multicolumn{5}{l}{{\footnotesize {of the bootstrap sample into all the
possible sub-histories of length $T$.}}}%
\end{tabular}

\end{table}

\begin{table}
\caption{\textbf{Fixed Effects Estimates of ATPs and AMEs} \label{table_6}}
\centering
\begin{tabular}[c]{r|c|c|c|c|c|c|c|c}
\hline \hline
& \multicolumn{4}{c}{\textbf{$T=6$ sub-histories}} &
\multicolumn{4}{|c}{\textbf{$T=8$ sub-histories}}\\
& \textit{Pers} & \textit{ATP} & \textit{ATE} & \textit{UHet} & \textit{Pers}
& \textit{ATP} & \textit{ATE} & \textit{UHet}\\
\textit{\ } & \textit{(}s.e.) & \textit{(s.e.)} & \textit{(}s.e.) &
\textit{(s.e.)} & \textit{(}s.e.) & \textit{(s.e.)} & \textit{(}s.e.) &
\textit{(s.e.)}\\ \hline
\textit{Heinz} & $0.1230$ & $0.6744$ & $0.0079$ & $0.1151$ & $0.1230$ &
$0.6708$ & $0.0043$ & $0.1187$\\
\textit{\ } & $(0.0033)$ & $(0.0057)$ & $(0.0066)$ & $(0.0068)$ & $(0.0033)$ &
$(0.0062)$ & $(0.0067)$ & $(0.0069)$\\
\textit{Hunt's} & $0.1667$ & $0.1752$ & $0.0189$ & $0.1478$ & $0.1667$ &
$0.1788$ & $0.0225$ & $0.1442$\\
\textit{\ } & $(0.0077)$ & $(0.0075)$ & $(0.0107)$ & $(0.0109)$ & $(0.0077)$ &
$(0.0072)$ & $(0.0106)$ & $(0.0109)$\\
\textit{Del Monte} & $0.2188$ & $0.1324$ & $0.0105$ & $0.2183$ & $0.2188$ &
$0.1345$ & $0.0126$ & $0.2062$\\
\textit{\ } & $(0.0090)$ & $(0.0067)$ & $(0.0112)$ & $(0.0115)$ & $(0.0090)$ &
$(0.0062)$ & $(0.0110)$ & $(0.0113)$\\
\textit{Store} & $0.2166$ & $0.0736$ & $0.0183$ & $0.1983$ & $0.2166$ &
$0.0805$ & $0.0252$ & $0.1914$\\
\textit{\ } & $(0.0062)$ & $(0.0071)$ & $(0.0094)$ & $(0.0099)$ & $(0.0062)$ &
$(0.0072)$ & $(0.0094)$ & $(0.0099)$ \\
\hline \hline
\multicolumn{9}{l}{{\footnotesize {(1) \textit{Pers} is brand choice persistence, $\mathbb{P}_{j|j}-\mathbb{P}_{j}$, as measured at the bottom line of Table \ref{table_4}.}}} \\
\multicolumn{9}{l}{{\footnotesize {(2) \textit{ATP} is the brand's Average
Transition Probability, $\Pi_{jj}$.}}}\\
\multicolumn{9}{l}{{\footnotesize {(3) $ATE = ATE_{jj} = \Pi_{jj}-\mathbb{E}%
(\mathbbm{1}\{y_{it}=j\})$.}}}\\
\multicolumn{9}{l}{{\footnotesize {(4) \textit{UHet} is defined as
$\mathbb{P}_{j|j}-\Pi_{jj}$. By construction, \textit{Pers = AME + UHet}.}}}\\
\multicolumn{9}{l}{{\footnotesize {(5) Standard errors $(s.e)$ are obtained
using the same bootstrap method as for the estimates in Table \ref{table_6}.}}}
\end{tabular}

\end{table}

\bigskip

Table \ref{table_6} presents Fixed Effect estimates of ATPs and a decomposition of brand choice persistence into the contributions of state dependence and unobserved heterogeneity. The estimation of the ATPs $\Pi_{jj}$ is based on equation \eqref{eq:iden_PI_jj} in Proposition \ref{new_prop_7_mnl}. We plug the CML estimates of $\beta_{jj}$ parameters and frequency estimates of probabilities of choice histories in this equation and obtain standard errors using a bootstrap method.

In Table \ref{table_6}, the column labeled \textit{Pers} provides brand choice persistence as measured by the difference between the transition probability $\mathbb{P}_{j|j}$ and the unconditional probability $\mathbb{P}_{j}$. The estimates of ATPs (in the columns labeled \textit{ATP}) are very precise and similar for $T=6$ and $T=8$. The column labelled $AME$ presents $ATE_{jj} = \Pi_{jj}-\mathbb{E}(\mathbbm{1}\{y_{it}=j\})$. This $AME$ is a measure of the contribution of state dependence to brand choice persistence. This contribution is quite small for all the brands: between $1$ and $2$ percentage points. In fact, for \textit{Heinz} and \textit{Del Monte}, we cannot reject the null hypothesis that this $AME$ is zero at a $5\%$ significance level. The \textit{Store brand} is the one with the largest contribution of state dependence. The column labeled $UHet$ presents the contribution of consumer taste heterogeneity to brand choice persistence, as measured by the difference between brand choice persistence and $ATE_{jj}$. This heterogeneity accounts for most of the persistence of brand choice. This finding contrasts with results found in studies using similar models and data but with a Random Effects specification of consumer unobserved taste heterogeneity (e.g., \citeauthor{keane_1997}, \citeyear{keane_1997}).

\section{Conclusion \label{sec:conclusion}}

Average marginal effects are informative parameters that represent causal effects. They depend on the model's structural parameters and the unobserved heterogeneity distribution. In fixed effects nonlinear panel data models with short panels, the distribution of the unobserved heterogeneity is not identified, and this problem has been
associated with the common belief that AMEs are not identified.

In the context of dynamic logit models, we prove the identification of AMEs associated with changes in lagged dependent variables and duration variables. Our proofs of the identification results are constructive and provide simple closed-form expressions for the AMEs in terms of frequencies of choice histories that can be obtained from the data. We illustrate our identification results using simulated and real-world consumer scanner data in a dynamic demand model with state dependence.

In this paper, we have derived identification results only for logit models, but the procedure that arises from necessary and sufficient conditions may work beyond the logistic. In particular, it may work for any function that
shares with the logistic the property of having terms in which the fixed effect appears multiplicatively separated from other parameters of the model so that it is feasible to form polynomials of functions of the fixed effect. We leave this for future research.

\newpage

\bibliographystyle{econometrica}
\bibliography{references}

@article{aguirregabiria_gu_2021,
	author = {V Aguirregabiria and J Gu and Y Luo},
	title = {Sufficient statistics for unobserved heterogeneity in dynamic structural logit models},
	journal = {Journal of Econometrics},
	year = {2021},
	volume = {223},
	number = {2},
	pages = {280-311}
	}

@article{andersen_1970,
	author = {E Andersen},
	title = {Asymptotic properties of conditional maximum likelihood estimators},
	journal = {Journal of the Royal Statistical Society, Series B},
	year = {1970},
	volume = {32},
	number = {},
	pages = {283-301}
	}

@article{arellano_bonhomme_2017,
	author = {M Arellano and S Bonhomme},
	title = {Nonlinear panel data methods for dynamic heterogeneous agent models},
	journal = {Annual Review of Economics},
	year = {2017},
	volume = {9},
	number = {},
	pages = {471-496}
	}

@article{botosaru_muris_2024,
	author = {I Botosaru and C Muris},
	title = {Identification of time-varying counterfactual parameters in nonlinear panel models},
	journal = {Journal of Econometrics Journal},
	year = {2024},
	volume = {In Press}
	}

@article{bonhomme_2011,
	author = {S Bonhomme},
	title = {Panel Data, Inverse Problems, and the Estimation of Policy Parameters},
	year = {2011},
        journal = {Unpublished Manuscript}
	}

@article{chamberlain_1980,
	author = {G Chamberlain},
	title = {Analysis of covariance with qualitative data},
	journal = {Review of Economic Studies},
	year = {1980},
	volume = {47},
	number = {1},
	pages = {225-238}
	}

@incollection{chamberlain_1984,
	author = {G Chamberlain},
	title = {Panel data},
	booktitle = {Handbook of Econometrics, Vol. 2},
	year = {1984},
	pages = {},
	editors = {Z Griliches and M Intriligator},
	publisher = {North-Holland Press}
	}

@incollection{chamberlain_1985,
	author = {G Chamberlain},
	title = {Heterogeneity, omitted variable bias, and duration dependence},
	year = {1985},
	booktitle = {Longitudinal Analysis of Labor Market Data},
	pages = {},
	editors = {JJ Heckman and B Singer},
	publisher = {Cambridge University Press}
	}

@article{chernozhukov_fernandez_2013,
	author = {V Chernozhukov and I Fernandez-Val and J Hahn and W Newey},
	title = {Average and quantile effects in nonseparable panel models},
	journal = {Econometrica},
	year = {2013},
	volume = {81},
	number = {2},
	pages = {535-580}
	}

@article{dobronyi_gu_2021,
	title={Identification of Dynamic Panel Logit Models with Fixed Effects},
	author={C Dobronyi and J Gu and K Kim},
    journal={arXiv preprint arXiv:2104.04590},
	year={2021}
    }

@article{erdem_imai_2003,
  title={Brand and quantity choice dynamics under price uncertainty},
  author={T Erdem and S Imai and M Keane},
  journal={Quantitative Marketing and Economics},
  volume={1},
  number={1},
  pages={5-64},
  year={2003},
  publisher={Springer}
}

@article{hahn_2001,
	author = {J Hahn},
	title = {Comment: Binary regressors in nonlinear panel-data models with fixed effects},
	journal = {Journal of Business \& Economic Statistics},
	year = {2001},
	volume = {19},
	number = {1},
	pages = {16-17}
	}

@incollection{heckman_1981,
	author = {J Heckman},
	year = {1981},
	title = {The incidental parameters problem and the problem of initial conditions in estimating a discrete time - discrete data stochastic process},
	booktitle = {Structural Analysis of Discrete Data with Econometric Applications},
	pages = {},
	editors = {C Manski and D McFadden},
	publisher = {MIT Press}
	}

@article{hoderlein_white_2012,
  title={Nonparametric identification in nonseparable panel data models with generalized fixed effects},
  author={S Hoderlein and H White},
  journal={Journal of Econometrics},
  volume={168},
  number={2},
  pages={300--314},
  year={2012},
  publisher={Elsevier}
}

@article{honore_kyriazidou_2000,
	author = {B Honor\'{e} and E Kyriazidou},
	title = {Panel data discrete choice models with lagged dependent variables},
	journal = {Econometrica},
	year = {2000},
	volume = {68},
	number = {4},
	pages = {839-874}
	}

@article{honore_muris_2021,
  title={Dynamic Ordered Panel Logit Models},
  author={Honor{\'e}, Bo E and Muris, Chris and Weidner, Martin},
  journal={arXiv preprint arXiv:2107.03253},
  year={2021}
}

@article{honore_weidner_2020,
  title={Dynamic Panel Logit Models with Fixed Effects},
  author={B Honor\'{e} and M Weidner},
  year = {2020},
  journal={arXiv preprint arXiv:2005.05942}
  }

@article{lancaster_2000,
	author = {T Lancaster},
	title = {The incidental parameter problem since 1948},
	journal = {Journal of Econometrics},
	year = {2000},
	volume = {95},
	number = {2},
	pages = {391-413}
	}

@article{neyman_scott_1948,
	author = {J Neyman and E Scott},
	title = {Consistent Estimates Based on Partially Consistent Observations},
	journal = {Econometrica},
	year = {1948},
	volume = {16},
	number = {1},
	pages = {1-32}
	}

@article{rasch_1961,
	author = {G Rasch},
	title = {On General Laws and the Meaning of Measurement in Psychology},
	journal = {Proceedings of the Fourth Berkeley Symposium on Mathematical Statistics and Probability},
	year = {1961},
	volume = {4},
	number = {},
	pages = {321-333}
	}

@article{davezies_2022,
      title={Identification and Estimation of Average Marginal Effects in Fixed Effects Logit Models}, 
      author={Laurent Davezies and Xavier D'Haultfoeuille and Louise Laage},
      year={2022},
      journal={arXiv preprint arXiv:2105.00879}
}

@article{pakel_weidner_2023,
  title={Bounds on average effects in discrete choice panel data models},
  author={Pakel, Cavit and Weidner, Martin},
  journal={arXiv preprint arXiv:2309.09299},
  year={2023}
}

@article{honore_depaula_2021,
  title={Identification in simple binary outcome panel data models},
  author={Honor{\'e}, Bo E and De Paula, {\'A}ureo},
  journal={The Econometrics Journal},
  volume={24},
  number={2},
  pages={C78--C93},
  year={2021},
  publisher={Oxford University Press}
}

@article{abrevaya_hsu_2021,
  title={Partial effects in non-linear panel data models with correlated random effects},
  author={Abrevaya, Jason and Hsu, Yu-Chin},
  journal={The econometrics journal},
  volume={24},
  number={3},
  pages={519--535},
  year={2021},
  publisher={Oxford University Press}
}

@article{hausman_1978,
  title = {Specification tests in econometrics},
  author = {J Hausman},
  journal = {Econometrica},
  pages = {1251--1271},
  year = {1978},
  publisher={JSTOR}
}

@article{hausman_taylor_1981,
  title = {Panel data and unobservable individual effects},
  author = {J Hausman and W Taylor},
  journal = {Econometrica},
  pages = {1377--1398},
  year = {1981},
  publisher={JSTOR}
}

@article{erdem_1996,
  title = {A dynamic analysis of market structure based on panel data},
  author = {T Erdem},
  journal = {Marketing Science},
  volume = {15},
  number = {4},
  pages = {359--378},
  year = {1996},
  publisher={INFORMS}
}

@article{keane_1997,
  title = {Modeling heterogeneity and state dependence in consumer choice behavior},
  author = {M Keane},
  journal = {Journal of Business \& Economic Statistics},
  volume = {15},
  number = {3},
  pages = {310--327},
  year = {1997},
  publisher = {Taylor \& Francis Group}
}

@article{roy_chintagunta_1996,
  title = {A framework for investigating habits,“The Hand of the Past,” and heterogeneity in dynamic brand choice},
  author = {R Roy and P Chintagunta and S Haldar},
  journal = {Marketing Science},
  volume = {15},
  number = {3},
  pages = {280--299},
  year = {1996},
  publisher = {INFORMS}
}

@article{dube_hitsch_2010,
  title = {State dependence and alternative explanations for consumer inertia},
  author = {JP Dub{\'e} and G Hitsch and P Rossi},
  journal={The RAND Journal of Economics},
  volume={41},
  number={3},
  pages={417--445},
  year={2010},
  publisher={Wiley Online Library}
}

@article{seetharaman_2004,
  title={Modeling multiple sources of state dependence in random utility models: A distributed lag approach},
  author={PB Seetharaman},
  journal={Marketing Science},
  volume={23},
  number={2},
  pages={263--271},
  year={2004},
  publisher={INFORMS}
}

@article{seetharaman_ainslie_1999,
  title={Investigating household state dependence effects across categories},
  author={PB Seetharaman and A Ainslie and P Chintagunta},
  journal={Journal of Marketing Research},
  volume={36},
  number={4},
  pages={488--500},
  year={1999},
  publisher={SAGE Publications Sage CA: Los Angeles, CA}
}

@article{osborne_2011,
  title={Consumer learning, switching costs, and heterogeneity: A structural examination},
  author={M Osborne},
  journal={Quantitative Marketing and Economics},
  volume={9},
  number={1},
  pages={25--70},
  year={2011},
  publisher={Springer}
}

@article{pakes_porter_2021,
	author={A Pakes and J Porter and M Shepard and S Calder-Wang},
	year={2021},
	title={Unobserved Heterogeneity, State Dependence, and Health Plan Choices},
	journal = {National Bureau of Economic Research}
  	}

@article{viard_2007,
  title={Do switching costs make markets more or less competitive? The case of 800-number portability},
  author={B Viard},
  journal={The RAND Journal of Economics},
  volume={38},
  number={1},
  pages={146--163},
  year={2007},
  publisher={Wiley Online Library}
}

@article{magnac_2000,
  title={Subsidised training and youth employment: distinguishing unobserved heterogeneity from state dependence in labour market histories},
  author={T Magnac},
  journal={The Economic Journal},
  volume={110},
  number={466},
  pages={805--837},
  year={2000},
  publisher={Wiley Online Library}
}

@article{magnac_2004,
  title={Panel binary variables and sufficiency: generalizing conditional logit},
  author={T Magnac},
  journal={Econometrica},
  volume={72},
  number={6},
  pages={1859--1876},
  year={2004},
  publisher={Wiley Online Library}
}

\end{doublespacing}

\newpage

\begin{onehalfspacing}

\appendix

\section{Appendix}

\subsection{Proof of Lemma \ref{new_lemma_1} \label{appendix_proof_lemma_1}}

For notational simplicity, define $\alpha_{i,\mathbf{x}} \equiv \alpha_{i} + \mathbf{x}^{\prime} \boldsymbol{\gamma}$. Using the definition of $\Delta(\alpha_{i}, \mathbf{x})$, we have:
\begin{equation}
\begin{array}[c]{ccl}
    \Delta(\alpha_{i}, \mathbf{x}) 
    & = & 
    \dfrac{\exp \{ \alpha_{i,\mathbf{x}} + \beta \}}{1+\exp \{ \alpha_{i,\mathbf{x}} +\beta \}} -\dfrac{\exp \{ \alpha_{i,\mathbf{x}} \}}
    {1+\exp \{ \alpha_{i,\mathbf{x}}\}} =
    \dfrac{\exp \{ \alpha_{i,\mathbf{x}}\} \text{ } 
    \left[  e^{\beta} -1 \right]}
    {\left[  1 +\exp \{ \alpha_{i,\mathbf{x}}\} \right]  \text{ }
    \left[  1+\exp \{ \alpha_{i,\mathbf{x}}+\beta \} \right]  }
    \\ &  & \\
    & = & 
    \left[  e^{\beta}  -1\right]  \text{ }
    \pi_{01}(\alpha_{i}, \mathbf{x}) \text{ }
    \pi_{10}(\alpha_{i}, \mathbf{x})
    \text{.}
\end{array}
\end{equation}
that give us equation \eqref{eq_lemma_1_condition_1} in Lemma \ref{new_lemma_1}. We also have that: 
\begin{equation}
    \begin{array}[c]{ccl}
        \dfrac{\pi_{11}(\alpha_{i}, \mathbf{x})}{\pi_{10}(\alpha_{i}, \mathbf{x})}
        \dfrac{\pi_{00}(\alpha_{i}, \mathbf{x})}{\pi_{01}(\alpha_{i}, \mathbf{x})} 
        & = & 
        \dfrac{\exp \{ \alpha_{i,\mathbf{x}} +\beta \}
        /[1+\exp \{\alpha_{i,\mathbf{x}}+\beta \}]}{1/[1+\exp \{ \alpha_{i,\mathbf{x}}+\beta\}]}
        \dfrac{1/[1+\exp \{\alpha_{i,\mathbf{x}}\}}
        {\exp \{ \alpha_{i,\mathbf{x}}\}
        /[1+\exp \{ \alpha_{i,\mathbf{x}}\}]}\\
        &  & \\
        & = & 
        \dfrac{\exp \{ \alpha_{i,\mathbf{x}}+\beta \}}{\exp \{ \alpha_{i,\mathbf{x}}\}} =
        e^{\beta}
        \text{.}
    \end{array}
\end{equation}
that corresponds to equation \eqref{eq_lemma_1_condition_2} in Lemma \ref{new_lemma_1}. $\qquad \blacksquare$

\subsection{Proof of Proposition \ref{prop_1_ident_AME}
\label{appendix_proof_prop_1} }

W.l.o.g., we consider $T=3$.\footnote{Given identification with $T=3$, it is obvious that there is also identification for any value of $T$ greater than $3$, as we can take
sub-histories with three periods.} For any choice sequence $(y_{1}, y_{2}, y_{3})$ and any sequence of covariates $\mathbf{x}^{\{1,3\}}$ with $\mathbf{x}_{2} = \mathbf{x}_{3} = \mathbf{x}$:, we have:
\begin{equation}
    \mathbb{P}_{y_{1} y_{2} y_{3} \text{ } \vert \text{ } \mathbf{x}^{\{1,3\}} } =
    {\displaystyle \int}
    p^{\ast}(y_{1} | \alpha_{i}, \mathbf{x}^{\{1,3\}})
    \text{ } 
    \pi_{y_{1} y_{2}}(\alpha_{i}, \mathbf{x})
    \text{ }
    \pi_{y_{2} y_{3}}(\alpha_{i}, \mathbf{x})
    \text{ }
    f_{\alpha| \mathbf{x}^{\{1,T\}}}(\alpha_{i} | 
    \mathbf{x}^{\{1,3\}})
    \text{ }
    d\alpha_{i}
    \label{eq_proof_prop_1}
\end{equation}
Applying equation \eqref{eq_lemma_1_condition_1} in Lemma 1 to equation \eqref{eq_proof_prop_1} for $\mathbb{P}_{010 \text{ } \vert \text{ } \mathbf{x}^{\{1,3\}} }$ and $\mathbb{P}_{101 \text{ } \vert \text{ } \mathbf{x}^{\{1,3\}} }$:
\begin{equation}
    \left \{
    \begin{array}[c]{ccl}
        \mathbb{P}_{010 \text{ } \vert \text{ } 
        \mathbf{x}^{\{1,3\}} }
        & = & 
        \dfrac{1}{e^{\beta}  -1}
        {\displaystyle \int}
        p^{\ast}(0|\alpha_{i}, \mathbf{x}^{\{1,3\}})
        \text{ } 
        \Delta(\alpha_{i}, \mathbf{x}) \text{ } 
        f_{\alpha| \mathbf{x}^{\{1,T\}}}(\alpha_{i} | 
        \mathbf{x}^{\{1,3\}})
        \text{ } d\alpha_{i}
        \\ &  & \\
        \mathbb{P}_{101 \text{ } \vert \text{ } \mathbf{x}^{\{1,3\}} } 
        & = & 
        \dfrac{1}{e^{\beta}  -1}
        {\displaystyle \int}
        p^{\ast}(1|\alpha_{i}, \mathbf{x}^{\{1,3\}})
        \text{ } 
        \Delta(\alpha_{i}, \mathbf{x}) \text{ } 
        f_{\alpha| \mathbf{x}^{\{1,T\}}}(\alpha_{i} | 
        \mathbf{x}^{\{1,3\}})
        \text{ } d\alpha_{i}
    \end{array}
    \right.
\end{equation}
Adding up these two equations, multiplying the resulting equation times $e^{\beta}  -1$, and taking into account that $p^{\ast}( 0 | \alpha_{i}, \mathbf{x}^{\{1,3\}}) + p^{\ast}(1 | \alpha_{i}, \mathbf{x}^{\{1,3\}})=1$, we have that $AME(\mathbf{x})=$ $[e^{\beta}  -1]$ $[\mathbb{P}_{010 \text{ } \vert \text{ } (\mathbf{x}_{1}, \mathbf{x}, \mathbf{x})}$ $+\mathbb{P}_{101 \text{ } \vert \text{ } (\mathbf{x}_{1}, \mathbf{x}, \mathbf{x})}]$ such that $AME(\mathbf{x}, \mathbf{x}^{\{1,3\}})$ is identified. $\qquad \blacksquare$

\subsection{{Proof of Proposition \ref{prop_2_nec_suf_cond}
\label{appendix_proof_prop_2}}}

In this proof, for notational simplicity but w.l.o.g.,  we omit $\mathbf{x}^{\{1,T\}}$ and $\boldsymbol{\theta}$ as arguments in all the functions. Remember equation \eqref{eq_prop_2_system_restrictions} in Proposition 2:
\begin{equation}
    {\displaystyle \sum 
    \limits_{\mathbf{y}^{\{2,T\}} \in \mathcal{Y}^{T-1}}}
    w_{y_{1},\mathbf{y}^{\{2,T\}}} \text{ }
    G\left(  \mathbf{y}^{\{2,T\}}| y_{1}, \alpha_{i} \right)  
    \text{ } = \text{ } 
    \Delta(\alpha_{i}).
\label{eq_proof_prop_2}
\end{equation}

\bigskip

\noindent \textbf{(A) Sufficient condition.} Multiplying \eqref{eq_proof_prop_2} times $p^{\ast}(y_{1}|\alpha)$ $f_{\alpha}(\alpha)$, integrating over $\alpha$, and taking into account that, as defined in \eqref{Observed Probability}, $\textstyle \int
G(\mathbf{y}^{\{2,T\}}|y_{1},\alpha)$ 
 $p^{\ast}(y_{1}|\alpha)$ $f_{\alpha}(\alpha)$ $d\alpha$ is equal to $P_{\mathbf{y}^{ \{1,T\} }}$, we obtain:
\begin{equation}
    {\displaystyle \sum 
    \limits_{\mathbf{y}^{\{2,T\}} \in \mathcal{Y}^{T-1}}}
    w_{y_{1},\mathbf{y}^{\{2,T\}}}
    \text{ } P_{\mathbf{y}^{ \{1,T\} }} 
    \text{ } = \text{ }
    {\displaystyle \int} 
    \Delta(\alpha) 
    \text{ }
    p^{\ast}(y_{1} | \alpha) \text{ }
    f_{\alpha}(\alpha) \text{ } 
    d\alpha. 
    \label{proof_lemma_3_equ_3}
\end{equation}
We can sum equation \eqref{proof_lemma_3_equ_3} over all the possible values of $y_{1}$ in $\mathcal{Y}$. Given that the sum of $p^{\ast} (y_{1}|\alpha)$ over all values of $y_{1}$ in $\mathcal{Y}$ is equal to $1$, the right-hand-side becomes $\textstyle \int \Delta(\alpha)$ $f_{\alpha}(\alpha)$ $d\alpha$, which is the definition of $AME$. Furthermore, the sum of the equation \eqref{proof_lemma_3_equ_3} over all the possible values of $y_{1}$ implies the following equation: 
\begin{equation}
    {\displaystyle \sum \limits_{\mathbf{y}\in \mathcal{D}\times \mathcal{Y}^{T}}}
    w_{\mathbf{y}} \text{ } 
    P_{\mathbf{y}^{ \{1,T\} }}
    \text{ } = \text{ } 
    AME,
\end{equation}
which is equation \eqref{eq_prop_2_weighted_sum} in Proposition \ref{prop_2_nec_suf_cond}.

\bigskip

\noindent \textbf{(B) Necessary condition.} The proof has two parts. In part (i), we prove that function $h(P_{\mathcal{Y}|\mathcal{X}})$ should be linear in $P_{\mathcal{Y}|\mathcal{X}}$. Then, in part (ii), we show that the linearity of the function $h(P_{\mathcal{Y}|\mathcal{X}})$ implies that equation \eqref{eq_proof_prop_2} should hold.

\bigskip

\noindent \textbf{Necessary -- Part (i)}. Equality $h(P_{\mathcal{Y}|\mathcal{X}}) =AME$ should hold for every distribution $f_{\alpha}$. Consequently, it should hold for the following three specific cases: (Case 1) a degenerate distribution where $\alpha_{i}=c$ with probability one, where $c$ is constant; (Case 2) a degenerate distribution where 
$\alpha_{i}=c^{\prime}$ with probability one, where $c^{\prime}$ is a constant different to $c$; and (Case 3) a distribution with $c$ and $c^{\prime}$ as the only two points of support, with $q \equiv f_{\alpha}(c)$. For each of these three cases, $AME$ has the following form: $AME= \Delta(c)$ in 
Case 1; $AME=\Delta(c^{\prime})$ in Case 2; and $AME= q \text{ } \Delta(c) + (1-q) \text{ } \Delta(c^{\prime})$ in Case 3. Therefore, function $h(P_{\mathcal{Y}|\mathcal{X}})$ should satisfy the following three restrictions:
\begin{equation}
    \left \{
    \begin{array}[c]{ccl}
        \text{Case 1} & : & 
        h\left(
            \boldsymbol{P}_{\mathcal{Y} | \mathcal{X}}
        \right)  =\Delta(\mathbf{c})\\
        \text{Case 2} & : & 
        h\left(     
            \boldsymbol{P}_{\mathcal{Y} | \mathcal{X}}^{(2)}
        \right)  =
        \Delta(\mathbf{c}^{\prime}) 
        \\
        \text{Case 3} & : & 
        h\left(  
            \boldsymbol{P}_{\mathcal{Y} | \mathcal{X}}^{(3)}
        \right)  = 
        q \text{ }
        \Delta(\mathbf{c}) + 
        \text{ } (1-q)\ text{ }
        \Delta(\mathbf{c}^{\prime})
    \end{array}
    \right.  
\label{Proof Prop 7 - conditions 3 cases}
\end{equation}
Where $P_{\mathcal{Y}|\mathcal{X}}$, $P_{\mathcal{Y}
|\mathcal{X}}^{(2)}$, and $P_{\mathcal{Y}|\mathcal{X}}^{(3)}$ represent the distributions of $\mathbf{y}^{ \{1,T\}}$ conditional on $\mathbf{x}^{ \{1,T\}}$ under the DGPs of cases 1, 2, and 3, respectively. Note that, by construction:
\begin{equation}
    P_{\mathcal{Y}|\mathcal{X}}^{(3)} =
    q \text{ } P_{\mathcal{Y} |\mathcal{X}} +
    \left(  1-q \right)   \text{ }
    P_{\mathcal{Y}|\mathcal{X}}^{(2)}
\end{equation}
This condition should hold for any arbitrary values of the constants $c$, $c^{\prime}$, and $q \in \left[  0,1\right]$. Multiplying equation \eqref{Proof Prop 7 - conditions 3 cases}(Case 1) times $q$, multiplying equation \eqref{Proof Prop 7 - conditions 3 cases}(Case 2) times $(1-q)$, adding up these two results, and then subtracting equation
\eqref{Proof Prop 7 - conditions 3 cases}(Case 3), we get that function $h\left(  \boldsymbol{P}_{\mathcal{Y}|\mathcal{X}}\right)$ should satisfy the following equation:
\begin{equation}
    q \text{ } 
    h\left(         
        \boldsymbol{P}_{\mathcal{Y} | \mathcal{X}}
    \right)
    + \left(  1-q \right)  \text{ }
    h\left(  
        \boldsymbol{P}_{\mathcal{Y} | \mathcal{X}}^{(2)}
    \right)  = 
    h\left(  
        q \text{ } \boldsymbol{P}_{\mathcal{Y} | \mathcal{X}} + \left(  1-q\right)  
        \text{ } \boldsymbol{P}_{\mathcal{Y} | \mathcal{X}}^{(2)}
    \right). 
    \label{igualdad}
\end{equation}
The only possibility that equation (\ref{igualdad}) holds for any arbitrary value of $c$, $c^{\prime}$, and $q\in \left[0,1\right]$ is that function $h\left(  \boldsymbol{P}_{\mathcal{Y}|\mathcal{X}}\right)$ is linear in $P_{\mathcal{Y} | \mathcal{X}}$, such that $h\left(  \boldsymbol{P}_{\mathcal{Y} |\mathcal{X}}\right)  =$ $\textstyle \sum \nolimits_{\mathbf{y}^{\{1,T\}} } 
w_{\mathbf{y}^{\{1,T\}}}$ $P_{\mathbf{y}^{\{1,T\}}}$.

\bigskip

\noindent \textbf{Necessary -- Part (ii)}.We need to prove that, if equation $\textstyle \sum \nolimits_{\mathbf{y}^{\{1,T\}} }  w_{\mathbf{y}^{\{1,T\}}}$ $P_{\mathbf{y}^{\{1,T\}}} =AME$ holds, then equation \eqref{eq_proof_prop_2} should hold for every value $\alpha$. The proof is by contradiction. Suppose that: (a) equation  $\textstyle \sum \nolimits_{\mathbf{y}^{\{1,T\}} }  w_{\mathbf{y}^{\{1,T\}}}$ $P_{\mathbf{y}^{\{1,T\}}} =AME$ holds for any distribution $f_{\alpha}$ in the DGP; and (b) there is a value $\alpha=c$ and a value $y_{1})$ of the initial condition such that equation \eqref{eq_proof_prop_2} does not hold: $\textstyle \sum_{\mathbf{y}^{\{2,T\}}} w_{y_{1},y^{\{2,T\}}} \text{ } G\left(  \mathbf{y}^{\{2,T\}}|y_{1}, c \right)  \text{ }  \neq \text{ } \Delta(c)$. We show below that condition (b) implies that there is a density function $f_{\alpha}$ (in fact, a continuum of density functions) such that condition (a) does not hold.

W.l.o.g., consider distributions of $\alpha$ with only two points support, $c$ and $c^{\prime}$ with $f_{\alpha}(c)=q$. Define the following function $d(y_{1},\alpha)$ that measures the extent in which equation \eqref{eq_proof_prop_2} is not satisfied:
\begin{equation}
    d(y_{1},\alpha) \equiv 
    {\displaystyle 
    \sum_{\mathbf{y}^{\{2,T\}}}}
    w_{y_{1},\mathbf{y}^{\{2,T\}}} \text{ }
    G\left( \mathbf{y}^{\{2,T\}} | y_{1}, \alpha \right)  
    \text{ } - \text{ }
    \Delta(\alpha)
\end{equation}
Condition (b) implies that $d(y_{1},c) \neq 0$. For
notational simplicity but w.l.o.g., consider that the initial condition $y_{1})$ has binary support $\{0,1\}$. Applying the same operations as in the proof of the sufficient condition, we get:
\begin{equation}
    \begin{array}[c]{l}
    {\displaystyle \sum 
    \limits_{\mathbf{y}^{\{1,T\}}} }  
    w_{\mathbf{y}^{\{1,T\}}} \text{ }
    \mathbb{P}_{\mathbf{y}^{\{1,T\}}} - 
    AME = 
    \\
    q \text{ } 
    \left[  
        p^{\ast}(0|c) \text{ } d(0,c) + 
        p^{\ast}(1| c) \text{ } d(1,c)
    \right]  
    \text{ } + \text{ }
    (1-q) \text{ }
    \left[  
        p^{\ast}(0|c^{\prime}) 
        \text{ } d(0,c^{\prime}) + 
        p^{\ast}(1|c^{\prime}) 
        \text{ } d(1,c^{\prime})
    \right]
    \end{array}
\label{proof Prop 7(B)(ii)}
\end{equation}
By definition, each value $d(y_{1},\alpha)$ is for a
particular value of $\alpha$, and therefore, it does not depend on the distribution $f_{\alpha}$. More specifically, $d(y_{1},\alpha)$ does not depend on the value of $q$. Therefore, there always exists (a continuum of) values of $q$ such that the right-hand side of \eqref{proof Prop 7(B)(ii)} is different from zero, and condition (a) does not hold. $\qquad \blacksquare$

\subsection{Proof of Proposition \ref{prop_3_iden_ame} \label{appendix_proof_prop_3} }

\subsubsection{Part A: Polynomial in $e^{\alpha}$}

In this proof, for notational simplicity but w.l.o.g., we omit $\mathbf{x}$ and $\boldsymbol{\theta}$ as arguments in all the functions. Using the structure of the function
$G$ in the binary choice model, as presented in equation \eqref{eq_g_function_binary}, and the expression for $\Delta(\alpha)$ in Lemma \ref{new_lemma_1}, we can rewrite equation \eqref{eq_prop_2_system_restrictions} as follows:
\begin{equation}
    {\displaystyle \sum_{ \mathbf{y}^{ \{2,T\} }} }
    w_{y_{1},\mathbf{y}^{ \{2,T\} } } \text{ }
    { \displaystyle \prod_{t=2}^{T} }
    \frac{ e^{ y_{t} [\alpha + \beta y_{t-1}]} }
    {1 + e^{\alpha + \beta y_{t-1}}} =
    \left( e^{\beta} - 1 \right)
    \frac{e^{\alpha}}{1+e^{\alpha}} \text{ }
    \frac{1}{1+e^{\alpha + \beta}}
\label{proof_proposition_3_part_A_eq1}
\end{equation}
Multiplying this equation times $[1+e^{\alpha}]^{T-1} [1+e^{\alpha + \beta}]^{T-1}$ to eliminate the denominators, and using the Binomial Theorem to expand the terms $\left[  1+ e^{\alpha} \right]^{n}$ as $\sum_{k=0}^{n}\dbinom{n}{k} [e^{\alpha}]^{k}$, and the terms $\left[  1+ e^{\alpha + \beta} \right]^{n}$ as $\sum_{k=0}^{n}\dbinom{n}{k} [e^{\beta}]^{k} \text{ } [e^{\alpha}]^{k}$, we can represent this equation as a polynomial in $e^{\alpha}$. Therefore, this system of equations holds for every value $\alpha \in \mathbb{R}$ if and only if the coefficients multiplying each
monomial term in the polynomial are equal to zero. This result defines a finite system of equations. More specifically, the coefficients multiplying each monomial term are linear functions of the weights $w_{y_{1}, y^{\{2,T\}}}$. The finite system of equations that makes the monomial coefficients equal to zero is linear in the weights $w_{y_{1},y^{\{2,T\}}}$. $\qquad \blacksquare$

\subsubsection{Part B: Identification -- Model with covariates}

W.l.o.g. we consider $T=3$ and $t=3$. In this case, equation \eqref{eq_prop_2_system_restrictions} takes the following form:
\begin{equation}
    {\displaystyle \sum_{y_{2},y_{3}} }
    w_{ y_{1},y_{2},y_{3} | \mathbf{x}^{\{1,3\}}} \text{ }
    \frac{ e^{ y_{2} [\alpha + \beta y_{1} + \mathbf{x}_{2}^{\prime} \boldsymbol{\gamma}]} }
    {1 + e^{\alpha + \beta y_{1} + \mathbf{x}_{2}^{\prime} \boldsymbol{\gamma}}} 
    \frac{ e^{ y_{3} [\alpha + \beta y_{2} + \mathbf{x}_{3}^{\prime} \boldsymbol{\gamma}]} }
    {1 + e^{\alpha + \beta y_{2} + \mathbf{x}_{3}^{\prime} \boldsymbol{\gamma}}} 
    =
    \frac{\left( e^{\beta} - 1 \right) \text{ } e^{\alpha + \mathbf{x}_{3}^{\prime} \boldsymbol{\gamma}}} 
    { \left( 1+e^{\alpha + \mathbf{x}_{3}^{\prime} \boldsymbol{\gamma}} \right)
    \left(1+e^{\alpha + \beta + \mathbf{x}_{3}^{\prime} \boldsymbol{\gamma}} \right)
    } 
\end{equation}
Multiplying this equation by the factor that eliminates the denominators, we get the following equality of polynomials in $e^{\alpha}$:
\begin{equation}
\begin{array}[c]{rcl}
    &  & 
    w_{000} + w_{000} \text{ } 
    e^{\mathbf{x}_{3}^{\prime} \boldsymbol{\gamma}} e^{\beta} \text{ } e^{\alpha} +
    w_{001} \text{ } e^{\mathbf{x}_{3}^{\prime} \boldsymbol{\gamma}} \text{ } e^{\alpha} +
    w_{001} \text{ } (e^{\mathbf{x}_{3}^{\prime} \boldsymbol{\gamma}})^{2} e^{\beta} \text{ } (e^{\alpha})^{2} 
    \\
    & + &
    w_{010} \text{ }
    e^{\mathbf{x}_{2}^{\prime} \boldsymbol{\gamma}}
    \text{ } e^{\alpha} + 
    w_{010} \text{ } e^{\mathbf{x}_{2}^{\prime} \boldsymbol{\gamma}} e^{\mathbf{x}_{3}^{\prime} \boldsymbol{\gamma}} \text{ } (e^{\alpha})^{2}
    \\
    & + &
    w_{011} \text{ } e^{\mathbf{x}_{2}^{\prime} \boldsymbol{\gamma}} e^{\mathbf{x}_{3}^{\prime} \boldsymbol{\gamma}} e^{\beta} \text{ } (e^{\alpha})^{2} + 
    w_{011} \text{ } e^{\mathbf{x}_{2}^{\prime} \boldsymbol{\gamma}} (e^{\mathbf{x}_{3}^{\prime} \boldsymbol{\gamma}})^{2} e^{\beta} \text{ } (e^{\alpha})^{3} 
    \\
    & = &
    e^{\mathbf{x}_{3}^{\prime} \boldsymbol{\gamma}} (e^{\beta} -1) \text{ } e^{\alpha} + 
    e^{\mathbf{x}_{2}^{\prime} \boldsymbol{\gamma}} e^{\mathbf{x}_{3}^{\prime} \boldsymbol{\gamma}}
    (e^{\beta} -1) \text{ } (e^{\alpha})^{2}
\end{array}
\end{equation}
To ensure that a solution to this condition holds for every $\alpha$, the coefficients of each monomial term must be equal on both sides of the equation. Therefore, the solution implies the following four equations:
\begin{equation}
\begin{array}[c]{rcl}
    w_{000}  &  = & 0\\
    w_{000} \text{ }
    e^{\mathbf{x}_{3}^{\prime} \boldsymbol{\gamma}} 
    e^{\beta} + 
    w_{001} \text{ }
    e^{\mathbf{x}_{3}^{\prime} \boldsymbol{\gamma}} +
    w_{010} \text{ }
    e^{\mathbf{x}_{2}^{\prime} \boldsymbol{\gamma}} 
    &  = &
    e^{\mathbf{x}_{3}^{\prime} \boldsymbol{\gamma}} 
     (e^{\beta} -1)
    \\
    w_{001} \text{ }
    (e^{\mathbf{x}_{3}^{\prime} \boldsymbol{\gamma}})^{2} e^{\beta} +
    w_{010} \text{ }
    e^{\mathbf{x}_{2}^{\prime} \boldsymbol{\gamma}} 
    e^{\mathbf{x}_{3}^{\prime} \boldsymbol{\gamma}}+
    w_{011} \text{ }
    e^{\mathbf{x}_{2}^{\prime} \boldsymbol{\gamma}} 
    e^{\mathbf{x}_{3}^{\prime} \boldsymbol{\gamma}}
    e^{\beta}
    &  = &
    e^{\mathbf{x}_{2}^{\prime} \boldsymbol{\gamma}} 
    e^{\mathbf{x}_{3}^{\prime} \boldsymbol{\gamma}} 
     (e^{\beta} -1)
    \\
    w_{011} \text{ }
    e^{\mathbf{x}_{2}^{\prime} \boldsymbol{\gamma}} 
    (e^{\mathbf{x}_{3}^{\prime} \boldsymbol{\gamma}})^{2}
    e^{\beta}
    &  = & 0
\end{array}
\end{equation}
The unique solution to this system is $w_{000} = w_{011} = 0$, $w_{001} = e^{[\mathbf{x}_{2}-\mathbf{x}_{3}]^{\prime} \boldsymbol{\gamma}} -1$, and $w_{010} = e^{[\mathbf{x}_{3}-\mathbf{x}_{2}]^{\prime} \boldsymbol{\gamma}} e^{\beta} -1$.

We can proceed in the same way for the case of $y_{1}=1$ to obtain the following unique solution for the weights: 
$w_{100} = w_{111} = 0$, $w_{101} = e^{[\mathbf{x}_{2}-\mathbf{x}_{3}]^{\prime} \boldsymbol{\gamma}} e^{\beta} -1$, and $w_{110} = e^{[\mathbf{x}_{3}-\mathbf{x}_{2}]^{\prime} \boldsymbol{\gamma}} -1$.

Putting these pieces together, we have that:
\begin{equation}
    AME(\mathbf{x}_{3},\mathbf{x}^{\{1,3\}}) = 
        \begin{array}[c]{rr}
        & w_{0,0,1,\mathbf{x}^{\{1,3\}}}
        \text{ } 
        \mathbb{P}_{ 0,0,1 | \mathbf{x}^{ \{1,3\} } } +
        w_{0,1,0,\mathbf{x}^{\{1,3\}}}
        \text{ } 
        \mathbb{P}_{ 0,1,0 | \mathbf{x}^{ \{1,3\} } } 
        \\
        + & 
        w_{1,0,1,\mathbf{x}^{\{1,3\}}}
        \text{ } 
        \mathbb{P}_{ 1,0,1 | \mathbf{x}^{ \{1,3\} } } +
        w_{1,1,0,\mathbf{x}^{\{1,3\}}}
        \text{ } 
        \mathbb{P}_{ 1,1,0 | \mathbf{x}^{ \{1,3\} } }
        \end{array}
\end{equation}
with
\begin{equation}
    \begin{array}[c]{ll}
        w_{0,0,1,\mathbf{x}^{\{1,3\}}} = 
        -1 +
        e^{ 
            [\mathbf{x}_{2}-\mathbf{x}_{3}]^{\prime} 
            \boldsymbol{\gamma} 
            } 
        \text{ } ; \text{ } 
        w_{0,1,0,\mathbf{x}^{\{1,3\}}} = 
        -1 +
        e^{ 
            \beta + 
            [\mathbf{x}_{3}-\mathbf{x}_{2}]^{\prime} 
            \boldsymbol{\gamma} 
            } 
        \text{ } ; \text{ } 
        & \\ & \\
        w_{1,0,1,\mathbf{x}^{\{1,3\}}} = 
        -1 +
        e^{ 
            \beta + 
            [\mathbf{x}_{2}-\mathbf{x}_{3}]^{\prime} 
            \boldsymbol{\gamma} 
            } 
        \text{ } ; \text{ }
        w_{1,1,0,\mathbf{x}^{\{1,3\}}} = 
        -1 +
        e^{ 
            [\mathbf{x}_{3}-\mathbf{x}_{2}]^{\prime} 
            \boldsymbol{\gamma} 
            } 
        \text{.} &
    \end{array}
\end{equation} 

In the model without covariates, the vector of parameters $\gamma$ equals zero. Consequently, the weights $w_{001}$ and $w_{110}$ are also zero, and we get $AME=\left[ e^{\beta}-1\right]  $ $\left[  \mathbb{P}_{0,1,0}+\mathbb{P}_{1,0,1}\right]$. $\qquad \blacksquare$ 

\subsubsection{Over-identification when T is greater than three \label{appendix_proof_prop_3_Tgt3} }

The identification result using only $3$ periods proves identification for any $T\geq3$ because, with more than $3$ periods, we can always take $3$ periods. Nonetheless, it is possible to obtain close form expression for higher values of $T$ using the same procedure based on Proposition \ref{prop_2_nec_suf_cond}. This expression will use all $T$ periods without having to combine several $3$-periods estimates.

For $T>3$, as said, there is overidentification, so more than one combination of the probability of histories exists. One way to choose one of them is to focus on the probabilities of the sufficient statistics used to identify $\beta$ in the CMLE. Specifically, for this model and other logit models, the log-probability of a choice history has the following structure:
\begin{equation}
\ln \mathbb{P}\left(  \mathbf{y}_{i}|\alpha_{i}\mathbf{,\beta}\right)
=\mathbf{s}(\mathbf{y}_{i})^{\prime}\text{ }\mathbf{g}(\alpha_{i}%
)+\mathbf{c}(\mathbf{y}_{i})^{\prime}\text{ }\mathbf{\beta}
\label{prob_choice_history}
\end{equation}
where $s(y_{i})$ and $c(y_{i})$ are vectors of
statistics (functions of $y_{i}$), and $g(\alpha_{i})$ is a vector of functions $\alpha_{i}$. $s(y)$ is a sufficient statistic for $\alpha_{i}$ because $P(y_{i}|\alpha_{i},\beta
,s_{i})=$ $P(y_{i}|\beta,s_{i})$.\footnote{See
\citeauthor{aguirregabiria_gu_2021} (\citeyear{aguirregabiria_gu_2021}) for
further details on this decomposition of the probability choice and sufficient statistics for discrete choice logit models.} Let $S_{T}$ be the set of possible values of $s(y)$, let $P_{\mathbf{s}}$ be the probability of a value $s$ of $s(y)$, and let $P_{\mathbf{s}}\equiv \{P_{\mathbf{s}}:s\in S_{T}\}$ be the probability distribution of this statistic. Given $\theta$, the empirical distribution $P_{\mathbf{s}}$ contains all the information in the data about the distribution of $\alpha_{i}$, and therefore, about AMEs. Taking into account the structure of the probability of a choice history in the equation
(\ref{prob_choice_history}), the model implies:
\begin{equation}
\mathbb{P}_{\mathbf{s}}={\displaystyle \sum \limits_{\mathbf{y}:\text{
}\mathbf{s}(\mathbf{y})=\mathbf{s}}}\left[  {\displaystyle \int}\exp \{
\mathbf{s}^{\prime}\text{ }\mathbf{g}(\alpha_{i})+\mathbf{c}(\mathbf{y}%
)^{\prime}\text{ }\mathbf{\theta}\} \text{ }f_{\alpha}(\alpha_{i})\text{
}d\alpha_{i}\right]
\end{equation}

If two sequences, say $k$ and $l$, have the same
$s(y_{j})$, the ratio of the probabilities of these two sequences is equal to $\exp \left[  c(\mathbf{y}_{k})^{\prime} \text{ } \beta-c(\mathbf{y}_{l})^{\prime} \text{ } \beta \right]$, which is not a function of $\alpha_{i}$. This includes the case in which $P(y_{j}$  $|\beta,\alpha_{i})$ is the same for both sequences. Therefore, the set of sequences with the same or proportional $P(y_{j}$  $|\beta,\alpha_{i})$ is the set of sequences with the same value of the sufficient statistic $s(y_{j})$. This result leads to an infinite number of combinations of these sequences, with the only restriction
being that all the combinations have to sum up to the same number (overall weight). We choose the combination in which all these sequences have the same weight $w$, and, therefore, look for combinations of $P_{\mathbf{s}}$ instead of $P_{\mathbf{y}}$.

In the BC-AR1 model the sufficient statistics $s(y_{i})$ is the vector $\left(  y_{i1},y_{iT},\sum \limits_{t=2}^{T}y_{it}\right)  ^{\prime}$ --see \citeauthor{aguirregabiria_gu_2021}
(\citeyear{aguirregabiria_gu_2021})-- and it can take $4T-4$ different values, $2T-2$ values with $y_{i1}=0$ and $2T-2$ values with $y_{i1}=1$. The conditions from Proposition \ref{prop_2_nec_suf_cond} are:
\begin{equation}
\left.
\begin{array}[c]{ccc}%
{\displaystyle \sum \limits_{j=1}^{2T-2}}
w_{j}\text{ }\mathbb{P}\left(  \mathbf{\mathbf{s}_{j}}\text{ }|\text{ }%
y_{j1}=0,\beta,\alpha_{i}\right)  & = & \Delta(\alpha_{i})\\%
{\displaystyle \sum \limits_{j=2T-2+1}^{4T-4}}
w_{j}\text{ }\mathbb{P}\left(  \mathbf{\mathbf{s}_{j}}\text{ }|\text{ }%
y_{j1}=1,\beta,\alpha_{i}\right)  & = & \Delta(\alpha_{i})
\end{array}
\right \}  \text{ for every }\alpha_{i}\in%
\mathbb{R}
\text{,}
\end{equation}
where $P\left(  \mathbf{\mathbf{s}_{j}}\text{ }|\text{ } y_{j1}=0,\beta,\alpha_{i}\right)  =\sum_{\mathbf{y:s(y)=s}_{j}}$%
$P\left(  \mathbf{y}^{\{2,T\}}|y_{1}=0,\beta,\alpha_{i}\right)$

Proceeding similarly, we obtain the weights for $AME$ in the binary choice AR(1) model for different values of $T$. These are in Tables \ref{table_7} and \ref{table_8}.

\bigskip

\begin{table}
\caption{\textbf{Weights for histories with $y_{1}=0$} \label{table_7}}
\centering
\begin{tabular}[c]{l|l|l|l|l}
\hline \hline
$(y_{1},y_{T},{\textstyle \sum \limits_{t=2}^{T}}y_{t})$ & $T=4$ & $T=5$ & $T=6
$ & $T=7$\\ \hline
$\left(  0,0,0\right)  $ & $0$ & $0$ & $0$ & $0$\\ \hline
$\left(  0,0,1\right)  $ & $\tfrac{e^{\beta}-1}{2}$ & $\tfrac{e^{\beta}-1}{3}
$ & $\tfrac{e^{\beta}-1}{4}$ & $\tfrac{e^{\beta}-1}{5}$\\ \hline
$\left(  0,1,1\right)  $ & $0$ & $0$ & $0$ & $0$\\ \hline
$\left(  0,0,2\right)  $ & $0$ & $\tfrac{e^{\beta}-1}{1+2e^{\beta}}$ &
$\tfrac{2(e^{\beta}-1)}{3+3e^{\beta}}$ & $\tfrac{3(e^{\beta}-1)}{6+4e^{\beta}%
}$\\ \hline
$\left(  0,1,2\right)  $ & $\tfrac{e^{\beta}-1}{1+e^{\beta}}$ & $\tfrac
{e^{\beta}-1}{2+e^{\beta}}$ & $\tfrac{e^{\beta}-1}{3+e^{\beta}}$ &
$\tfrac{e^{\beta}-1}{4+e^{\beta}}$\\ \hline
$\left(  0,0,3\right)  $ & Not possible & $0$ & $\tfrac{e^{\beta}%
-1}{2+2e^{\beta}}$ & $\tfrac{(e^{\beta}-1)(1+2e^{\beta})}{1+6e^{\beta
}+3e^{2\beta}}$\\ \hline
$\left(  0,1,3\right)  $ & $0$ & $\tfrac{e^{\beta}-1}{2+e^{\beta}}$ &
$\tfrac{(e^{\beta}-1)(1+e^{\beta})}{1+4e^{\beta}+e^{2\beta}}$ & $\tfrac
{(e^{\beta}-1)(2+e^{\beta})}{3+6e^{\beta}+e^{2\beta}}$\\ \hline
$\left(  0,0,4\right)  $ & Not possible & Not possible & $0$ & $\tfrac
{e^{\beta}-1}{3+2e^{\beta}}$\\ \hline
$\left(  0,1,4\right)  $ & Not possible & $0$ & $\tfrac{e^{\beta}%
-1}{3+e^{\beta}}$ & $\tfrac{(e^{\beta}-1)(2+e^{\beta})}{3+6e^{\beta}%
+e^{2\beta}}$\\ \hline
$\left(  0,0,5\right)  $ & Not possible & Not possible & Not possible &
$0$\\ \hline
$\left(  0,1,5\right)  $ & Not possible & Not possible & $0$ & $\tfrac
{e^{\beta}-1}{4+e^{\beta}}$\\ \hline
$\left(  0,1,6\right)  $ & Not possible & Not possible & Not possible &
$0$\\ \hline \hline
\end{tabular}

\end{table}

\bigskip

\begin{table}
\caption{\textbf{Weights for histories with $y_{1}=1$} \label{table_8}}
\centering
\begin{tabular}[c]{l|l|l|l|l}
\hline \hline
$(y_{1},y_{T},{\textstyle \sum \limits_{t=2}^{T}}y_{t})$ & $T=4$ & $T=5$ & $T=6
$ & $T=7$\\ \hline
$\left(  1,0,0\right)  $ & $0$ & $0$ & $0$ & $0$\\ \hline
$\left(  1,0,1\right)  $ & $\tfrac{e^{\beta}-1}{1+e^{\beta}}$ & $\tfrac
{e^{\beta}-1}{2+e^{\beta}}$ & $\tfrac{e^{\beta}-1}{3+e^{\beta}}$ &
$\tfrac{e^{\beta}-1}{4+e^{\beta}}$\\ \hline
$\left(  1,1,1\right)  $ & $0$ & $0$ & $0$ & $0$\\ \hline
$\left(  1,0,2\right)  $ & $0$ & $\tfrac{e^{\beta}-1}{2+e^{\beta}}$ &
$\tfrac{(e^{\beta}-1)(1+e^{\beta})}{1+4e^{\beta}+e^{2\beta}}$ & $\tfrac
{(e^{\beta}-1)(2+e^{\beta})}{3+6e^{\beta}+e^{2\beta}}$\\ \hline
$\left(  1,1,2\right)  $ & $\tfrac{e^{\beta}-1}{2}$ & $\tfrac{e^{\beta}%
-1}{1+2e^{\beta}}$ & $\tfrac{e^{\beta}-1}{2+2e^{\beta}}$ & $\tfrac{e^{\beta
}-1}{3+2e^{\beta}}$\\ \hline
$\left(  1,0,3\right)  $ & Not possible & $0$ & $\tfrac{e^{\beta}%
-1}{3+e^{\beta}}$ & $\tfrac{(e^{\beta}-1)(2+e^{\beta})}{3+6e^{\beta}%
+e^{2\beta}}$\\ \hline
$\left(  1,1,3\right)  $ & $0$ & $\tfrac{e^{\beta}-1}{3}$ & $\tfrac
{2(e^{\beta}-1)}{3+e^{\beta}}$ & $\tfrac{(e^{\beta}-1)(1+2e^{\beta}%
)}{1+6e^{\beta}+3e^{2\beta}}$\\ \hline
$\left(  1,0,4\right)  $ & Not possible & Not possible & $0$ & $\tfrac
{e^{\beta}-1}{4+e^{\beta}}$\\ \hline
$\left(  1,1,4\right)  $ & Not possible & $0$ & $\tfrac{e^{\beta}-1}{4}$ &
$\tfrac{3(e^{\beta}-1)}{6+4e^{\beta}}$\\ \hline
$\left(  1,0,5\right)  $ & Not possible & Not possible & Not possible &
$0$\\ \hline
$\left(  1,1,5\right)  $ & Not possible & Not possible & $0$ & $\tfrac
{e^{\beta}-1}{5}$\\ \hline
$\left(  1,1,6\right)  $ & Not possible & Not possible & Not possible &
$0$\\ \hline \hline
\end{tabular}

\end{table}

\subsection{Proof of Proposition \ref{new_prop_4_iden_tranprob} \label{appendix_proof_prop_4} }

We omit $\mathbf{x}$ as an argument throughout this proof for notational simplicity. However, one should understand that the probability of the initial conditions $p^{\ast}$, the density function of $\boldsymbol{\alpha}_{i}$, the empirical
probabilities of choice histories and the average transition probabilities are all conditional on $\mathbf{x}_{i}^{\{1,3\}} = [\mathbf{x}_{1},\mathbf{x}
,\mathbf{x}]$. 

By definition of $\Pi_{00}$, and taking into account that $p^{\ast}(0|\alpha_{i}) + p^{\ast}(1|\alpha_{i}) = 1$, we have that:
\begin{equation}
    \Pi_{00}=
    {\displaystyle \int}
    \left[  
        p^{\ast}(0|\alpha_{i}) + 
        p^{\ast}(1|\alpha_{i}) 
    \right]  \text{ } 
    \pi_{00}(\alpha_{i}) \text{ } 
    f_{\alpha}(\alpha_{i}) \text{ } d\alpha_{i} 
\end{equation}
This expression includes the term ${\textstyle \int}
p^{\ast}(0|\alpha_{i})$ $\pi_{00}(\alpha_{i})$ $f_{\alpha}(\alpha_{i})$
$d\alpha_{i}$, which is equal to the choice history probability $\mathbb{P}_{00}$. However, it also includes the "counterfactual" ${\textstyle \int}
p^{\ast}(1|\alpha_{i})$ $\pi_{00}(\alpha_{i})$ $f_{\alpha}(\alpha_{i})$ $d\alpha_{i}$. Denote this counterfactual as $\delta_{100}$. Given that $\pi_{10}(\alpha_{i}) + \pi_{11}(\alpha_{i}) = 1$, we can represent this counterfactual as:
\begin{equation}
       \delta_{100} = 
        {\displaystyle \int}
        p^{\ast}(1|\alpha_{i}) \text{ }
        \left[  
            \pi_{10}(\alpha_{i}) + 
            \pi_{11}(\alpha_{i}) 
        \right]  
        \text{ } \pi_{00}(\alpha_{i}) 
        \text{ }
        f_{\alpha}(\alpha_{i}) 
\end{equation}
This equation shows that $\delta_{100}$ is the sum of two terms. The first term  is ${\textstyle \int}
p^{\ast}(1|\alpha_{i})$ $\pi_{10}(\alpha_{i})$ $\pi_{00}(\alpha_{i})$ $f_{\alpha}(\alpha_{i})$, which is equal to the choice history probability $\mathbb{P}_{100}$. The second term is ${\textstyle \int} p^{\ast}(1|\alpha_{i})$ $\pi_{11}(\alpha_{i})$ $\pi_{00}(\alpha_{i})$ $f_{\alpha}(\alpha_{i})$, which in principle is a counterfactual. However, by Lemma \ref{new_lemma_1}, we have that $\pi_{11}(\alpha_{i}) \pi_{00}(\alpha_{i})$ is equal to $e^{\beta}$ $\pi_{10}(\alpha_{i}) \pi_{01}(\alpha_{i})$. Therefore, the counterfactual ${\textstyle \int} p^{\ast}(1|\alpha_{i})$ $\pi_{11}(\alpha_{i})$ $\pi_{00}(\alpha_{i})$ $f_{\alpha}(\alpha_{i})$ is equal to $e^{\beta}$ ${\textstyle \int} p^{\ast}(1|\alpha_{i})$ $\pi_{10}(\alpha_{i})$ $\pi_{01}(\alpha_{i})$ $f_{\alpha}(\alpha_{i})$, and in turn this is equal to $e^{\beta}$ $\mathbb{P}_{101}$.

Putting all the pieces together, we have that:
\begin{equation}
    \Pi_{00} = 
    \mathbb{P}_{00} + 
    \mathbb{P}_{100} + 
    e^{\beta} \text{ } \mathbb{P}_{101}
\end{equation}
Using the same procedure, we can show that $\Pi_{11} = \mathbb{P}_{11} + \mathbb{P}_{011} + 
e^{\beta} \text{ } \mathbb{P}_{010}$. $\mathit{\qquad}\blacksquare$

\subsection{Proof of Lemma \ref{new_lemma_2} \label{appendix_proof_lemma_2_nperiod} }

For clarity in notation, we refrain from explicitly including $\mathbf{x}$ as an argument throughout this proof. It is important to note, however, that all probabilities and expectations in this proof are conditioned on $\mathbf{x}_{i}^{\{1,T\}} = [\mathbf{x}_{1}, \mathbf{x}, ..., \mathbf{x}]$, where $\mathbf{x}_{1}$ is free, and $\mathbf{x}_{2} = ... = \mathbf{x}_{T} = \mathbf{x}$. 

Using the Markov structure of the model and the chain rule, we have that:
\begin{equation}
    \begin{array}[c]{ccl}
        \mathbb{E}(y_{i,t+n} 
        \text{ } | \text{ } \alpha_{i}, y_{it}) 
        & = & 
        \mathbb{P} \left(  
            y_{i,t+n-1} = 0 \text{ } | \text{ } \alpha_{i} ,y_{it}
        \right)  \text{ }
        \pi_{01}(\alpha_{i}) + 
        \mathbb{P} \left(  
            y_{i,t+n-1} = 1 \text{ } | \text{ }
            \alpha_{i}, y_{it}
        \right)  \text{ }
        \pi_{11}(\alpha_{i}) 
        \\ &  & \\
        & = & 
        \pi_{01}(\alpha_{i}) + 
        \mathbb{E}\left(  
            y_{i,t+n-1} \text{ } | \text{ } 
            \alpha_{i}, y_{it}
        \right)  \text{ }
        \left[  
            \pi_{11}(\alpha_{i}) - 
            \pi_{01}(\alpha_{i})
        \right]
\end{array}
\label{eq_recursive_for_Lemma_1}
\end{equation}
Given the definition of $\Delta^{(n)}(\alpha_{i})$ as $\mathbb{E}(y_{i,t+n}$ $|$ $\alpha_{i},y_{it}=1)-\mathbb{E}(y_{i,t+n}|\alpha_{i},y_{it}=0)$, and
applying equation \eqref{eq_recursive_for_Lemma_1}, we have that:
\begin{equation}
    \begin{array}[c]{ccl}
        \Delta^{(n)}(\alpha_{i}) 
        & = & \
            \left[  
                \mathbb{E}\left(  
                    y_{i,t+n-1} \text{ } | \text{ }
                    \alpha_{i},y_{it}=1
                \right)  -
                \mathbb{E}\left(  
                    y_{i,t+n-1} \text{ } | \text{ }
                    \alpha_{i},y_{it}=0
                \right)  
            \right]  \text{ }
            \left[  
                \pi_{11}(\alpha_{i}) - 
                \pi_{01}(\alpha_{i})
            \right] 
            \\ &  & \\
            & = & 
            \Delta^{(n-1)}(\alpha_{i}) \text{ } 
            \left[  
                \pi_{11}(\alpha_{i}) - 
                \pi_{01}(\alpha_{i})
            \right]
\end{array}
\end{equation}
Applying this expression recursively, we obtain that $\Delta^{(n)}(\alpha_{i})=$ $[\pi_{11}(\alpha_{i})-\pi_{01}(\alpha_{i})]^{n}=$ $[\Delta (\alpha_{i})]^{n}$. Finally, as established in Lemma \ref{new_lemma_1},  is that $\Delta(\alpha_{i}) = \left[  e^{\beta}-1\right]  $ $\pi_{10}(\alpha_{i})$ $\pi_{01}(\alpha_{i})$. Thus, we have that $\Delta^{(n)}(\alpha_{i}) =$ $\left[  e^{\beta} - 1 \right]  ^{n}$ $\left[ \pi_{10}(\alpha_{i}) \right] ^{n}$ $\left[  \pi_{01}(\alpha_{i}) \right]^{n}$. $\qquad \blacksquare$

\subsection{Proof of Proposition \ref{new_prop_5}
 \label{appendix_proof_prop_5_AMEn} }

Similarly as in the proof of Lemma \ref{new_lemma_2} above, we omit $\mathbf{x}$ as an argument, but one should understand that all the probabilities in this proof are conditioned on $\mathbf{x}_{i}^{\{1,T\}} = [\mathbf{x}, \mathbf{x}, ..., \mathbf{x}]$. W.l.o.g., we consider that
$T=2n+1$. Given the definition of histories $(0,\widetilde{\mathbf{10}}^{n})$
and $(\widetilde{\mathbf{10}}^{n},1)$, it is straightforward to see that:
\begin{equation}
    \left \{
    \begin{array}[c]{ccl}
        \mathbb{P}_{0,\widetilde{\mathbf{10}}^{n}} 
        & = & 
        {\displaystyle \int}p^{\ast}(0|\alpha_{i}) 
        \text{ } 
        \left[  \pi_{10}(\alpha_{i}) \right]^{n} \text{ }
        \left[  \pi_{01}(\alpha_{i})\right]^{n} \text{ } 
        f_{\alpha}(\alpha_{i}) \text{ } d\alpha_{i} 
        \\ &  & \\
        \mathbb{P}_{\widetilde{\mathbf{10}}^{n},1} 
        & = & 
        {\displaystyle \int} 
        p^{\ast}(1|\alpha_{i}) \text{ }
        \left[ \pi_{10}(\alpha_{i}) \right]^{n} \text{ }
        \left[ \pi_{01}(\alpha_{i}) \right]^{n} \text{ }
        f_{\alpha}(\alpha_{i}) \text{ } d\alpha_{i}
    \end{array}
    \right.
\end{equation}
Applying equation \eqref{eq_lemma_indiv_nper} from Lemma
\ref{new_lemma_2}, we have that:
\begin{equation}
    \left \{
    \begin{array}[c]{ccl}
        \mathbb{P}_{0,\widetilde{\mathbf{10}}^{n}} 
        & = & 
        \dfrac{1}{\left[  \exp \{\beta \}-1\right]^{n}}{\displaystyle \int}p^{\ast}(0|\alpha_{i})
        \text{ } \Delta^{(n)}(\alpha_{i}) 
        \text{ } f_{\alpha}(\alpha_{i}) \text{ } d\alpha_{i}\\
        &  & \\
        \mathbb{P}_{\widetilde{\mathbf{10}}^{n},1} 
        & = & \dfrac{1}{\left[  \exp \{\beta \}-1\right]  ^{n}}{\displaystyle \int}p^{\ast}(1|\alpha_{i}) 
        \text{ } \Delta^{(n)}(\alpha_{i})
        \text{ } f_{\alpha}(\alpha_{i})
        \text{ } d\alpha_{i}
    \end{array}
    \right.
\end{equation}
Adding up these two equations, multiplying the resulting equation times $\left[  e^{\beta}-1\right]  ^{n}$, and taking into account that $p^{\ast}(0|\alpha_{i})+p^{\ast}(1|\alpha_{i})=1$, we have that $AME^{(n)}$ $=[e^{\beta}-1]^{n}$ $[\mathbb{P}_{0,\widetilde{\mathbf{10}}^{n}%
}+\mathbb{P}_{\widetilde{\mathbf{10}}^{n},1}]$ such that $AME^{(n)}$ is identified. $\qquad \blacksquare$

\subsection{Proof of Lemma \ref{new_lemma_3} \label{appendix_proof_lemma_3_mnl} }

Given the expression for the choice probabilities in the logit model, it is simple to verify that:
\begin{equation}
    \frac{\pi_{k \ell}(\boldsymbol{\alpha}_{i}, \mathbf{x})}
    {\pi_{k j}(\boldsymbol{\alpha}_{i}, \mathbf{x})} = 
    \exp \{
        \alpha_{i}(\ell)-\alpha_{i}(j) +
        \mathbf{x}^{\prime}
        \left(  \gamma_{\ell}-\gamma_{j}\right)  
        - \beta_{j} \mathbbm{1}\{ k=j\} 
    \}
\end{equation}
and
\begin{equation}
    \frac{\pi_{j j}(\boldsymbol{\alpha}_{i}, \mathbf{x})}
    {\pi_{j \ell}(\boldsymbol{\alpha}_{i}, \mathbf{x})} = 
    \exp \{
        \alpha_{i}(j)-\alpha_{i}(\ell) +
        \mathbf{x}^{\prime}
        \left(  \gamma_{j}-\gamma_{\ell} \right)  
        + \beta_{j} - \beta_{j} \mathbbm{1}\{ \ell = j\} 
    \}
\end{equation}
The product of these two expressions is equation 
\eqref{eq_lemma_3_mnl}. $\qquad \blacksquare$

\subsection{Proof of Proposition \ref{new_prop_6_duration} -- Model with duration dependence \label{sec:proof_prop_5} }

Here, we prove the identification of $AME_{d}(1)$. Using a similar argument, we can establish the identification of $AME_{d}(d)$ for any $d \geq 1$. First, we write the expression of the probabilities of choice histories conditional on $\alpha_{i}$, $P_{y_{1},y_{2},y_{3},y_{4} \text{ } | \text{ } \alpha_{i}}$, implied by the model. Second, for each of these probabilities, we multiply the equation times the weights $w_{y_{1},y_{2},y_{3},y_{4}}$ in the enunciate of Proposition \ref{new_prop_6_duration}. For the probabilities with non-zero weights, we have:
\begin{equation}
    \begin{array}[c]{rcl}
        \displaystyle \frac{e^{\beta + \delta}-1}{2}
        \left[  
            \mathbb{P}_{0,0,1,0 \text{ } | \text{ }
            \alpha_{i}} + 
            \mathbb{P}_{0,1,0,0 \text{ } | \text{ }
            \alpha_{i}}
        \right]
        & = & 
        p^{\ast}(0|\alpha_{i}) \text{ }
        \displaystyle 
        \frac{\left(  e^{\beta + \delta}-1 \right)  e^{\alpha_{i}}}{\left(  1+e^{\alpha_{i}+\beta + \delta}\right)
        \left(  1+e^{\alpha_{i}}\right)  ^{2}}
        \\ \\ 
        \displaystyle \frac{e^{\beta + \delta}-1}
        {e^{\beta + \delta}}
        \text{ } 
        \mathbb{P}_{0,0,1,1 \text{ } | \text{ } \alpha_{i}} & = & 
        p^{\ast}(0|\alpha_{i})
        \text{ } 
        \displaystyle \frac{\left(  e^{\beta + \delta}-1\right)  e^{\alpha_{i}}
        e^{\alpha_{i}}}
        {\left(  1+e^{\alpha_{i}+\beta + \delta}\right)  \left(
        1+e^{\alpha_{i}}\right)  ^{2}}
        \\ \\ 
        \left(  e^{\beta + \delta}-1\right)  
        \left[  
            \mathbb{P}_{1,0,1,0 \text{ } |\text{ } 
            \alpha_{i}} + 
            \mathbb{P}_{1,0,1,1 \text{ } |\text{ }
            \alpha_{i}} 
        \right]  
        & = &
        p^{\ast}(1|\alpha_{i}) \text{ }
        \displaystyle 
        \frac{\left(  e^{\beta + \delta}-1 \right)  e^{\alpha_{i}}
        \left(  e^{\alpha_{i}+\beta + \delta} +1 \right)}
        {\left(  1+e^{\alpha_{i}+\beta + \delta}\right)  ^{2}\left(  1+e^{\alpha_{i}}\right)  }
    \end{array}
\end{equation}
Third, we sum up these three equations. Simplifying factors and taking into account that $p^{\ast}(0|\alpha_{i})+$  $p^{\ast}(1|\alpha_{i})=1$, we get:
\begin{equation}
    \begin{array}[c]{ccc}
        \displaystyle 
        \frac{e^{\beta + \delta}-1}{2}
        \left[  
               \mathbb{P}_{0,0,1,0 \text{ } | \text{ }
               \alpha_{i}} + 
               \mathbb{P}_{0,1,0,0 \text{ } | \text{ }
               \alpha_{i}}
            \right]  +
            \displaystyle \frac{e^{\beta + \delta}-1}{e^{\beta + \delta}}
            \mathbb{P}_{0,0,1,1 \text{ } | \text{ } 
            \alpha_{i}} + 
            \left(  e^{\beta + \delta}-1\right)  
            \left[
                \mathbb{P}_{1,0,1,0 \text{ } | \text{ }
                \alpha_{i}} + 
                \mathbb{P}_{1,0,1,1 \text{ } | \text{ }
                \alpha_{i}}
            \right]  
            &  & 
            \\ \\ 
            = 
            \displaystyle \frac{\left(  e^{\beta + \delta}-1\right)  
            e^{\alpha_{i}}}
            {\left(1+e^{\alpha_{i}+\beta + \delta}\right)  \left(  1+e^{\alpha_{i}}\right)} = 
            \displaystyle \frac{e^{\alpha_{i}+\beta + \delta}}{\left(  1+e^{\alpha_{i}+\beta + \delta}\right)  } - 
            \displaystyle \frac{e^{\alpha_{i}}}
            {\left(1+e^{\alpha_{i}}\right) } = \Delta_{d}(\alpha_{i},1) 
            &  &
\end{array}
\end{equation}
Finally, we integrate the two sides of this equation over the
distribution of $\alpha_{i}$ to obtain
\begin{equation}
    \displaystyle 
    \frac{e^{\beta + \delta}-1}{2}
    \left[ 
        \mathbb{P}_{0,0,1,0} + \mathbb{P}_{0,1,0,0} 
    \right]  + 
    \displaystyle \frac{e^{\beta + \delta}-1}
    {e^{\beta + \delta}}
    \mathbb{P}_{0,0,1,1} + 
    \left(  e^{\beta + \delta}-1\right)  
    \left[
        \mathbb{P}_{1,0,1,0} + \mathbb{P}_{1,0,1,1}
    \right]  = 
    AME_{d}(1)
\end{equation}
such that $AME_{d}(1)$ is identified. 

We can proceed similarly to prove the identification of $AME_{d}(d)$ for any value $d \geq 1$. For instance, we can prove that:
\begin{equation}
    \begin{array}[c]{lll}
    AME_{d}(2)
    & = & 
    \displaystyle \frac{e^{\beta + 2 \delta}-1}{2}
    \left[
        \mathbb{P}_{0,0,1,0}+\mathbb{P}_{0,1,0,0}
    \right]  +
    \displaystyle 
    \frac{e^{\beta + 2 \delta}-1}{e^{\beta + \delta}}
    \text{ } \mathbb{P}_{0,0,1,1} 
    \\ \\
    & + & 
    \left(  \displaystyle 
    \frac{e^{\beta + 2 \delta}\left(  1-e^{\beta + 2 \delta} \right)}
    {e^{\beta + \delta}}+e^{\beta + 2 \delta}-1\right)  \mathbb{P}_{0,1,1,0}
    \\  \\
    & + & 
    \left(  e^{\beta + \delta}-\displaystyle 
    \frac{e^{\beta + \delta}}{e^{\beta + 2 \delta}}\right)  
    \left[  
        \mathbb{P}_{1,0,1,0} + \mathbb{P}_{1,0,1,1}
    \right]  +
    \left(
        \displaystyle \frac{e^{\beta + 2 \delta}-1}
        {e^{\beta + \delta}}-1 + 
        \displaystyle \frac{1}{e^{\beta + 2 \delta}} \right)  
        \mathbb{P}_{1,1,0,0} 
        \qquad \blacksquare
\end{array}
\label{AME_d=0_to_d=2}
\end{equation}

\subsection{Proof of Proposition \ref{new_prop_7_mnl} \label{appendix_prop_prop_7_mnl_tranprob} }

We omit $\mathbf{x}$ as an argument throughout this proof for notational simplicity. However, one should understand that the probability of the initial conditions $p^{\ast}$, the density function of $\boldsymbol{\alpha}_{i}$, the empirical
probabilities of choice histories and the average transition probabilities are all conditional on $\mathbf{x}_{i}^{\{1,3\}} = [\mathbf{x},\mathbf{x}
,\mathbf{x}]$. We can write $\Pi_{jj}$ as:
\begin{equation}
    \Pi_{jj}=
    {\displaystyle \int}
    \left[  p^{\ast}(0|\alpha_{i})+p^{\ast}(1|\alpha_{i})...+p^{\ast}(J|\alpha_{i})\right]  
    \text{ } \pi_{jj}(\alpha_{i}) \text{ } 
    f_{\alpha}(\alpha_{i}) \text{ } 
    d\alpha_{i} 
\label{Prop 7 step 1}
\end{equation}
This expression includes the term ${\textstyle \int}
p^{\ast}(j|\alpha_{i})$ $\pi_{jj}(\alpha_{i})$ $f_{\alpha}(\alpha_{i})$
$d\alpha_{i}$ that is equal to the choice history probability $\mathbb{P}%
_{j,j}$. However, it also includes the "counterfactuals" $\delta_{k,j,j} \equiv {\textstyle \int}
p^{\ast}(k|\alpha_{i})$ $\pi_{jj}(\alpha_{i})$ $f_{\alpha}(\alpha_{i})$ $d\alpha_{i}$ for $k\neq j$. We can represent each of these counterfactuals as:
\begin{equation}
    \delta_{k,j,j}=
    {\displaystyle \int}
    p^{\ast}(k|\alpha_{i})\text{ }\left[  \pi_{k0}(\alpha_{i})+\pi_{k1}(\alpha_{i})+...+\pi_{kJ}(\alpha_{i})\right]  \text{ }\pi_{jj}(\alpha_{i}) 
    \text{ }
    f_{\alpha}(\alpha_{i}) 
\label{Prop 7 step 2}
\end{equation}
That is, we have that $\delta_{k,j,j}=\sum_{\ell=0}^{J}\delta
_{k,\ell,j,j}^{(2)}$, with $\delta_{k,\ell,j,j}^{(2)} \equiv {\textstyle \int}
p^{\ast}(k|\alpha_{i})$ $\pi_{k\ell}(\alpha_{i})$ $\pi_{jj}(\alpha_{i})$
$f_{\alpha}(\alpha_{i})$. For $\ell=j$, we have that $\delta_{k,j,j,j}^{(2)}$
corresponds to the choice history probability $\mathbb{P}_{k,j,j}$. For the
rest of the terms $\delta_{k,\ell,j,j}^{(2)}$, we apply Lemma 3. According to
Lemma 3, we have that $\pi_{k\ell}(\alpha_{i})$ $\pi_{jj}(\alpha_{i})=\exp \{
\beta_{k\ell}-\beta_{kj}-\beta_{j\ell}\}$ $\pi_{kj}(\alpha_{i})$ $\pi_{j\ell
}(\alpha_{i})$. Finally, note that ${\textstyle \int}
p^{\ast}(k|\alpha_{i})$ $\pi_{kj}(\alpha_{i})$ $\pi_{j\ell}(\alpha_{i})$
$f_{\alpha}(\alpha_{i})$ is the choice history probability $\mathbb{P}_{k,j,\ell}$. Putting all the pieces together, we have the expression in
equation (\ref{eq:iden_PI_jj}). $\mathit{\qquad}\blacksquare$

\subsection{Proof of Proposition \ref{new_prop_8_mnl_noiden} \label{appendix_proof_prop_8_mnl_noiden} }

We omit $\mathbf{x}$ as an argument throughout this proof for notational simplicity. Equation \eqref{eq_prop_2_system_restrictions}, from Proposition \ref{prop_2_nec_suf_cond}, provides the necessary and sufficient condition for identifying an AME. Applying this condition to the model defined by equation \eqref{eq_MNL_model} and Assumption 1-MNL, with $J+1=3$, $T=3$, and $y_{1} =0$, we get:
\begin{equation}
    \begin{array}[c]{rrl}
        & w_{0,0,0} \text{ } 
        \mathbb{P}_{0,0,0 \text{ } | \text{ }(y_{1}=0,\alpha_{i})} +
        w_{0,0,1} \text{ } 
        \mathbb{P}_{0,0,1 \text{ } | \text{ }(y_{1}=0,\alpha_{i})} + 
        w_{0,0,2} \text{ } 
        \mathbb{P}_{0,0,2 \text{ } | \text{ }(y_{1}=0,\alpha_{i})} 
        &
        \\ \\ 
        + & w_{0,1,0} \text{ } 
        \mathbb{P}_{0,1,0 \text{ } | \text{ }(y_{1}=0,\alpha_{i})} + 
        w_{0,1,1} \text{ } 
        \mathbb{P}_{0,1,1 \text{ } | \text{ }(y_{1}=0,\alpha_{i})} + 
        w_{0,1,2} \text{ } 
        \mathbb{P}_{0,1,2 \text{ } | \text{ }(y_{1}=0,\alpha_{i})} 
        &
        \\ \\
        + & w_{0,2,0} \text{ } 
        \mathbb{P}_{0,2,0 \text{ } | \text{ }(y_{1}=0,\alpha_{i})} + 
        w_{0,2,1} \text{ } 
        \mathbb{P}_{0,2,1 \text{ } | \text{ }(y_{1}=0,\alpha_{i})} + 
        w_{0,2,2} \text{ } 
        \mathbb{P}_{0,2,2 \text{ } | \text{ }(y_{1}=0,\alpha_{i})} 
        & = 
        \pi_{10}(\alpha_{i})
    \end{array}
\label{no_iden_1}
\end{equation}

Let's denote $d_{j} \equiv 1 + e^{\beta_{j} \text{ } 1\{j=1\} + \alpha_{i}(1)} + e^{\beta_{j} \text{ } 1\{j=2\} + \alpha_{i}(2)}$, for $j=0,1,2$. Replacing the probabilities by their expression based on the logistic CDF:
\begin{equation}
    \begin{array}[c]{rrl}
    & \displaystyle
    w_{0,0,0} \frac{1}{d_{0}^{2}} + 
    w_{0,0,1} \frac{ e^{\alpha_{i}(1)} }{d_{0}^{2}} +
    w_{0,0,2} \frac{ e^{\alpha_{i}(2)} }{d_{0}^{2}} 
    &
    \\ \\
    + & \displaystyle
    w_{0,1,0} \frac{ e^{\alpha_{i}(1)} } {d_{0}d_{1}} +
    w_{0,1,1} \frac{ e^{\alpha_{i}(1)} \text{ }
    e^{\beta_{1} + \alpha_{i}(1)} }{d_{0}d_{1}} +
    w_{0,1,2} \frac{ e^{\alpha_{i}(1)} \text{ }
    e^{\alpha_{i}(2)}  }{d_{0}d_{1}} 
    &
    \\ \\
    + & \displaystyle
    w_{0,2,0} \frac{ e^{\alpha_{i}(2)} }{d_{0}d_{2}} +
    w_{0,2,1} \frac{  e^{\alpha_{i}(1)} \text{ }
    e^{\alpha_{i}(2)} }{d_{0}d_{2}} + 
    w_{0,2,2} \frac{ e^{\alpha_{i}(2)} \text{ } 
    e^{\beta_{2} + \alpha_{i}(2)} }{d_{0}d_{2}} 
    & \displaystyle = \frac{1}{d_{1}}
    \end{array}
\end{equation}
After some algebra to undo the fractions on both sides:
\begin{equation}
    \begin{array}[c]{rrl}
    & w_{0,0,0} \text{ } d_{1}d_{2} + 
    w_{0,0,1} \text{ } e^{\alpha_{i}(1)} d_{1}d_{2} +
    w_{0,0,2} \text{ } e^{\alpha_{i}(2)} d_{1}d_{2}
    & 
    \\ \\
    + & 
    w_{0,1,0} \text{ } e^{\alpha_{i}(1)} d_{0}d_{2} +
    w_{0,1,1} \text{ } e^{\alpha_{i}(1)}
    e^{\beta_{1} + \alpha_{i}(1)} d_{0}d_{2} + 
    w_{0,1,2} \text{ } 
    e^{\alpha_{i}(1)} e^{\alpha_{i}(2)} d_{0}d_{2}
    &
    \\ \\
    + & 
    w_{0,2,0} \text{ } e^{\alpha_{i}(2)} d_{0} d_{1} + 
    w_{0,2,1} \text{ } e^{\alpha_{i}(1)} 
    e^{\alpha_{i}(2)} d_{0}d_{1} +
    w_{0,2,2} \text{ } e^{\beta_{2} + \alpha_{i}(2)} 
    e^{\alpha_{i}(2)} d_{0}d_{1}
    & =d_{0}^{2}d_{2}
    \end{array}
\end{equation}

Expanding this equation by doing the products of $d_{j}$ and of the exponential, we obtain on both sides of the equality a polynomial in $(e^{\alpha_{i}(1)})^{h}$ $(e^{\alpha_{i}(2)})^{\ell}$, where the minimum value
of $h$ and $\ell$ is $0$, and the maximum value is
$4$. Following Lemma \ref{new_lemma_3}, equating the coefficient of each monomial in both sides of the equality, we get a system of linear
equation whose unknowns are the weights $w_{0,0,0},...,w_{0,2,2}$. This condition on the monomials of 
$(e^{\alpha_{i}(1)})^{2}$ and $(e^{\alpha_{i}(1)})^{3}$ imply, respectively: 

\begin{equation}
    \begin{array}[c]{rcl}
        w_{0,0,1} + w_{0,1,0} + 
        w_{0,0,1} \text{ } e^{\beta_{1}} + w_{0,1,0} 
        & = & 1
        \\ \\
        w_{0,0,1} \text{ } e^{\beta_{1}} + w_{0,1,0}
        & = & 0
    \end{array}
\end{equation}
which leads to 
\begin{equation}
    w_{0,0,1} + w_{0,1,0} = 1.
\label{no_iden_7}
\end{equation}
At the same time, the condition on the monomial of $(e^{\alpha_{i}(1)})^{1}$ implies:
\begin{equation}
    w_{0,0,1} + w_{0,1,0} = 2,
\end{equation}
which is incompatible with the condition $w_{0,0,1} + w_{0,1,0} = 1$. Therefore, there no exist weights that solve the system of equations in \ref{no_iden_1}. By Proposition \ref{prop_2_nec_suf_cond}, this implies that $\Pi_{10}$ is not point identified. $\qquad \qquad \blacksquare$

\subsection{Proof of Proposition \ref{iden_ordered_logit} -- Ordered Logit \label{proof_iden_ordered} }

The weights in the statement of Proposition \ref{iden_ordered_logit} were obtained using the general procedure established in Proposition 
\ref{prop_2_nec_suf_cond}. Here, we present the proof for $\Pi_{20}$, but it proceeds the same for other $\Pi_{kj}$. By definition, $\Pi_{20}=$ $\displaystyle  \int \pi_{20}(\alpha_{i})$ $f_{\alpha}(\alpha_{i})$  $d\alpha_{i}$, and according to equation \eqref{Ordered_probability}, $\displaystyle \pi_{20}(\alpha_{i})=\frac{1}{1+\exp \left(  \beta_{2} -\lambda_{0}+\alpha_{i}\right)  }$. We start with the probabilities of choice histories conditional on $\alpha_{i}$, that is, $P_{(y_{1},y_{2},y_{3}) \text{ } | \text{ } \alpha_{i}}$. We write
the expression for these model probabilities as functions of parameters $\boldsymbol{\beta}$,  $\boldsymbol{\lambda}$, and $\alpha_{i}$.
For each of these probabilities, we multiply the equation times the weights $w_{y_{1},y_{2},y_{3}}$ that appear in the statement of  Proposition \ref{iden_ordered_logit}. For the probabilities with non-zero
weights for $\Pi_{20}$, we have:
\begin{equation}
    \begin{array}[c]{rcl}
        \mathbb{P}_{0,0,0 
        \text{ } | \text{ } \alpha_{i}} + 
        \mathbb{P}_{0,0,1 
        \text{ } | \text{ } \alpha_{i}} 
        & = & 
        \displaystyle
        p^{\ast}(0|\alpha_{i}) \text{ } 
        \frac{1}
        {
        \left(
            1 + e^{\beta_{0}-\lambda_{1}+\alpha_{i}}
        \right)
        \left(
            1 + e^{\beta_{0}-\lambda_{0}+\alpha_{i}}
        \right)  
        }
        \\ \\
        e^{\lambda_{1}-\lambda_{0}} \text{ } 
        \mathbb{P}_{0,0,2 
        \text{ } | \text{ } \alpha_{i}} 
        & = & 
        \displaystyle
        p^{\ast}(0|\alpha_{i}) \text{ }
        \frac{
            e^{\beta_{0}-\lambda_{0}+\alpha_{i}}
        }
        {
        \left(  
            1 + e^{\beta_{0}-\lambda_{1}+\alpha_{i}}
        \right)  
         \left(  
            1 + 
            e^{\beta_{0}-\lambda_{0}+\alpha_{i}}
        \right)  
        }
        \\ \\
        \displaystyle
        \left(  
            1 - \frac{e^{\beta_{2}-\lambda_{0}} }
            {e^{\beta_{0}-\lambda_{1}} }
        \right)  
        \mathbb{P}_{0,2,0
        \text{ } | \text{ } \alpha_{i}}  
        & = & 
        \displaystyle
        p^{\ast}(0|\alpha_{i}) \text{ }
        \frac
        {
        e^{\beta_{0}-\lambda_{1}+\alpha_{i}}
        - e^{\beta_{2}-\lambda_{0}+\alpha_{i}}
        }
        {
        \left(  
            1 + 
            e^{\beta_{0}-\lambda_{1}+\alpha_{i}}
        \right)  
        \left(  
            1 + 
            e^{\beta_{2}-\lambda_{0}+\alpha_{i}}
        \right)  
        }
        \\ \\
        \displaystyle 
        \sum \limits_{k=0}^{l}
        \sum \limits_{\ell=0}^{J-1}
        \mathbb{P}_{1,k, \ell
        \text{ } | \text{ } \alpha_{i}} 
        & = & 
        \displaystyle
        p^{\ast}(1|\alpha_{i}) 
        \left(  
            \frac{1}
            {1+e^{\beta_{2}-\lambda_{0}+\alpha_{i}}} 
            - 
            \frac{1- 
            e^{\beta_{2}-\lambda_{0}+\alpha_{i}} }
            {\left(  
                    1 + 
                    e^{\beta_{1}-\lambda_{1}+\alpha_{i}}
                \right)  
                \left(  
                    1 + 
                    e^{\beta_{2}-\lambda_{0} + \alpha_{i}}
                \right)
            }
            \right) 
        \\ \\
        \displaystyle 
        \left(  
            1 - 
            \frac{ e^{\beta_{2}-\lambda_{0}} }
            { e^{\beta_{1}-\lambda_{1}} }
        \right)  
        \mathbb{P}_{1,2,0 
        \text{ } | \text{ } \alpha_{i}} 
        & = & 
        \displaystyle 
        p^{\ast}(1|\alpha_{i}) \text{ }
        \frac{
            1 - 
            e^{\beta_{2}-\lambda_{0}+\alpha_{i}}
        }
        {
        \left(
            1 + 
            e^{\beta_{1}-\lambda_{1}+\alpha_{i}}
        \right)
        \left(
            1 +
            e^{\beta_{2}-\lambda_{0}+\alpha_{i}}
        \right)
        }
        \\ \\ 
        \displaystyle 
        \sum \limits_{k=0}^{J-1}
        \mathbb{P}_{2,0,k
        \text{ } | \text{ } \alpha_{i}} 
        & = & 
        \displaystyle 
        p^{\ast}(2|\alpha_{i}) \text{ }
        \frac{1}
        {1 + e^{\beta_{2}-\lambda_{0}+\alpha_{i}} }
    \end{array}
\end{equation}
Summing up these equations, simplifying factors,  and taking into account that $p^{\ast}(0|\alpha_{i}
)+$  $p^{\ast}(1|\alpha_{i})+p^{\ast}(2|\alpha_{i})=1$, we get:
\begin{equation}
    \begin{array}[c]{rcl}
        \displaystyle 
        \mathbb{P}_{0,0,0 
        \text{ } | \text{ } \alpha_{i}} + 
        \mathbb{P}_{0,0,1
        \text{ } | \text{ } \alpha_{i}} +
        e^{\lambda_{1}-\lambda_{0}}
        \mathbb{P}_{0,0,2 
        \text{ } | \text{ } \alpha_{i}} + 
        \left(  
            1 - 
            \frac{e^{\beta_{2}-\lambda_{0}} }
            { e^{\beta_{0}-\lambda_{1}} }
        \right)
        \mathbb{P}_{0,2,0 
        \text{ } | \text{ } \alpha_{i}}
        & &
        \\ 
        \displaystyle
        + \sum \limits_{k=0}^{l}
        \sum \limits_{\ell=0}^{J}
        \mathbb{P}_{1,k, \ell
        \text{ } | \text{ } \alpha_{i}} +
        \left(
            1 - 
            \frac{e^{\beta_{2}-\lambda_{0}} } 
            { e^{\beta_{1}-\lambda_{1}} }
        \right)  
        \mathbb{P}_{1,2,0 
        \text{ } | \text{ } \alpha_{i}} + 
        \sum \limits_{k=0}^{J}
        \mathbb{P}_{2,0,k 
        \text{ } | \text{ } \alpha_{i}} 
        & = & \pi_{20}(\alpha_{i})
    \end{array}
\end{equation}
Finally, we integrate the two sides of this equation over the distribution of $\alpha_{i}$ to obtain the expression for $\Pi_{20}$ in equation \eqref{eq_proposition_ordered_logit}. 
$\qquad \blacksquare$

\end{onehalfspacing}

\end{document}